\pdfoutput=0 
\documentclass[12pt,onecolumn,journal]{IEEEtran}
\ifCLASSOPTIONonecolumn
\usepackage{setspace}
\else
\fi

\usepackage{amsmath,amssymb,amsthm}
\usepackage[noadjust]{cite}
\usepackage{gensymb}

\usepackage[font=footnotesize]{caption}
\usepackage{subcaption}
\usepackage{graphicx,fancyhdr,epstopdf}
\graphicspath{{./figures/}}
\usepackage{enumitem}
\usepackage[table]{xcolor}
\usepackage{multirow}
\usepackage{comment}
\usepackage{steinmetz} 
\usepackage{gensymb} 
\usepackage{mathrsfs}
\usepackage{siunitx}
\usepackage{algorithm}
\usepackage{algpseudocode}

\usepackage{array}
\usepackage{hhline}



\begin{document}
%

\title{RIS-Enabled Wireless Channel Equalization: Adaptive RIS Equalizer and Deep Reinforcement Learning
\thanks{The authors are with the Nhu Department of Electrical Engineering and Computer Science,
University of California, Irvine.}
\thanks{This work is partially supported by NSF grant 2030029.}
}

\ifCLASSOPTIONonecolumn
\author{\IEEEauthorblockN{Gal Ben-Itzhak, {\em Graduate Student Member, IEEE}}
\IEEEauthorblockN{\ and Ender Ayanoglu, {\em Fellow, IEEE}}\\
}
\fi

\maketitle
\begin{abstract}
    Reconfigurable Intelligent Surfaces (RISs) offer a promising means of reshaping the wireless propagation environment, yet practical methods for configuring large passive arrays to achieve reliable signal equalization remain limited. Equalization is essential in wideband links to counteract multipath-induced pulse distortion that otherwise degrades symbol recovery. This work investigates RIS-assisted pulse response equalization and signal boosting using both classical adaptive filtering and model-free deep reinforcement learning (DRL). We develop a steepest descent (SD) method that exploits cascaded BS-RIS-UE channel information to configure RIS coefficients for multipath mitigation and SNR enhancement, and we show that the tradeoffs between SD and DRL primarily arise from the extensive channel estimation required for accurate equalization with passive RIS hardware. Unlike traditional adaptive filtering, which updates delayed filter coefficients after signal reception, our approach uses the RIS positioned within the cascaded channel to perform equalization without delay elements, prior to reception at the UE. In this framework, the channel is estimated before equalization, forming the basis of what we term adaptive RIS equalization (ARISE). To overcome the reliance on channel estimation required for ARISE, we explore several DRL algorithms -- DDPG, TD3, and SAC -- that optimize RIS coefficients directly from the received pulse response without explicit channel estimation. Through extensive simulations across diverse channel conditions and RIS sizes, we show that SAC achieves fast, stable convergence and equalization performance comparable to ARISE while offering significantly lower implementation complexity. These results highlight the potential of DRL as a practical and scalable solution for real-time RIS control in future wireless systems.
\end{abstract}
\section{INTRODUCTION}

Reconfigurable Intelligent Surfaces (RISs) have emerged as a promising technology for next-generation wireless networks, offering the ability to reshape the propagation environment through nearly passive, low-power hardware \cite{9475160, 9086766}. By adjusting the phase of incident signals, RISs can enhance coverage, mitigate interference, and improve link reliability between the base station (BS) and user equipment (UE) without requiring active radio frequency (RF) chains. However, realizing these benefits in practice requires efficient methods for configuring large RIS arrays under realistic channel conditions. A central challenge is achieving reliable signal equalization and amplification without relying on full channel state information (CSI), which is difficult to obtain due to the cascaded BS-RIS-UE channel structure and the high dimensionality of RIS-assisted links \cite{9847080}. These limitations become even more pronounced in dynamic or richly scattered environments, where rapid adaptation is essential for maintaining link quality \cite{9328501}. These challenges motivate the need for approaches that can optimize RIS configurations without relying on full CSI acquisition or repeated pilot transmissions, particularly in dynamic environments.

A substantial body of work has explored RIS optimization using classical signal processing and model-based techniques \cite{zhou2023surveymodelbasedheuristicmachine, 9847080, 8741198, 9140329, 9180053, 9551980, 9769997, 10742896}. Many approaches assume access to accurate CSI and focus on maximizing received power, improving signal-to-noise ratio (SNR), maximizing sum rates, or enabling beamforming and focusing \cite{9424177, 8982186, 9570143, 10319318, 9110889, 9398559}. While effective in controlled scenarios, these methods face scalability issues as RIS sizes grow and channel estimation overhead becomes prohibitive. To address this, researchers have proposed reduced-complexity estimation schemes, compressive sensing techniques, and analytical optimization frameworks \cite{9328485}. In parallel, learning-based methods -- particularly deep reinforcement learning (DRL) -- have gained traction as a model-free alternative \cite{9110869, 10283517, 9110869, 10137638}. DRL enables RIS configuration directly from observed performance metrics, bypassing the need for explicit CSI and offering robustness in environments where analytical models are unreliable. Algorithms such as Deep Deterministic Policy Gradient (DDPG) \cite{9110869}, Twin Delayed DDPG (TD3) \cite{10118891}, and Soft Actor-Critic (SAC) \cite{10283517} have been applied to RIS beamforming, resource allocation, and adaptive control, demonstrating strong potential for real-time deployment due to their scalability and support for continuous action spaces. Moreover, their model-free nature allows them to operate effectively even when RIS hardware exhibits nonlinearities, coupling effects, or frequency-dependent behavior that are difficult to capture analytically.

Despite these advances, RIS-based signal equalization remains relatively underexplored. Equalization remains essential in high-speed links to mitigate intersymbol interference (ISI) caused by multipath distortion, making RIS-assisted temporal shaping particularly relevant. Most existing RIS literature focuses on maximizing received power or SNR, while the temporal shaping of the received signal -- critical for mitigating multipath distortion and ensuring reliable symbol recovery -- has received limited attention. A few recent works have begun to address this gap. For example, \cite{9414612} demonstrates that RISs can be configured to suppress or align multipath components prior to reception, effectively performing spatial-domain equalization. Similarly, \cite{10501013} shows that RISs can compensate for channel distortion with minimal receiver-side processing, enabling ``over-the-air equalization'' in constrained environments. However, these studies represent early steps, and the practical constraints of passive RIS hardware introduce additional challenges: effective equalization may require large or finely tuned coefficient sets, and there is an inherent tradeoff between signal gain and equalization accuracy. In this work, we purpose the RIS as an ``over-the-air equalizer,'' by placing the RIS at an intermediate physical layer node -- the midpoint of the BS-RIS-UE wireless link -- to equalize the entire signal \textit{before} the signal reaches the receiver. The RIS compensates the composite channel by directly shaping the electromagnetic propagation environment. We depart from conventional equalization approaches which correct distortion \textit{after} reception using digital baseband processing methods or RF front-ends. For more details, see Section III, where we derive an unconventional version of the steepest descent (SD) algorithm \cite{1141414} that optimizes the RIS reflection coefficients based on estimated cascaded BS-RIS-UE channels to equalize the signal. Although proving itself fruitful, this SD implementation depends on accurate cascaded CSI, which introduces substantial communication overhead and delays before RIS optimization can begin. The difficulties involved with this algorithm implementation, as well as the hardware-related challenges discussed in the previous paragraph, leave open questions about how classical adaptive filtering techniques compare with modern DRL approaches when applied to RIS-assisted equalization.

To address these gaps, this work presents a comprehensive study of RIS-assisted signal equalization and amplification using both classical and learning-based methods. We first develop an SD-based approach that leverages estimated cascaded BS-RIS-UE channels to configure RIS coefficients for pulse response equalization and SNR enhancement. The algorithm we develop, although based on SD, is different in implementation than conventional SD. For that reason, we name it adaptive RIS equalization (ARISE). We analyze the inherent tradeoffs associated with applying ARISE to passive RISs, including the extensive channel estimation required for accurate equalization and the resulting complexity implications. We then investigate model-free DRL algorithms -- DDPG, TD3, and SAC -- that optimize RIS coefficients directly from the received pulse response without requiring channel estimation overhead. Through extensive simulations across diverse channel conditions and RIS sizes, we demonstrate that SAC consistently achieves fast, stable convergence and performance comparable to ARISE, while offering significantly lower implementation complexity. In particular, SAC begins optimizing the RIS immediately from raw pulse response observations, avoiding the CSI acquisition delay inherent to ARISE and enabling faster adaptation in time-varying channels.

This work makes the following key contributions:

\begin{itemize}
    \item \textbf{RIS-based equalization and amplification framework:} We formulate the problem of RIS-assisted pulse response equalization and signal boosting and analyze its challenges under realistic passive RIS constraints.
    \item \textbf{SD-based RIS optimization:} We derive an SD algorithm that uses cascaded channel estimates to configure RIS coefficients and characterize its performance, complexity, and tradeoffs. Because of its difference with respect to conventional SD, we call this algorithm ARISE.
    \item \textbf{Model-free DRL methods for RIS control:} We develop and evaluate DDPG, TD3, and SAC algorithms that optimize RIS coefficients directly from received signal observations, eliminating the need for channel estimation.
    \item \textbf{Comprehensive comparative analysis:} We compare ARISE and DRL approaches in terms of convergence behavior, computational complexity, communication complexity, equalization performance, and robustness across channel types and RIS sizes.
    \item \textbf{Identification of SAC as a practical solution:} We show that SAC achieves rapid, stable convergence and performance on par with ARISE, positioning it as a strong candidate for real-time RIS control in future wireless systems.
\end{itemize}

The paper is organized as follows: Section II presents the system and channel models, and defines the optimization problem. Section III derives the ARISE algorithm and provides simulation results to compare with baseline cases. Section IV introduces the three different DRL algorithms and discusses their complexities. Section V shows comparative simulation results among the ARISE and DRL algorithms across diverse environmental models. 
\section{SYSTEM MODEL AND PROBLEM FORMULATION}

\subsection{SYSTEM MODEL}

The RIS-assisted wireless communication network consists of a single-antenna BS, an $M$-element metasurface-based RIS, and a single-antenna UE. The RIS elements are metallic patches arranged in a periodic linear array structure and separated by a distance $d_x$ along the $x$ direction, electrically controlled by varactor diodes with continuous voltage biases to modulate the reflection phases. The reflection coefficient of each RIS element $m$ is given by $\Gamma_m=|\Gamma_m|e^{j\phi_m}$, with phase shift $\phi_m$. To enforce the passivity of the RIS, we assume for simplicity that $|\Gamma_m|=1$ for all $m=1,2,\ldots,M$.

\begin{figure}[!htbp]
    \centering
    \includegraphics[width=0.7\linewidth]{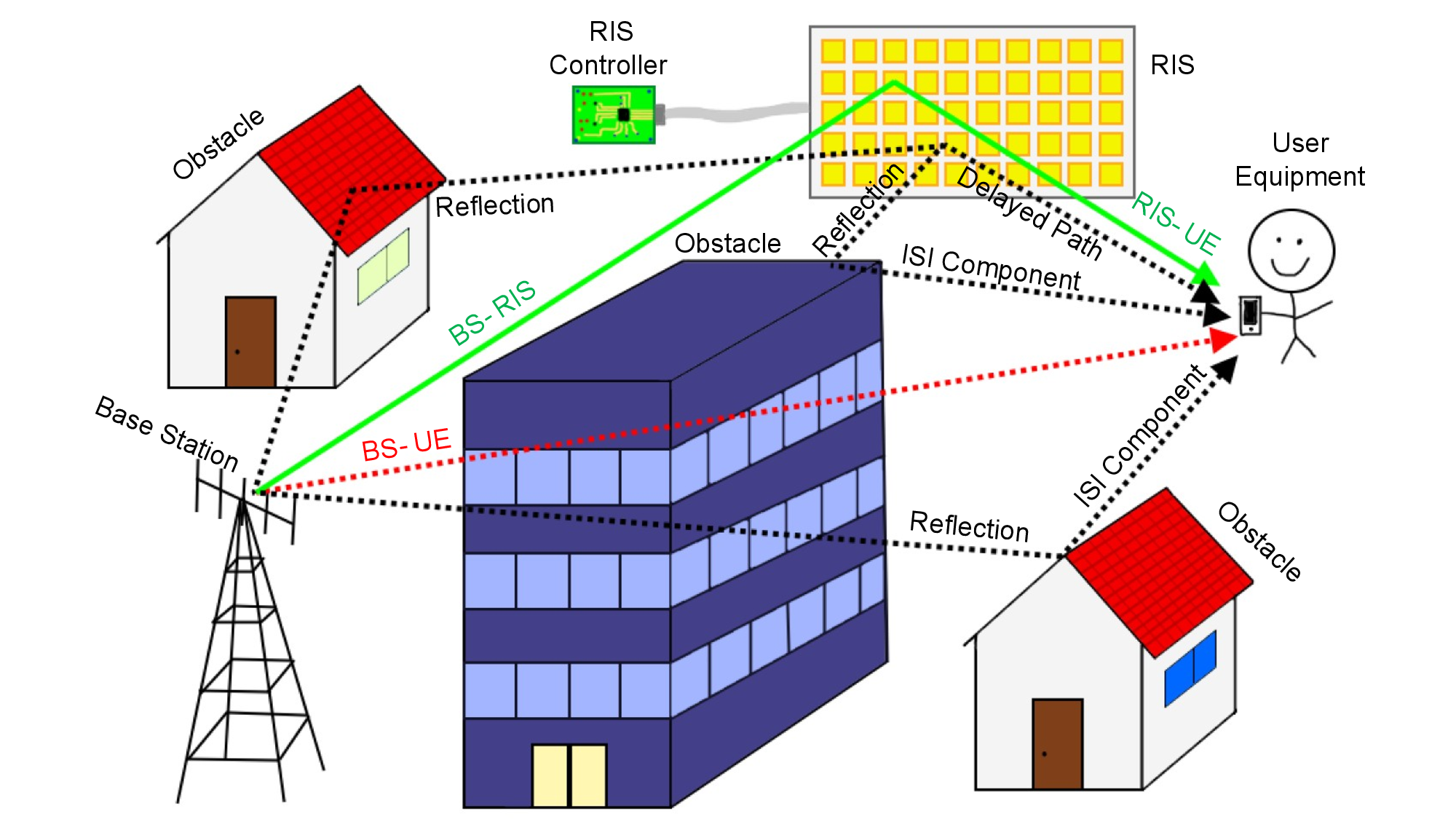}
    \vspace{3mm}
    \caption{Illustration of the downlink RIS scenario. LoS paths are marked by solid lines, while NLoS paths are marked by dotted lines. Delayed paths and ISI components are created by reflections from obstacles.}
    \label{fig:illustration}
\end{figure}
In the downlink scenario shown in Fig.~\ref{fig:illustration}, the UE receives two main rays of signals: the first coming directly from the BS and the second is the reflected signal from the BS through the RIS. All channels are assumed to be far-field and frequency-selective. We assume that there is no dominant LoS path between the BS and the UE due to obstacles, thus the channel impulse response is mainly characterized by Rayleigh coefficients
\begin{equation}
    h_{\text{BU}}(t)=(\beta_{\text{BU}})^{1/2} h_{\text{NLoS}}(t),
\end{equation}
where $\beta_{\text{BU}}$ is the path loss related to the distance between the BS and the UE, $\beta_{\text{BU}}=Gd_{\text{BU}}^{-\alpha}$. Here, $G$ accounts for transmitter gains, receiver gains, and normalized path losses, $d_{\text{BU}}$ is the distance between the BS and the UE, and $\alpha$ is the path loss exponent \cite{9414612, 7109864}. The NLoS components are distributed as standard circularly symmetric complex Gaussian random variables $h_{\text{NLoS}}(t) \in \mathcal{CN}(0,\sigma^2(t))$ whose variance $\sigma^2(t)$ depends on the arrival time of the signal at the receiver due to the delayed channel components from multipath effects, with $\sigma^2(0)=1$ \cite{1146527}.

The channels characterizing the reflection links between the BS and UE through the RIS elements are modeled by the impulse response vector
\begin{equation}
    \boldsymbol{h}_{\text{BRU}}(t)=(\beta_{\text{BRU}})^{1/2}\boldsymbol{R}^{1/2}\boldsymbol{\Gamma} \boldsymbol{h}(\theta_{\text{BRU}},t),
\end{equation}
where $\beta_{\text{BRU}}=G'(d_{\text{BR}}d_{\text{RU}})^{-\alpha}$ is the path loss related to the distances between the BS and the center of the RIS, and between the center of the RIS and the UE, denoted by $d_{\text{BR}}$ and $d_{\text{RU}}$, respectively \cite{9201413, 8990007}. We define the steering vector by the RIS for the LoS channels as a function of the azimuth angle $\theta$, $\boldsymbol{h}_{\text{LoS}}(\theta)=\left[1,e^{j\omega(\theta)},\ldots,e^{j(M-1)\omega(\theta)}\right]^T$ with spatial frequencies $\omega(\theta)=2\pi d_x (f_c/c) \sin(\theta)$, where $f_c$ is the carrier frequency and $c$ is the speed of light \cite{9847080}.  Combining the above, we express the fading parameters of the channels as the Rician coefficients
\begin{equation}
    \boldsymbol{h}(\theta,t)=
    \sqrt{\frac{\kappa}{\kappa+1}}\boldsymbol{h}_{\text{LoS}}(\theta) + \sqrt{\frac{1}{\kappa+1}}\boldsymbol{h}_{\text{NLoS}}(t),
\end{equation}
where $\kappa,\; 0\leq \kappa < \infty$, is the Rician factor. The BS illuminates the RIS with normal incidence and the azimuth angle between the BS, RIS, and UE is $\theta_{\text{BRU}}$. We assume the LoS channel component is only valid for the first signal sampling instance at $t=0$ where most of the signal energy is received. Afterwards, the delayed and distorted signal components are modeled by Rayleigh fading using $\boldsymbol{h}_{\text{NLoS}}(t)$. Additionally, the channels are correlated by the positive semidefinite spatial correlation matrix $\boldsymbol{R}$ with elements
\begin{equation}
    [\boldsymbol{R}]_{n,m}=\text{sinc} \left( \frac{2||u_n-u_m||}{c/f_c}\right) \quad n,m=1,\ldots,M,
\end{equation}
where $u_n$ and $u_m$ are the positions along $x$ of the RIS elements $n$ and $m$, respectively, and $\text{sinc}(x)=\sin(\pi x)/(\pi x)$ is the sinc function \cite{9300189, 9776512}.

For a transmitted signal $s(t)$, the received baseband signal at the UE is expressed as 
\begin{equation}
    y(t)=\left(h_{\text{BU}}(t)+ \sum_{m=1}^{M}{\Gamma_m h_{\text{BRU},m}(t)} \right)*s(t) + n(t),
    \label{eq:rx_sig}
\end{equation}
where $n(t)$ is additive white Gaussian noise (AWGN) and $*$ is the convolution operator.

\subsection{PROBLEM FORMULATION}
Our objective is to eliminate the ISI of the received signal while maximizing the received signal power. This is achieved by utilizing the passive RIS as both a spatial equalizer and a beamformer. Although the RIS does not have any delay components, unlike traditional equalizers such as the feed forward equalizer (FFE) or decision feedback equalizer (DFE), its phase shifts can be used with delays from the multipath propagation in the environment to orthogonalize the channel components that create the ISI.

Mathematically, we define the ISI $I$ of the received pulse response $y(t)$ as the delayed signal content arriving at multiples of the sampling period $T$. Assuming the peak of the signal arrives at the receiver at time $t=0$, the ISI is expressed as
\begin{equation}
    I=\sum_{k \neq 0}{y(t-kT)} |_{t=0},
\end{equation}
where the transmitted discrete-time baseband signal is the pulse given by
\begin{equation}
    s_k=
    \begin{cases}
        1 & k=0 \\
        0 & k\neq 0.
    \end{cases}
\end{equation}
The optimization problem is defined as
\begin{equation}
    (P1):\qquad \max_{\boldsymbol{\theta}} \quad \eta,
\end{equation}
where
\begin{equation}
    \eta = \text{sgn} \big(\Re\{y(0)\}\big) \cdot \big[\Re\{y(0)\}\big]^2 - \sum_{k \neq 0} \left| y(t-kT) \right|^2.
    \label{eq:eta}
\end{equation}

The metric $\eta$ is designed to explicitly reward a strong, correctly phased main tap of the received pulse response while penalizing residual ISI energy. The first term, $\text{sgn} \big(\Re\{y(0)\}\big) \cdot \big[\Re\{y(0)\}\big]^2$, promotes a large real-valued peak at $t=0$ (consistent with the real-valued transmitted pulse) and enforces phase alignment through the sign factor. Otherwise, the received symbols may be rotated over the real and imaginary axes, as will be seen in some of the plots in Figs.~\ref{fig:sim_9_qpsk} and~\ref{fig:sim_1_qpsk}. The second term, $- \sum_{k \neq 0} \left| y(t-kT) \right|^2$, suppresses the power of all the ISI components. 
\section{ADAPTIVE RIS EQUALIZATION}

\subsection{THE ARISE ALGORITHM}

\begin{figure}
    \centering
    \includegraphics[width=1.0\linewidth]{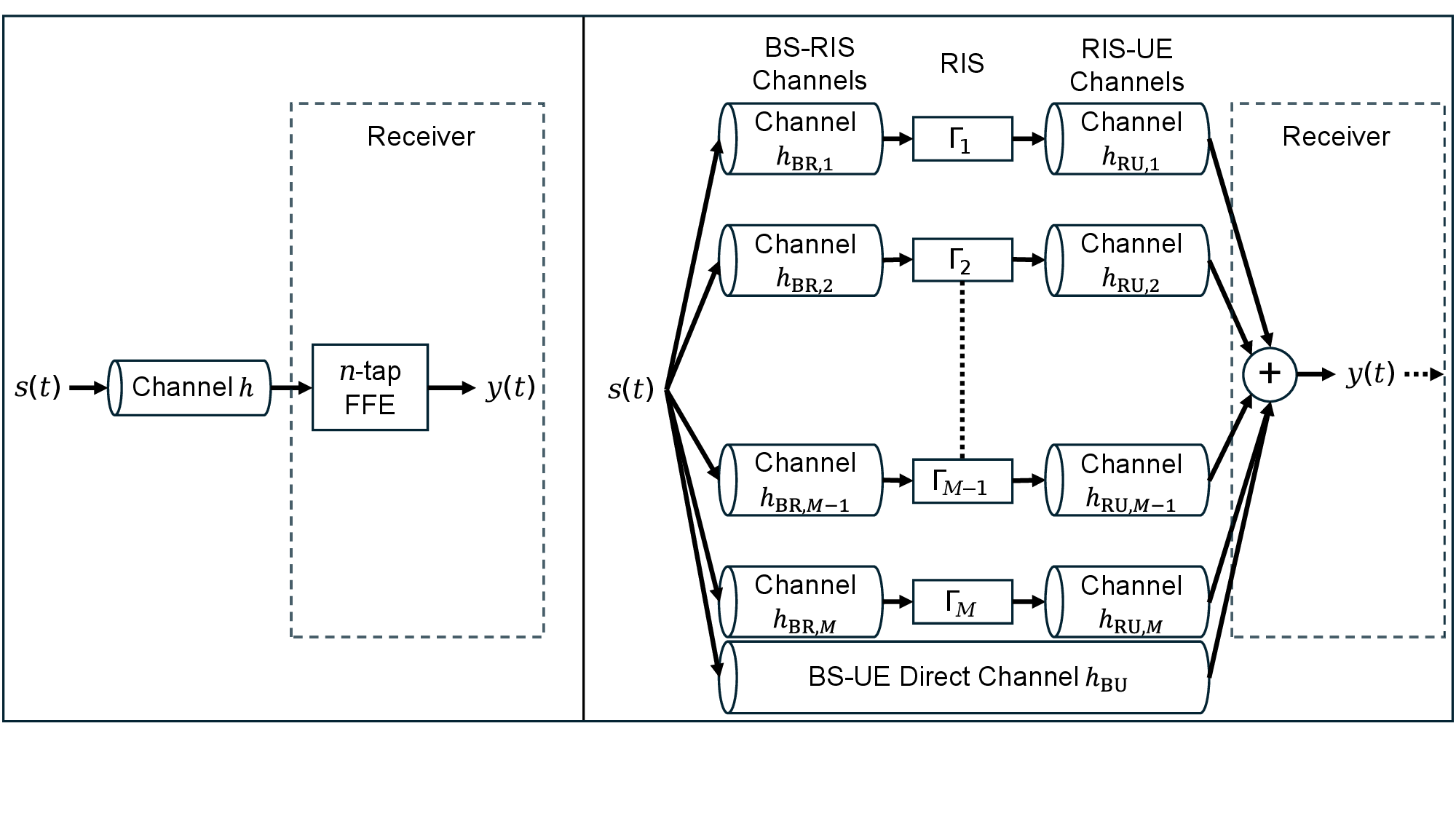}
    \caption{Left: conventional feedforward equalizer (FFE) at the receiver vs.~right: our proposed RIS-based equalizer. The conventional equalizer processes the baseband signal after reception using $n$ delay taps, whereas the RIS-based equalizer is at the midpoint of the cascaded BS-RIS-UE link and equalizes the signal before reception by using each one of its $M$ reflection coefficients and their corresponding channels.}
    \label{fig:ris_eq_SD}
\end{figure}

A popular approach to minimize the ISI of a signal, generally when designing equalizers in communication systems, is by using the SD algorithm to adaptively adjust the equalizer coefficients to produce a desired signal \cite{sayed2011adaptive}. This is done either by sending a pilot pattern, such as a pseudo-random binary sequence (PRBS), or with single-bit pulses to capture the pulse response of the channel for the given data rate \cite{1457566}. In the case of the RIS-based equalizer, the equalizer coefficients would be the RIS reflection phases, as demonstrated in Fig.~\ref{fig:ris_eq_SD}. Note the important distinction in Fig.~\ref{fig:ris_eq_SD} that while the conventional SD algorithm on the left hand side achieves channel equalization via the memory elements in the $n$-tap FFE, the structure we propose on the right hand side of Fig.~\ref{fig:ris_eq_SD} does not have any memory elements and uses $\Gamma_1, \Gamma_2, \ldots, \Gamma_M$ for the purposes of spatial equalization. In the conventional equalizer, an SD algorithm would only need to process the input signal at the receiver, $x(t)=s(t)*h(t)$, where $h(t)$ is the channel from the transmitter to the receiver. Assuming $s(t)$ is known, then the filter coefficients can be tuned directly to minimize the squared magnitude of the error signal $e(t)=s(t)-y(t)$ at the desired sampling instances $t=kT$, where $y(t)=x(t)*c(t)$ and $c(t)$ is the impulse response of the equalizer.

In the case of the RIS-based equalizer, the RIS reflection coefficients $\boldsymbol{\Gamma}$ are the equalizer coefficients that when multiplied by the cascaded channels $\boldsymbol{h}_{\text{BRU}}$, they result in phase and magnitude-shifted versions of the transmitted signal $s(t)$, each given by $x_m(t)=\Gamma_m h_{\text{BRU},m}(t)*s(t)$ for $m=1,2,\ldots,M$. The resulting signal at the receiver is a combination of the RIS-aided propagation and the direct channel from the BS to UE and can be written as $y(t)=s(t)*h_{\text{BU}}(t)+\sum_{m=1}^{M}{x_m(t)}$, similar to (\ref{eq:rx_sig}) but with the noise term omitted. By adjusting the coefficients correctly and combining all of these signals at the receiver, we may eliminate (or at least minimize) the signal distortion and ISI caused by multipath scattering associated with the direct BS-UE link and with the BS-RIS-UE links themselves. Because we have $M$ such signals due to the $M$ RIS elements, we will need to estimate $M$ cascaded BS-RIS-UE channels in order to compute the gradients correctly and minimize the resulting error signal, as demonstrated by the SD update rule derived and discussed in this section. This channel estimation introduces both additional computational complexity and communication complexity, as this SD algorithm, called ARISE, can only begin tuning the RIS \textbf{after} all the CSI for the $\boldsymbol{h}_{\text{BRU}}$ channel matrix have been obtained. This discussion is elaborated further in Section IV-E.

We assume the receiver samples $y(t)$ with period $T$ at integer time points $k$. Going forward, we use the discrete-time signal notation $y_k$ as the equivalent to $y(kT)$ for all $k$, and the array notation $\boldsymbol{y}$ as $[y_0,y_1,\ldots,y_L]$ where $L$ is the number of ISI components in the received pulse. For a given received pulse response $\boldsymbol{y}$, the ARISE algorithm needs to match it to the transmitted signal $\boldsymbol{s}$, given by the $1 \times (L+1)$ array $\boldsymbol{s}=[1,0,\ldots,0]$, to solve the optimization problem (P1). We define a cost function $J$ as the mean of the error magnitudes squared
\begin{equation}
    J=E[\epsilon_k\epsilon_k^*],
\end{equation}
with the error signal $\epsilon_k$ at each discrete time $k$, defined by
\begin{equation}
    \epsilon_k=s_k-y_k.
\end{equation}

By observation of the optimization problem (P1), we note that the RIS acts as both an equalizer and a signal booster, since it has the capability to beamform and increase the SNR of the signal, while also reducing the ISI. If the magnitude of $s_0$ is significantly larger than those of $\boldsymbol{y}$, the algorithm will see most of the error contribution at the main tap and thus will focus mostly on power maximization and barely on ISI minimization. On the contrary, if $s_0$ is too small, the algorithm will attempt to minimize the ISI while keeping the main tap small, because the magnitude of the ISI error in $\epsilon_{k},\;k>0,$ will become larger in proportion to the main tap error in $\epsilon_0$. Note that the received pulse at sampling time $t=0$ will go through a phase shift due to the channels and will include imaginary components. Furthermore, its real part may become negative. For the reasons above, we choose to normalize $\boldsymbol{y}$ based on its magnitude at $k=0$ to match the desired pulse $\boldsymbol{s}$. This accounts for the loss in signal power as it passes through the channel, while ensuring the ISI components are still relatively significant with respect to the main tap. Furthermore, to maximize the signal power by using the RIS as a beamformer, we scale the signal by the number of RIS elements to account for the maximum achievable gain by using the passive RIS. We define a scaling variable $\alpha_{\text{s}}$ and the scaling factor
\begin{equation}
    y_s=\alpha_{\text{s}}M|y_0|,
\end{equation}
using the main tap $y_0$ of the initially received pulse, to scale $\boldsymbol{y}$ during iterations of ARISE. The scaling may differ by choice of emphasizing either ISI cancellation or main tap amplification, as demonstrated later, and is accounted for by tuning $\alpha_{\text{s}}$.

To minimize the cost function $J$, the ARISE algorithm will directly modify the RIS reflection coefficients $\Gamma_m$ to obtain stable gradients, rather than the reflection phases $\theta_m$ which affect the received signal in a nonlinear manner. The objective function of ARISE thus becomes
\begin{equation}
    \min_{\boldsymbol{\Gamma}}{E[\epsilon_k\epsilon_k^*]},
\end{equation}
with the gradient descent update rule of the reflection coefficients $\boldsymbol{\Gamma}$ as
\begin{equation}
    \boldsymbol{\Gamma}_{t+1}=\boldsymbol{\Gamma}_t+\mu E \left[\sum_{l=0}^{L}{\boldsymbol{h}_{\text{BRU},l}^* s_{k-l}^*}\epsilon_k \right],
    \label{eq:gamma_update_SD}
\end{equation}
following the derivation in Appendix A. We take the expectation over $L+1$ signal taps -- essentially as a sliding window across the pulse response -- to calculate the ensemble average and therefore the ``true gradient'' during each time step \cite{sayed2011adaptive}. In practice, this can be achieved by either:
\begin{itemize}
    \item Updating the RIS configuration during each step and obtaining the new pulse response from the receiver.
    \item Calculating the expected pulse response offline assuming the RIS is updated, using the expression in (\ref{eq:rx_sig}).
\end{itemize}
Note the tradeoff between the two approaches above: the first requires live communication with the receiver while the RIS is optimized, whereas the second approach requires channel estimation for the direct $\boldsymbol{h}_{\text{BU}}$ vector for accurate computation of the expected pulse responses.

To enforce the passivity of the RIS, we normalize the magnitude of each reflection coefficient after every update using
\begin{equation}
    \Gamma_{t+1,m} \gets \frac{\Gamma_{t+1,m}}{|\Gamma_{t+1,m}|}
    \label{eq:ris_mag_norm}
\end{equation}
for $m=1,2,\ldots,M$.

The full ARISE algorithm is outlined in Algorithm~\ref{alg:SD}. The step size for gradient descent is given by
\begin{equation}
    \mu = \frac{\alpha_\mu (L+1)}{\sum_{k}{|y_k|}},
\end{equation}
with an arbitrary scaling factor $\alpha_{\mu}$ and signal components $|y_k|$ taken from the initially received pulse response. The stop condition for convergence depends on a threshold $\eta_{\text{th}}$ based on the absolute difference between normalized versions of $\eta$,
\begin{equation}
    \eta_{\text{n}} = \frac{\eta}{\sum_{k}|{y_k}|^2},
    \label{eq:eta_norm}
\end{equation}
over ten consecutive steps.
For an additional safety measure, we define $\eta_{\text{n,max}}$ to track the optimal objective function and reduce the step size in case the algorithm overshoots and diverges, using another arbitrary parameter $\eta_{\text{d}}$ to track the difference between the current objective and the best objective.

Although we are using a steepest descent approach by applying the expectation operator as in the update rule in (\ref{eq:gamma_update_SD}), we note that a stochastic gradient descent-based implementation of ARISE is also achievable, similar to the least mean squares (LMS) algorithm \cite{1454555}. The update rule becomes
\begin{equation}
    \boldsymbol{\Gamma}_{t+1}=\boldsymbol{\Gamma}_t+\mu \sum_{l=0}^{L}{\boldsymbol{h}_{\text{BRU},l}^* s_{k-l}^*}\epsilon_k.
    \label{eq:gamma_update_SGD}
\end{equation}
Then, instead of processing the pulse response via a sliding window and averaging the errors before each RIS update, we use the stochastic gradient obtained from processing the pulse (or any other desired pilot signal) during only a single window per time step. This approach would have less computational complexity per RIS update step, but would likely require more steps until convergence and result in noisy gradients. In the simulation results discussed in this work, we use the update rule with the expectation operator as in (\ref{eq:gamma_update_SD}) to take complete advantage of every sample in the received pulse response, thereby equalizing the signal using a smooth gradient.

\begin{algorithm}
    \caption{Adaptive RIS Equalization (ARISE)}
    \label{alg:SD}
    \begin{algorithmic}[1]
        \State {Initialize $\boldsymbol{\Gamma}, \boldsymbol{h}_{\text{BRU}}, \boldsymbol{s}, \eta_{\text{n,max}}$.}
        \State {Initialize $\boldsymbol{y}$ by transmitting $\boldsymbol{s}$ and recording the received signal from~(\ref{eq:rx_sig}).}
        \State{Initialize $\mu\gets \frac{\alpha_\mu (L+1)}{\sum_{k}{|y_k|}}$.}
        \State{Initialize $y_s\gets \alpha_{\text{s}}M|y_0|$.}
        \State{Initialize $i \gets 0$.}
        \Repeat
            \State {$\boldsymbol{\epsilon} \gets \boldsymbol{s}-\boldsymbol{y}/y_s$.}
            \State {Calculate $\eta$ using~(\ref{eq:eta}) and $\eta_{\text{n}}$ using~(\ref{eq:eta_norm}).}
            \If {$\eta_{\text{n}} > \eta_{\text{n,max}}$}
                \State {$\eta_{\text{n,max}} \gets \eta_{\text{n}}$.}
            \EndIf
            \State {Update $\boldsymbol{\Gamma}$ using~(\ref{eq:gamma_update_SD}).}
            \State {Normalize $\boldsymbol{\Gamma}$ using~(\ref{eq:ris_mag_norm}).}
            \State {Obtain new received pulse $\boldsymbol{y}'$ from~(\ref{eq:rx_sig}).}
            \State {Calculate new objective functions $\eta'$ using~(\ref{eq:eta}) and $\eta_{\text{n}}'$ using~(\ref{eq:eta_norm}).}
            \If {$\eta_{\text{n}}' < \eta_{\text{n,max}} - \eta_{\text{d}}$}
                \State {$\mu \gets \frac{\mu}{2}$.}
                \State {$\eta_{\text{n,max}} \gets -1.0$.}
            \EndIf
            \If {$|\eta_{\text{n}}'-\eta_{\text{n}}|<\eta_{\text{th}}$}
                \State {$i \gets i+1$.}
            \Else
                \State{$i \gets 0$.}
            \EndIf
            \If {$i \geq 10$}
                \State {\textbf{break}}
            \EndIf
        \Until{convergence}
    \end{algorithmic}
\end{algorithm}

\subsection{ARISE SIMULATION RESULTS}

The RIS environment consists of a single-antenna BS, single-antenna UE, and by default, an $M=100$-element RIS. The spacing between the RIS elements is $d_x=20$ mm. We use a sampling period $T =16.3$ ns, in accordance with the 3GPP standard for 5G New Radio (NR) \cite{3gpp38.211}. The normalized gain factors are $G=G'=-43$ dBm, as in \cite{9414612}, and the noise power is $P_n=-96$ dBm in the AWGN signal $n(t)$. We define a 2-dimensional coordinate system with the BS at the center with coordinates (in meters) $(0,0)$. The BS directly illuminates the RIS, located at $(10,0)$, at normal incidence. The starting location for the UE is at $(-20,-20)$ and the UE is allowed to move within the rectangular area bounded between the points $(-100, -100)$ and $(-10, 100)$. At the start of each channel coherence block, the NLoS channels are updated completely and the UE moves in a random walk fashion, each time in a random direction and a random distance up to $20$ m away from the previous point, both uniformly distributed. Considering the ISI paths between the BS and the RIS may also be scattered and delayed in the RIS-UE link, we define the relationship between the number of delayed paths  $n_{\text{r}}$ and number of ISI terms $L$ as $L=2n_{\text{r}}$.

We define two baseline RIS reflection coefficient schemes for comparison:
\begin{itemize}
    \item \textbf{Random Phases} -- the RIS reflection phases are sampled uniformly between $[-\pi,\pi]$.
    \item \textbf{Inverse Phases} -- the RIS reflection coefficients are the complex conjugates of the $M$ channel coefficients in $\boldsymbol{h}_{\text{BRU},k}$, evaluated at the sampling instance $k=0$ and normalized to unity magnitude, e.g.,
    \begin{equation}
        \boldsymbol{\Gamma}=\exp({-j\angle{\boldsymbol{h}_{\text{BRU},0}}}),
    \end{equation}
    intending to produce a coherent sum at the receiver for $y_0$ \cite{10742896}.
\end{itemize}

For the ARISE algorithm, we set the threshold parameter $\eta_{\text{th}}=10^{-5}$ and the overshoot parameter $\eta_{\text{d}}=0.5$. We scale the step size $\mu$ by $\alpha_{\mu}=10$ for rapid convergence.

\begin{figure*}[!htbp]
    \centering
    \begin{subfigure}[b]{0.24\linewidth}
        \centering
        \includegraphics[width=\linewidth]{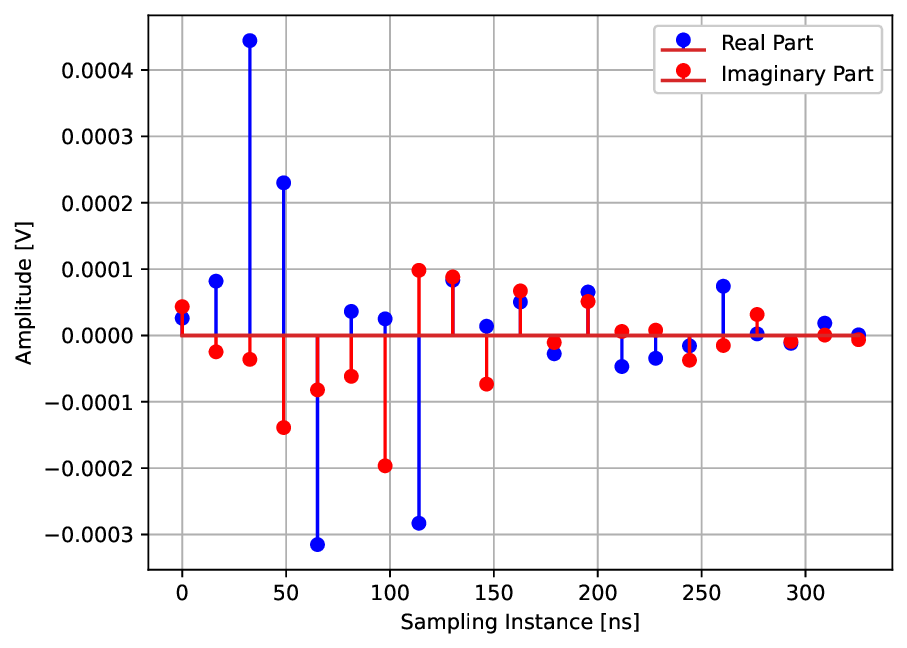}
        \caption{Random Phases, $\eta_{\text{n}}$ = -0.994}
        \label{fig:sim_8p_rnd}
    \end{subfigure}
    \begin{subfigure}[b]{0.24\linewidth}
        \centering
        \includegraphics[width=\linewidth]{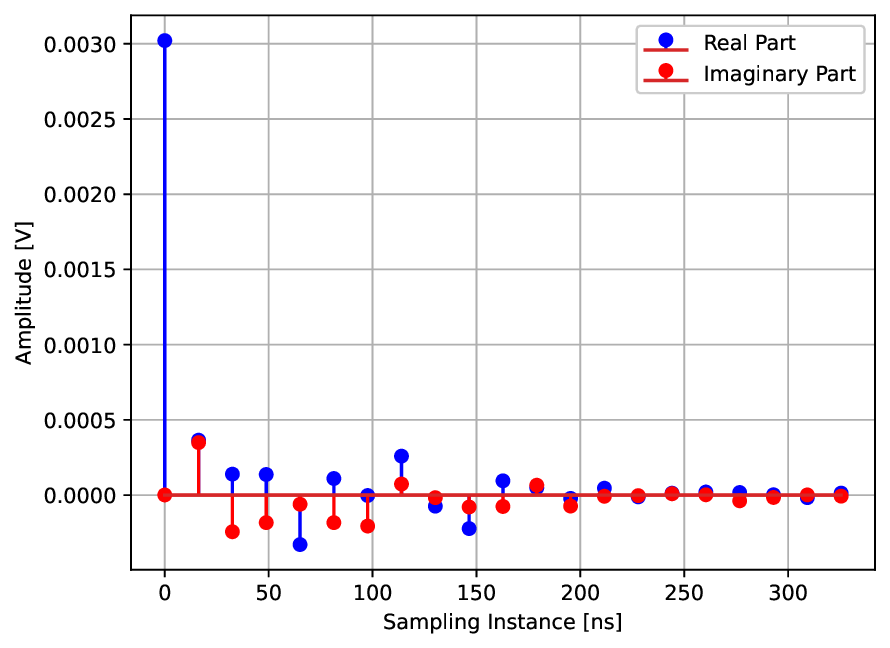}
        \caption{Inverse Phases, $\eta_{\text{n}}$ = 0.848}
        \label{fig:sim_8p_inv}
    \end{subfigure}
    \begin{subfigure}[b]{0.24\linewidth}
        \centering
        \includegraphics[width=\linewidth]{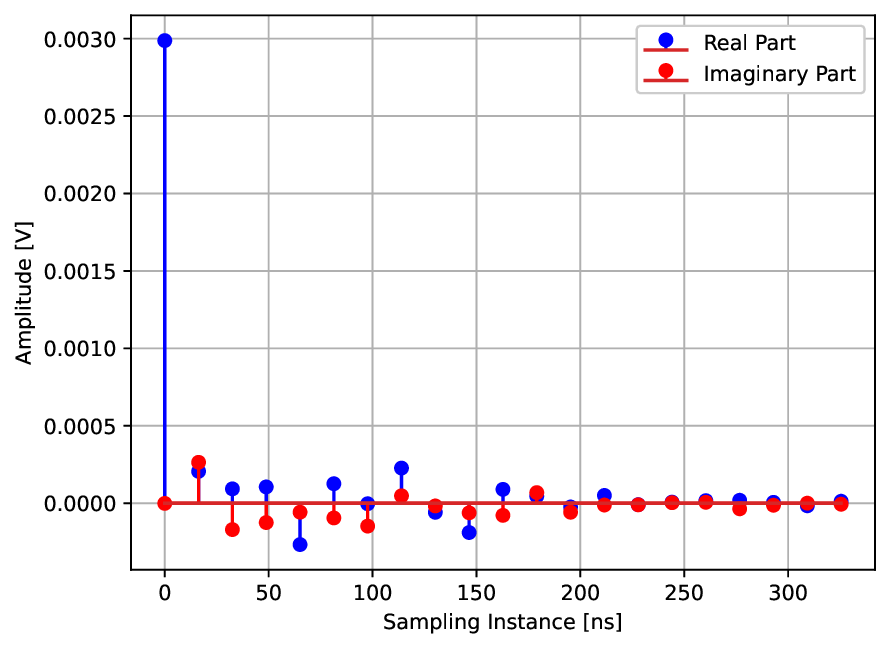}
        \caption{ARISE, $\alpha_{\text{s}}=1.00$, $\eta_{\text{n}}$ = 0.909}
        \label{fig:sim_8p_SD1}
    \end{subfigure}
    \begin{subfigure}[b]{0.24\linewidth}
        \centering
        \includegraphics[width=\linewidth]{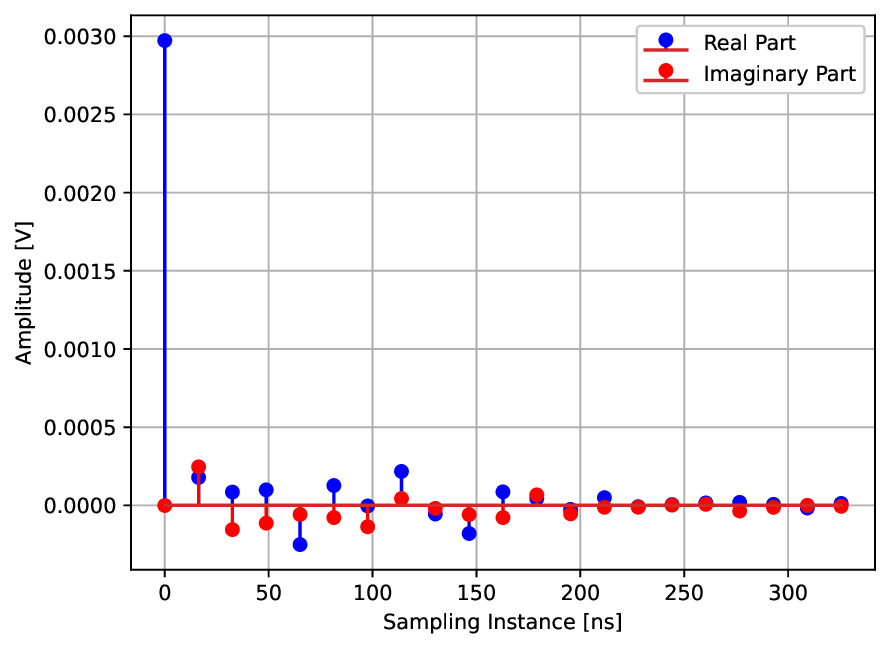}
        \caption{ARISE, $\alpha_{\text{s}}=0.90$, $\eta_{\text{n}}$ = 0.919}
        \label{fig:sim_8p_SD2}
    \end{subfigure}
    \begin{subfigure}[b]{0.24\linewidth}
        \centering
        \includegraphics[width=\linewidth]{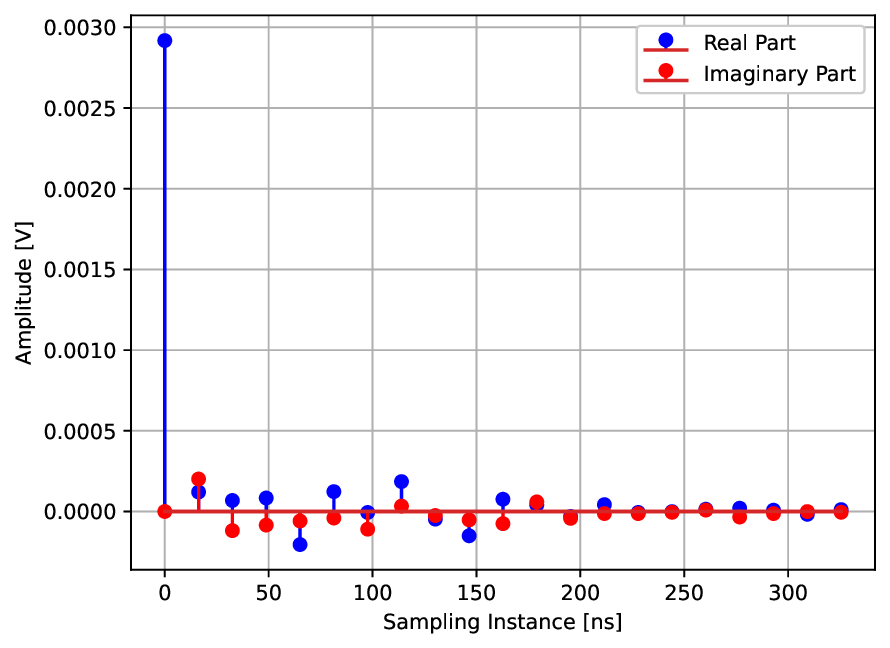}
        \caption{ARISE, $\alpha_{\text{s}}=0.75$, $\eta_{\text{n}}$ = 0.943}
        \label{fig:sim_8p_SD3}
    \end{subfigure}
    \begin{subfigure}[b]{0.24\linewidth}
        \centering
        \includegraphics[width=\linewidth]{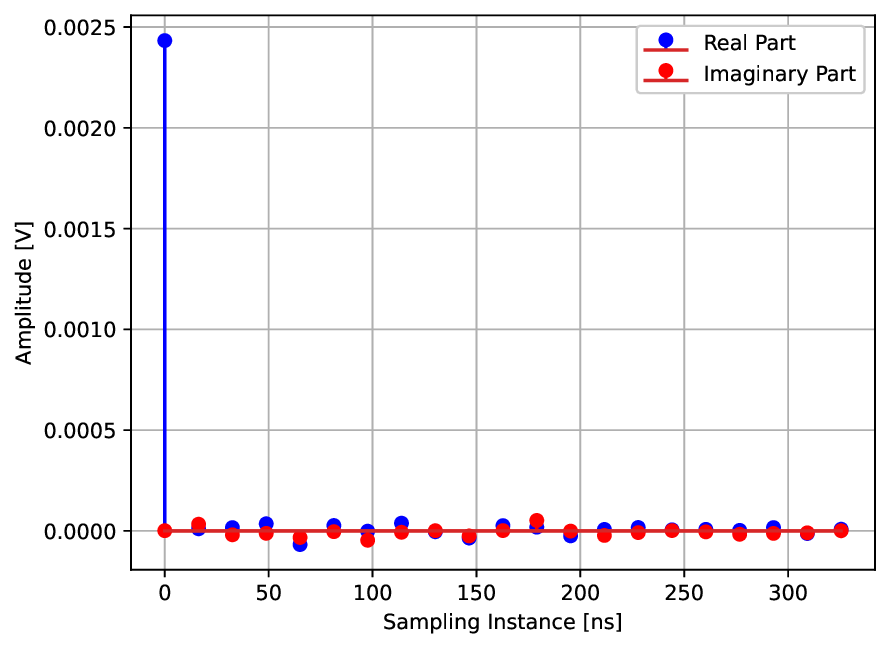}
        \caption{ARISE, $\alpha_{\text{s}}=0.50$, $\eta_{\text{n}}$ = 0.993}
        \label{fig:sim_8p_SD4}
    \end{subfigure}
    \begin{subfigure}[b]{0.24\linewidth}
        \centering
        \includegraphics[width=\linewidth]{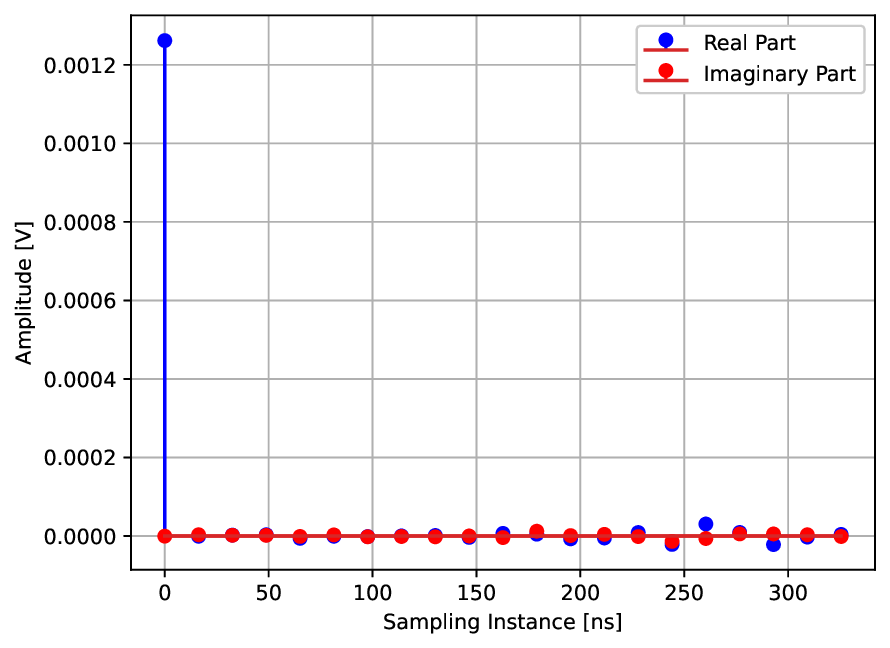}
        \caption{ARISE, $\alpha_{\text{s}}=0.25$, $\eta_{\text{n}}$ = 0.997}
        \label{fig:sim_8p_SD5}
    \end{subfigure}
    \begin{subfigure}[b]{0.24\linewidth}
        \centering
        \includegraphics[width=\linewidth]{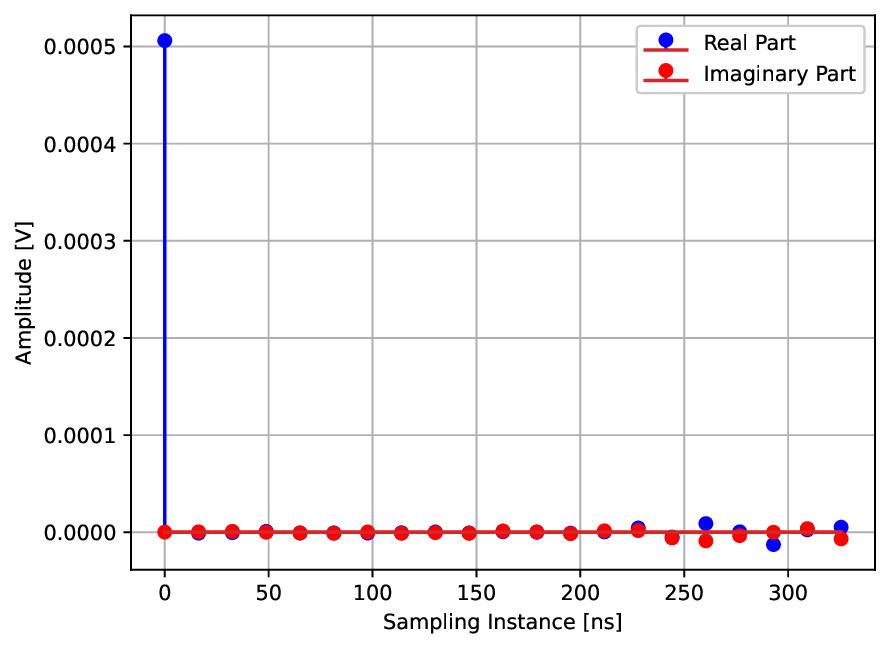}
        \caption{ARISE, $\alpha_{\text{s}}=0.10$, $\eta_{\text{n}}$ = 0.996}
        \label{fig:sim_8p_SD6}
    \end{subfigure}
    \caption{Converged pulse responses real and imaginary parts at UE location (-20, -20), $M=100, \kappa=10,n_{\text{r}}=10$.}
    \label{fig:sim_8_pulse}
\end{figure*}

\begin{figure*}[!htbp]
    \centering
    \begin{subfigure}[b]{0.24\linewidth}
        \centering
        \includegraphics[width=\linewidth]{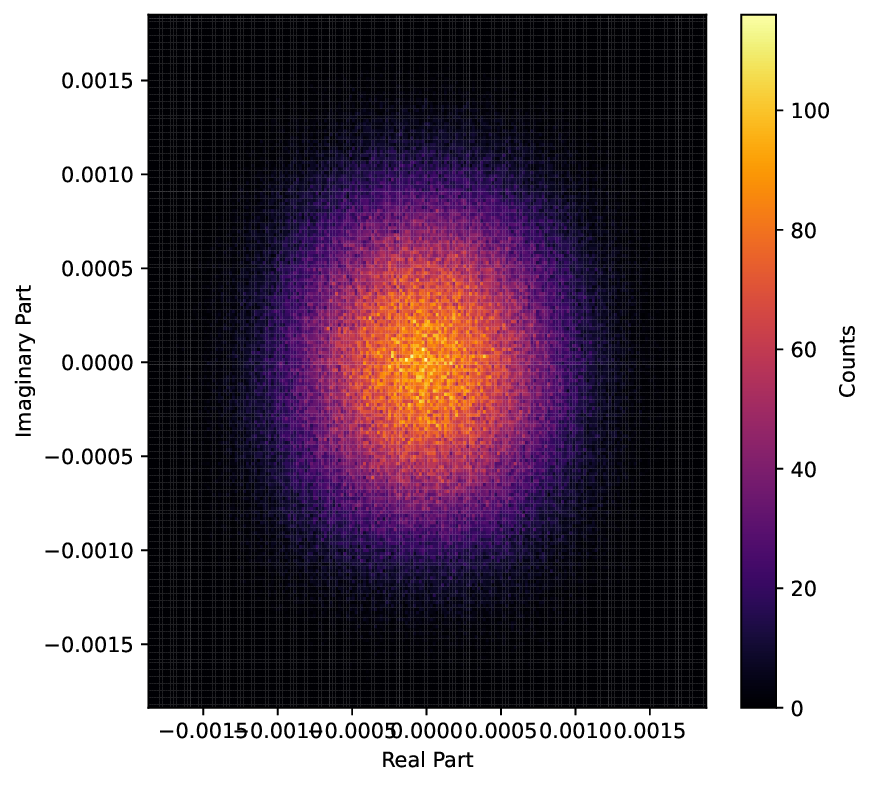}
        \caption{Random Phases,\\ SNR = 2.87 dB}
        \label{fig:sim_8q_rnd}
    \end{subfigure}
    \begin{subfigure}[b]{0.24\linewidth}
        \centering
        \includegraphics[width=\linewidth]{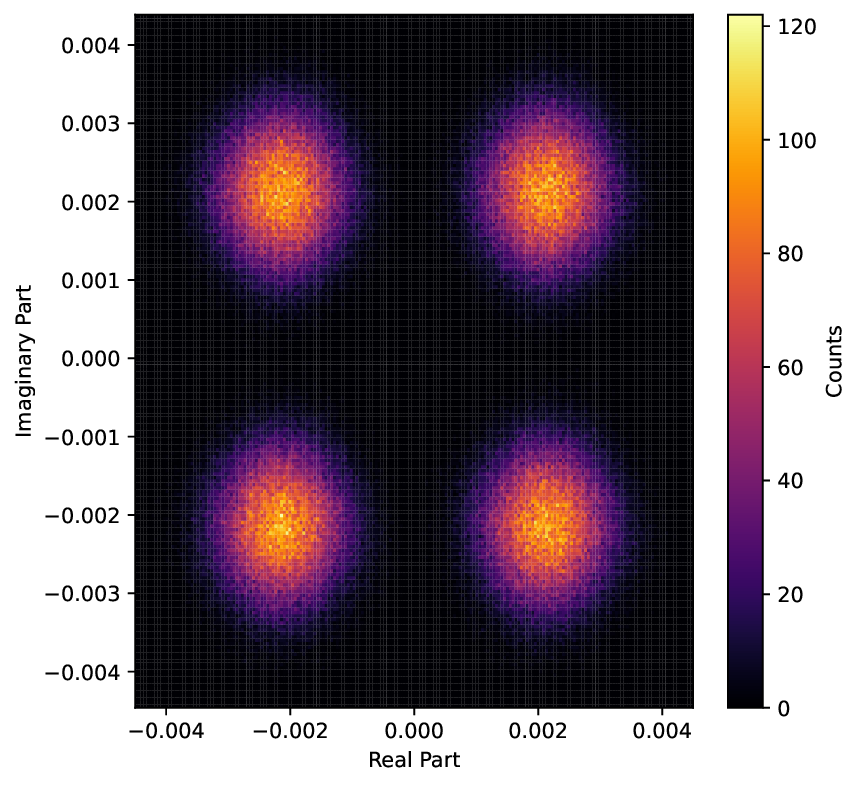}
        \caption{Inverse Phases,\\ SNR = 10.84 dB}
        \label{fig:sim_8q_inv}
    \end{subfigure}
    \begin{subfigure}[b]{0.24\linewidth}
        \centering
        \includegraphics[width=\linewidth]{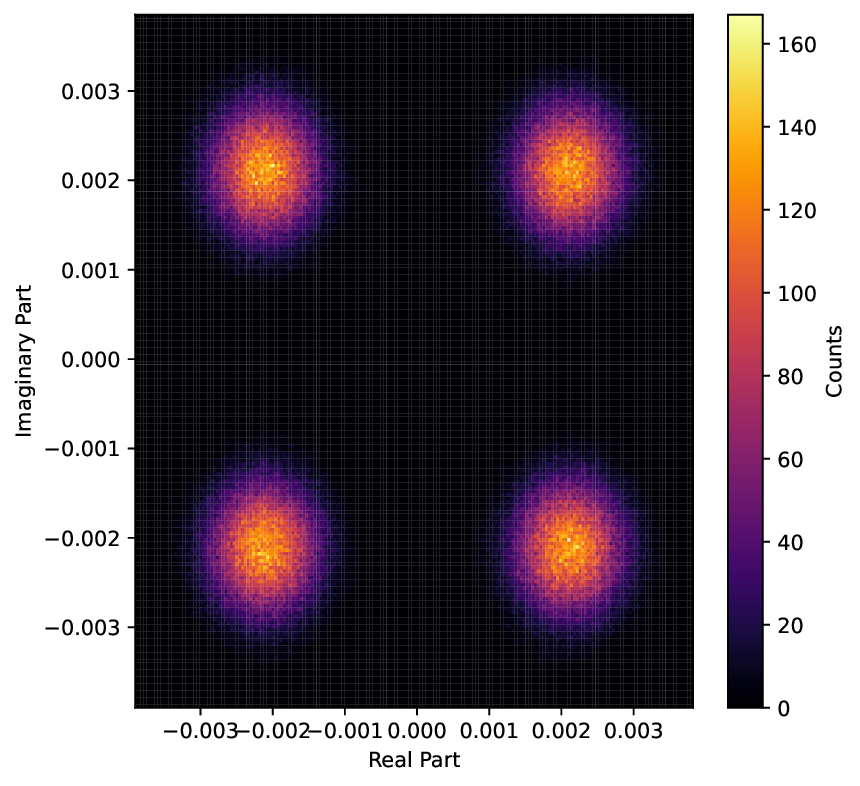}
        \caption{ARISE,\\ $\alpha_{\text{s}}=1.00$, SNR = 13.22 dB}
        \label{fig:sim_8q_SD1}
    \end{subfigure}
    \begin{subfigure}[b]{0.24\linewidth}
        \centering
        \includegraphics[width=\linewidth]{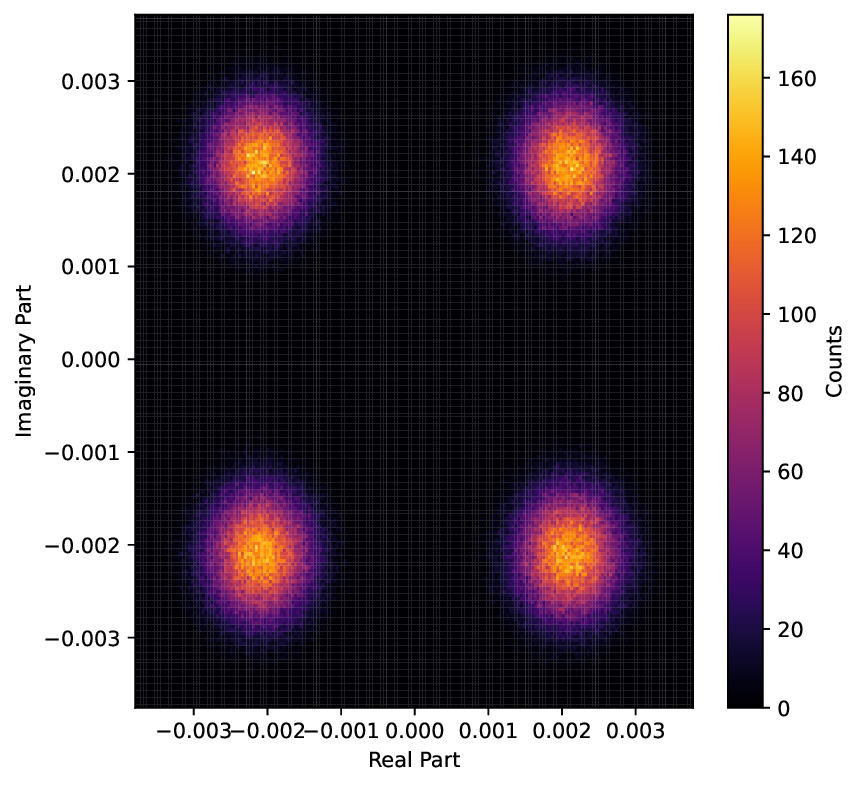}
        \caption{ARISE,\\ $\alpha_{\text{s}}=0.90$, SNR = 13.77 dB}
        \label{fig:sim_8q_SD2}
    \end{subfigure}
    \begin{subfigure}[b]{0.24\linewidth}
        \centering
        \includegraphics[width=\linewidth]{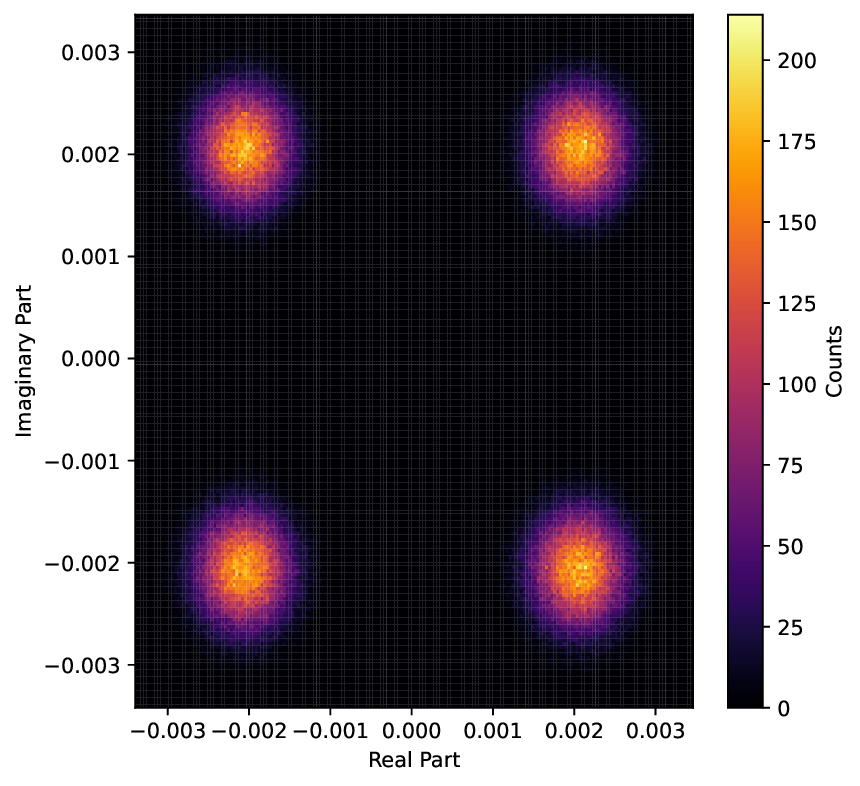}
        \caption{ARISE,\\ $\alpha_{\text{s}}=0.75$, SNR = 15.32 dB}
        \label{fig:sim_8q_SD3}
    \end{subfigure}
    \begin{subfigure}[b]{0.24\linewidth}
        \centering
        \includegraphics[width=\linewidth]{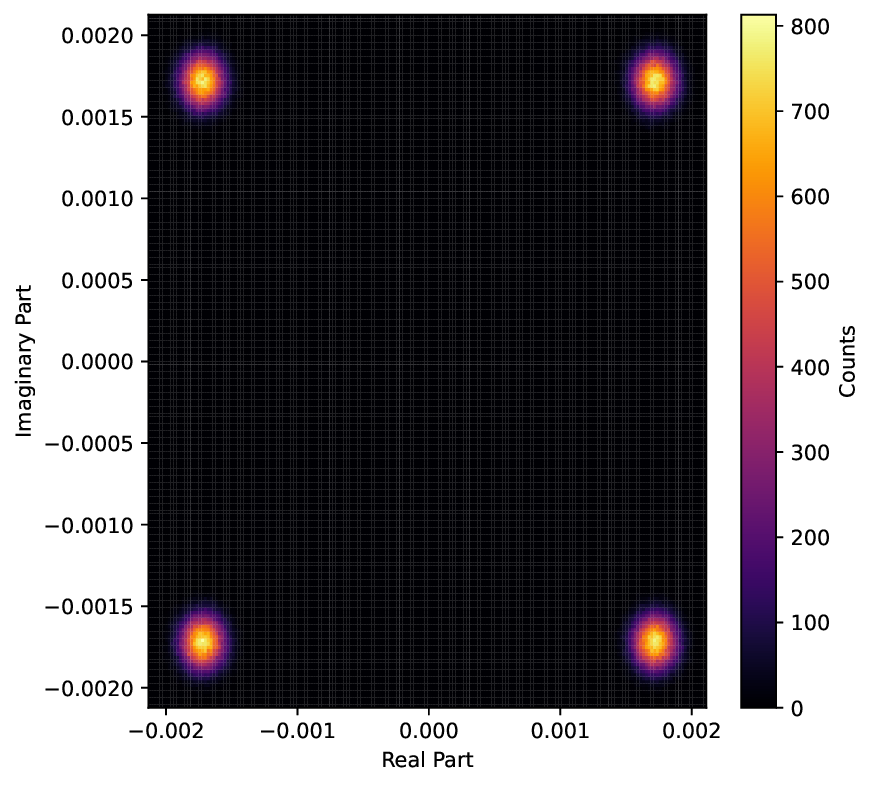}
        \caption{ARISE,\\ $\alpha_{\text{s}}=0.50$, SNR = 24.36 dB}
        \label{fig:sim_8q_SD4}
    \end{subfigure}
    \begin{subfigure}[b]{0.24\linewidth}
        \centering
        \includegraphics[width=\linewidth]{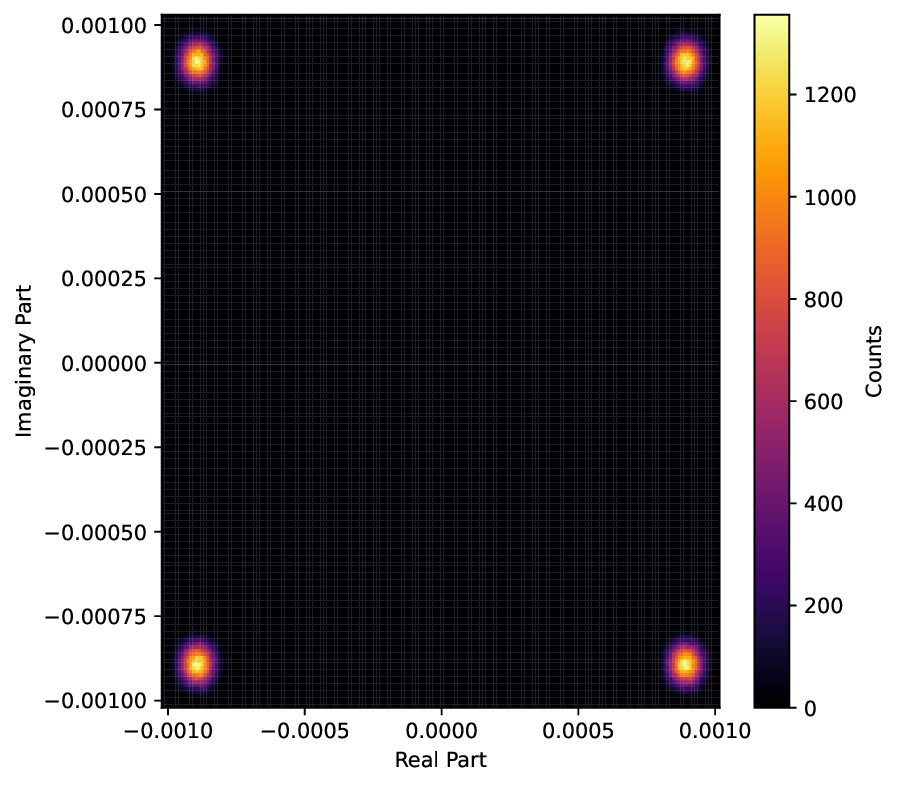}
        \caption{ARISE,\\ $\alpha_{\text{s}}=0.25$, SNR = 27.57 dB}
        \label{fig:sim_8q_SD5}
    \end{subfigure}
    \begin{subfigure}[b]{0.24\linewidth}
        \centering
        \includegraphics[width=\linewidth]{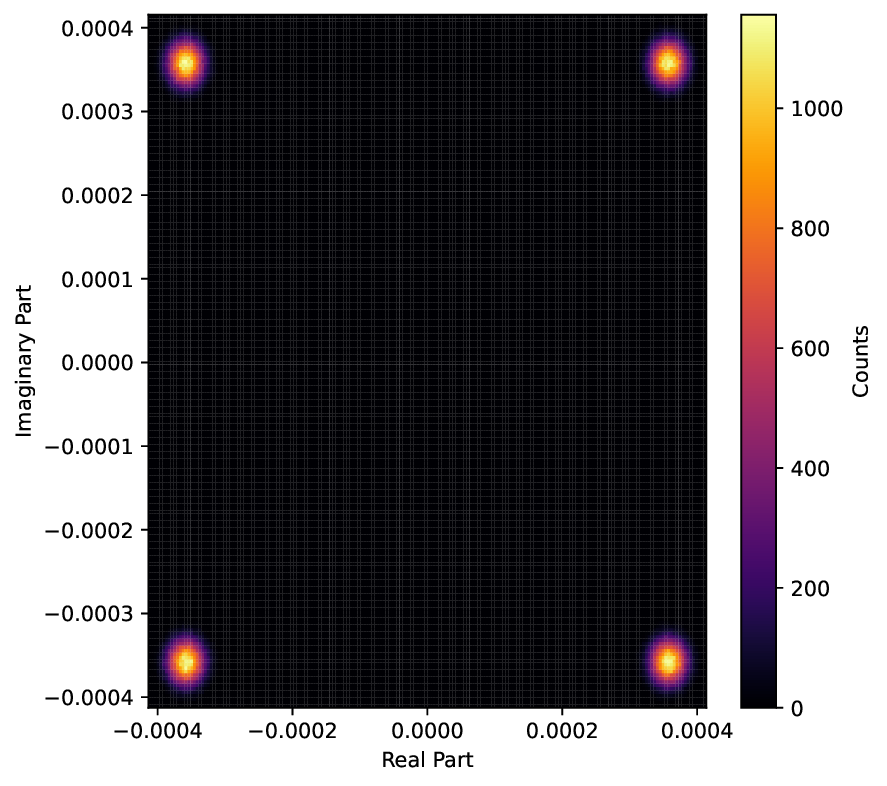}
        \caption{ARISE,\\ $\alpha_{\text{s}}=0.10$, SNR = 26.89 dB}
        \label{fig:sim_8q_SD6}
    \end{subfigure}
    \caption{Converged QPSK constellations at UE location (-20, -20), $M=100, \kappa=10,n_\text{r}=10$.}
    \label{fig:sim_8_qpsk}
\end{figure*}

\begin{figure*}[!htbp]
    \centering
    \begin{subfigure}[b]{0.49\linewidth}
        \centering
        \includegraphics[width=\linewidth]{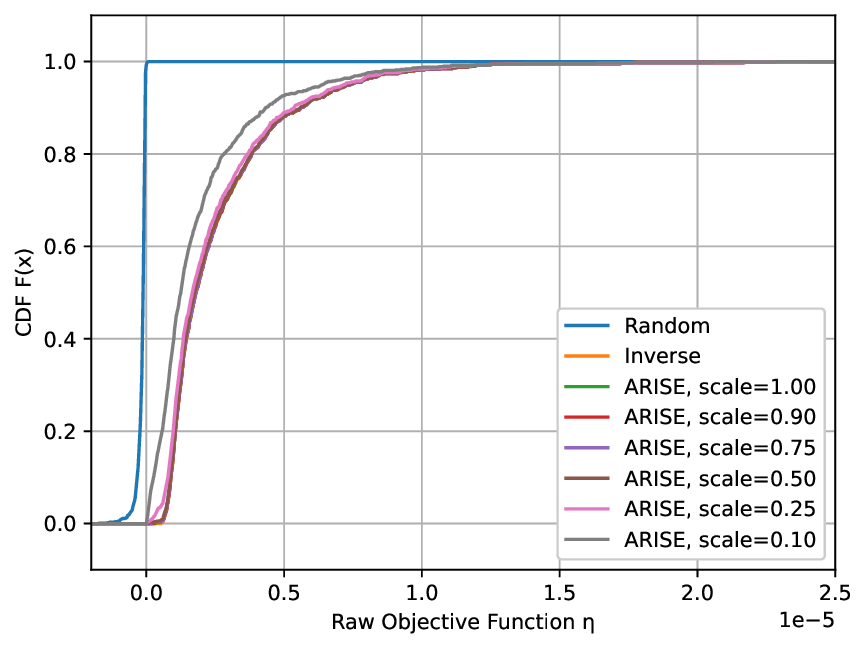}
        \caption{Converged $\eta$ values}
        \label{fig:sim_6_raw}
    \end{subfigure}
    \begin{subfigure}[b]{0.49\linewidth}
        \centering
        \includegraphics[width=\linewidth]{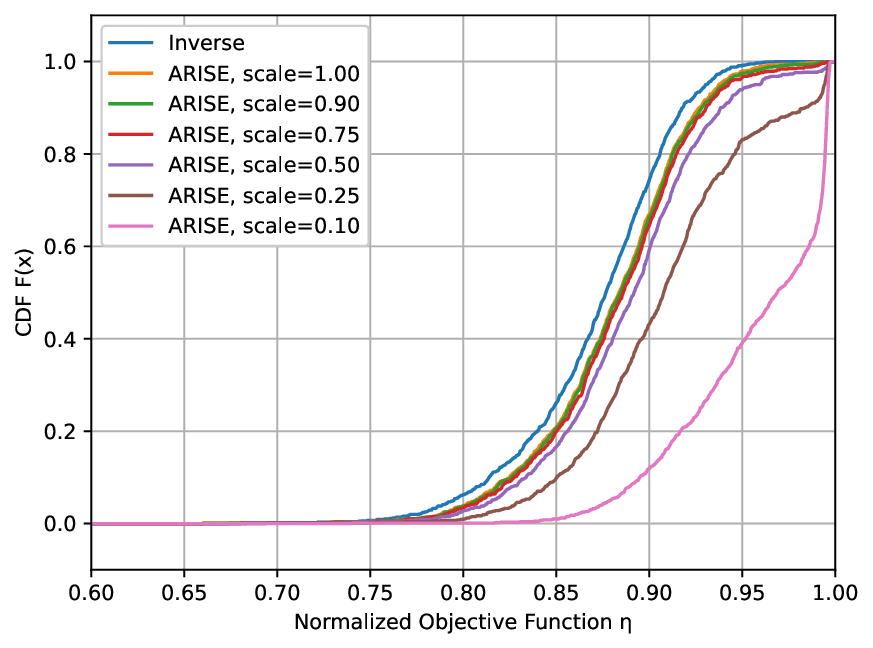}
        \caption{Converged normalized $\eta_{\text{n}}$ values}
        \label{fig:sim_6_norm}
    \end{subfigure}
    \caption{Converged CDFs over 1000 simulations, $M=100, \kappa=10,n_{\text{r}}=10$.}
    \label{fig:sim_6}
\end{figure*}

\begin{figure*}[!htbp]
    \centering
    \begin{subfigure}[b]{0.24\linewidth}
        \centering
        \includegraphics[width=\linewidth]{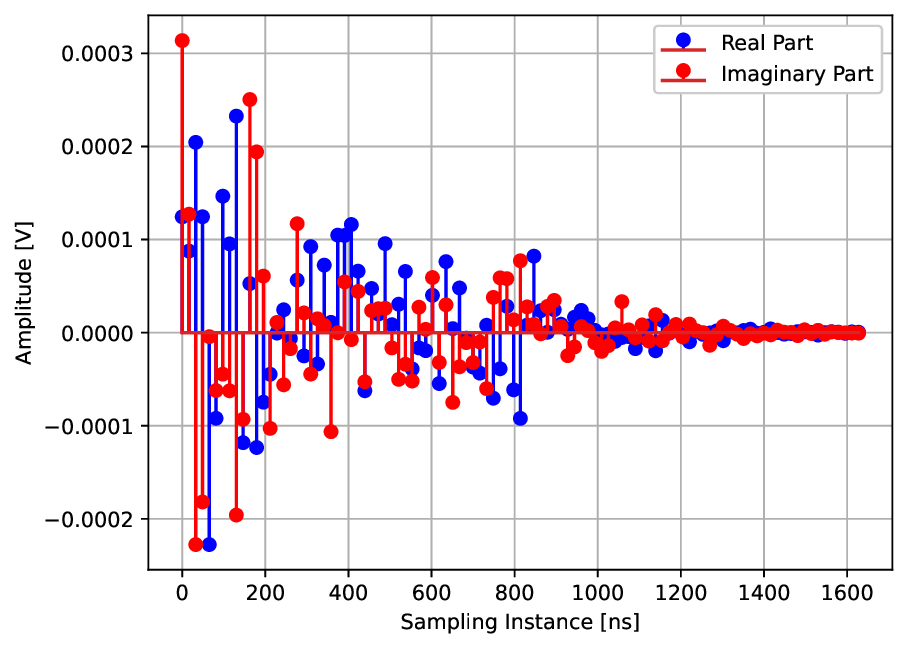}
        \caption{Random Phases, $\eta_{\text{n}}$ = -0.848}
        \label{fig:sim_9p_rnd}
    \end{subfigure}
    \begin{subfigure}[b]{0.24\linewidth}
        \centering
        \includegraphics[width=\linewidth]{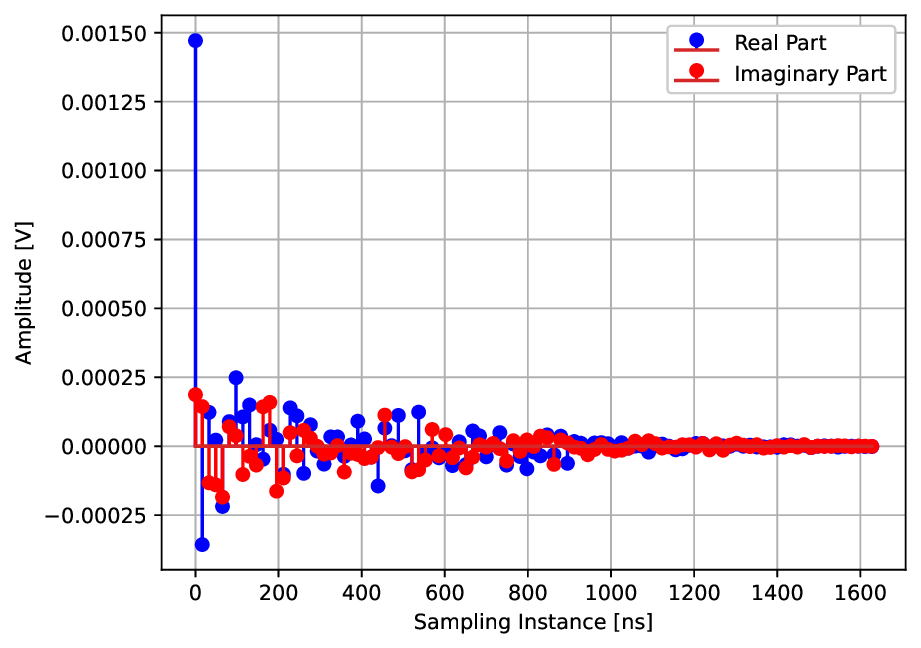}
        \caption{Inverse Phases, $\eta_{\text{n}}$ = 0.470}
        \label{fig:sim_9p_inv}
    \end{subfigure}
    \begin{subfigure}[b]{0.24\linewidth}
        \centering
        \includegraphics[width=\linewidth]{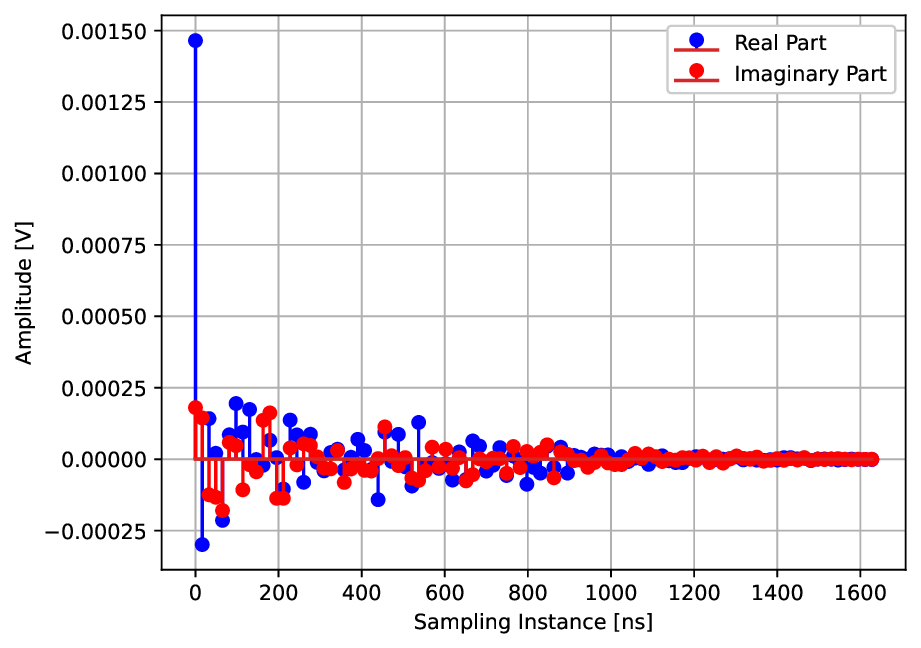}
        \caption{ARISE, $\alpha_{\text{s}}=1.00$, $\eta_{\text{n}}$ = 0.511}
        \label{fig:sim_9p_SD1}
    \end{subfigure}
    \begin{subfigure}[b]{0.24\linewidth}
        \centering
        \includegraphics[width=\linewidth]{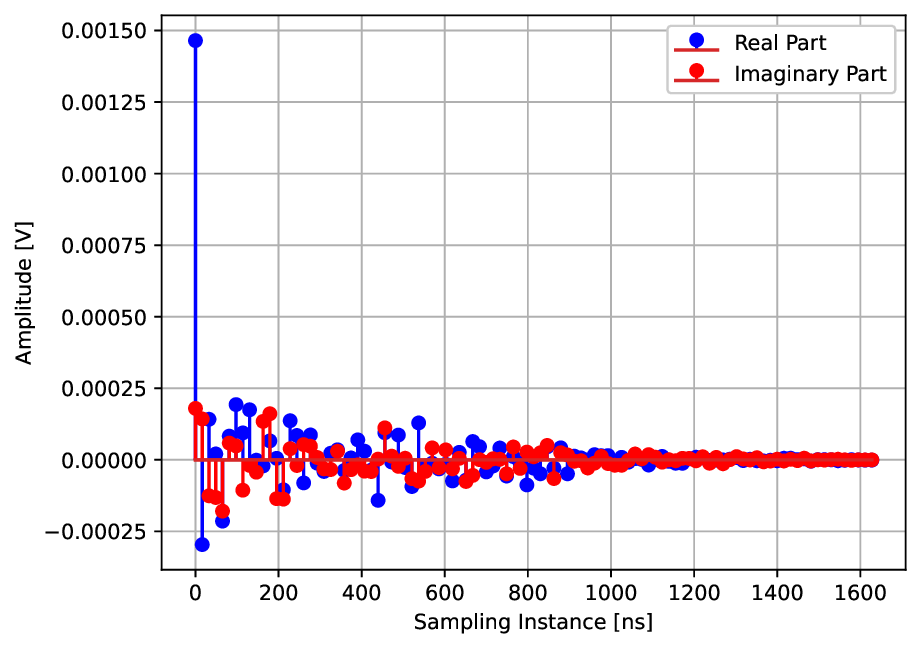}
        \caption{ARISE, $\alpha_{\text{s}}=0.90$, $\eta_{\text{n}}$ = 0.513}
        \label{fig:sim_9p_SD2}
    \end{subfigure}
    \begin{subfigure}[b]{0.24\linewidth}
        \centering
        \includegraphics[width=\linewidth]{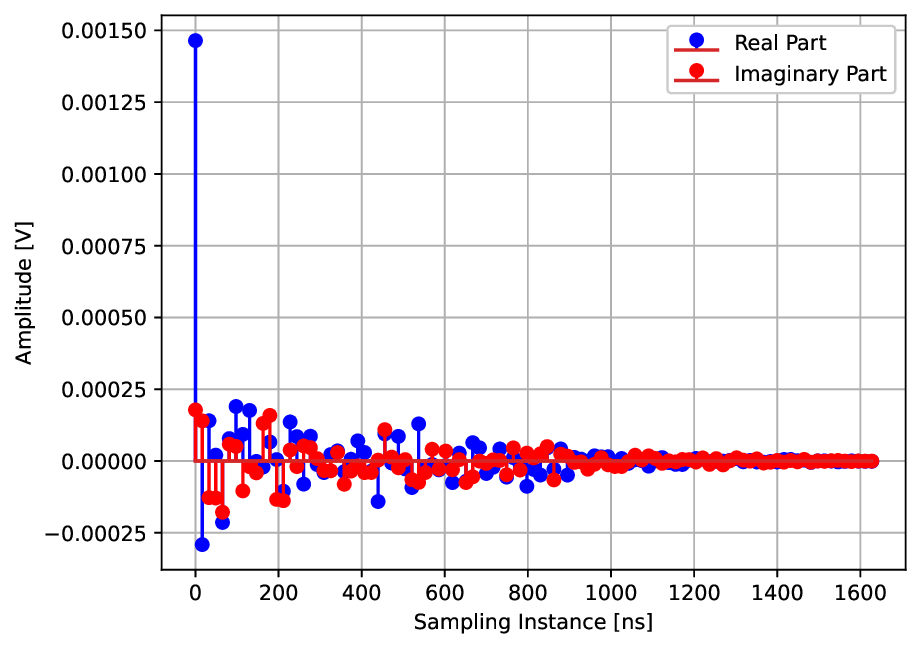}
        \caption{ARISE, $\alpha_{\text{s}}=0.75$, $\eta_{\text{n}}$ = 0.518}
        \label{fig:sim_9p_SD3}
    \end{subfigure}
    \begin{subfigure}[b]{0.24\linewidth}
        \centering
        \includegraphics[width=\linewidth]{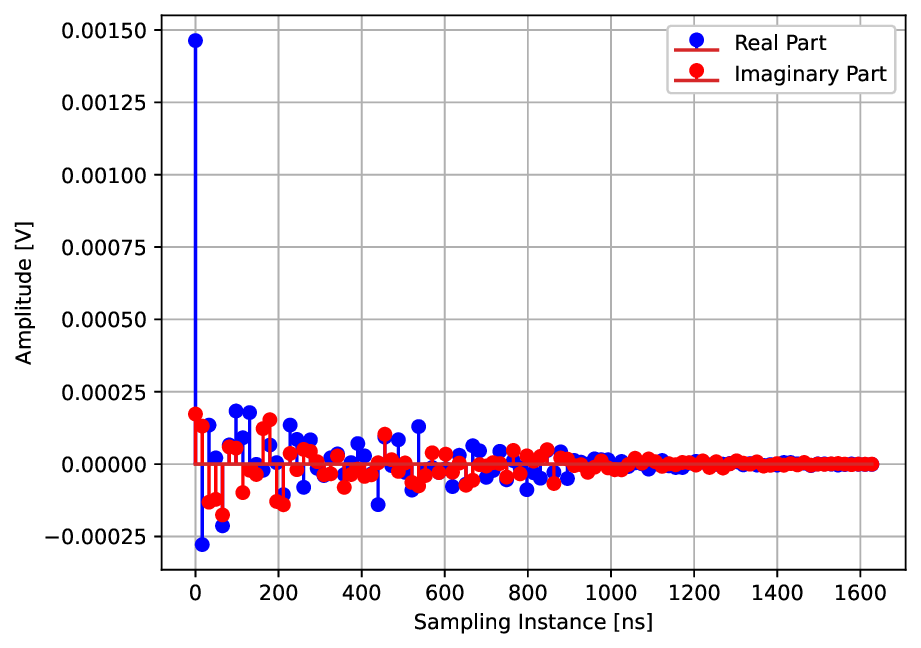}
        \caption{ARISE, $\alpha_{\text{s}}=0.50$, $\eta_{\text{n}}$ = 0.532}
        \label{fig:sim_9p_SD4}
    \end{subfigure}
    \begin{subfigure}[b]{0.24\linewidth}
        \centering
        \includegraphics[width=\linewidth]{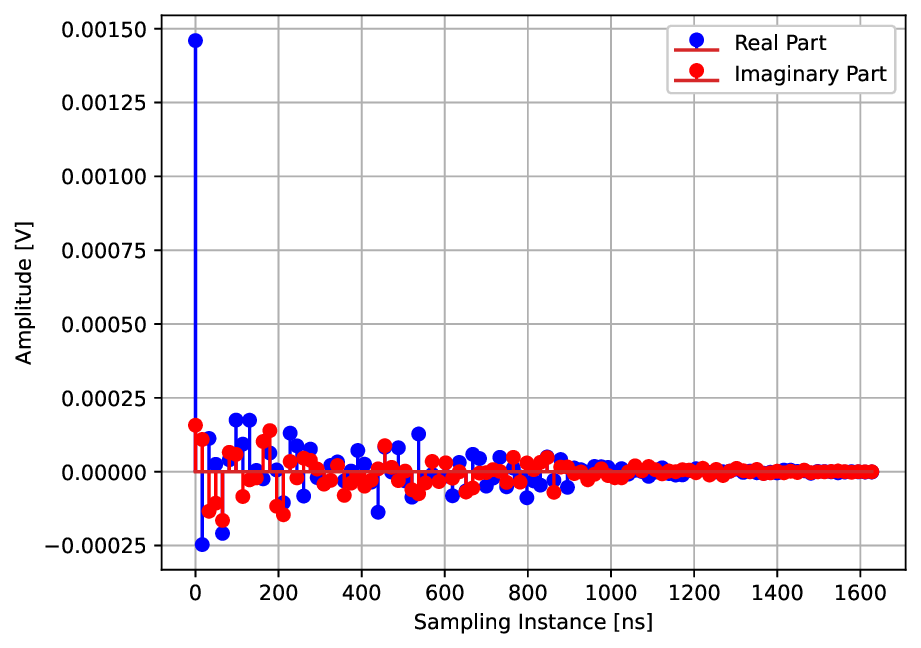}
        \caption{ARISE, $\alpha_{\text{s}}=0.25$, $\eta_{\text{n}}$ = 0.568}
        \label{fig:sim_9p_SD5}
    \end{subfigure}
    \begin{subfigure}[b]{0.24\linewidth}
        \centering
        \includegraphics[width=\linewidth]{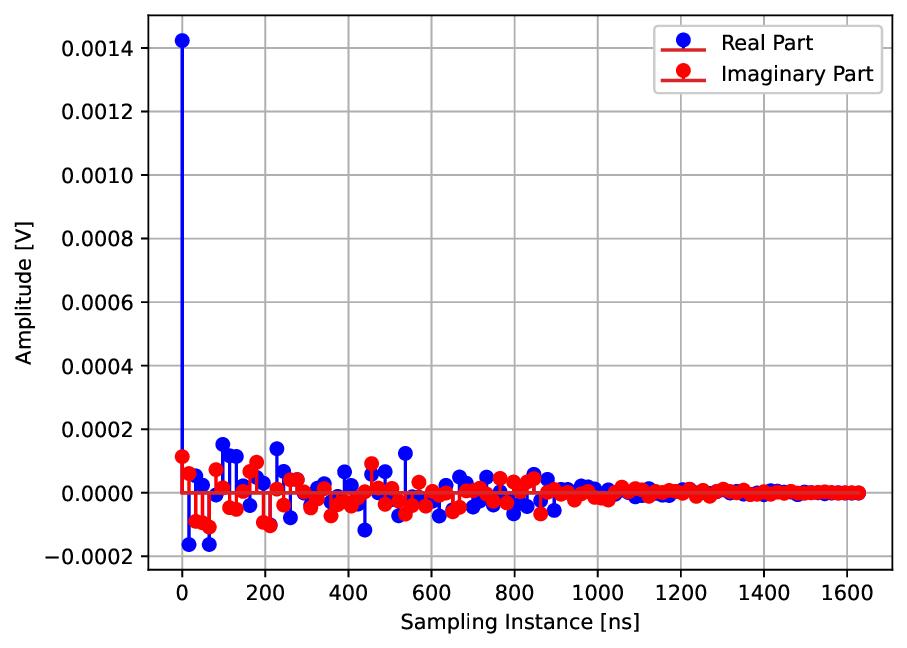}
        \caption{ARISE, $\alpha_{\text{s}}=0.10$, $\eta_{\text{n}}$ = 0.680}
        \label{fig:sim_9p_SD6}
    \end{subfigure}
    \caption{Converged pulse responses real and imaginary parts at UE location (-20, -20), $M=100, \kappa=0,n_{\text{r}}=50$.}
    \label{fig:sim_9_pulse}
\end{figure*}

\begin{figure*}[!htbp]
    \centering
    \begin{subfigure}[b]{0.24\linewidth}
        \centering
        \includegraphics[width=\linewidth]{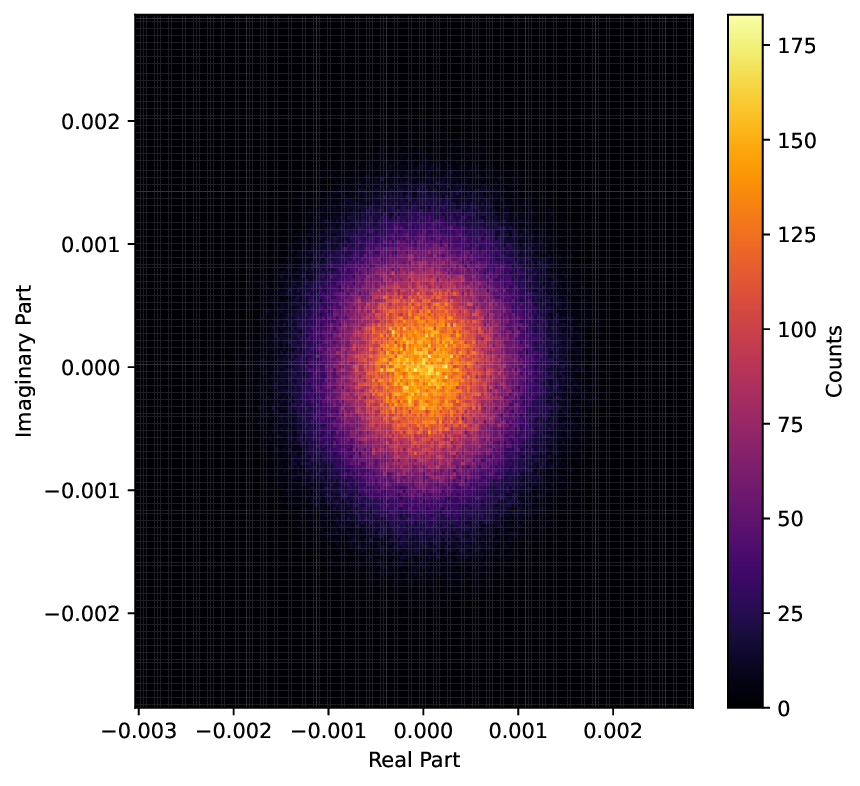}
        \caption{Random Phases,\\ SNR = 2.52 dB}
        \label{fig:sim_9q_rnd}
    \end{subfigure}
    \begin{subfigure}[b]{0.24\linewidth}
        \centering
        \includegraphics[width=\linewidth]{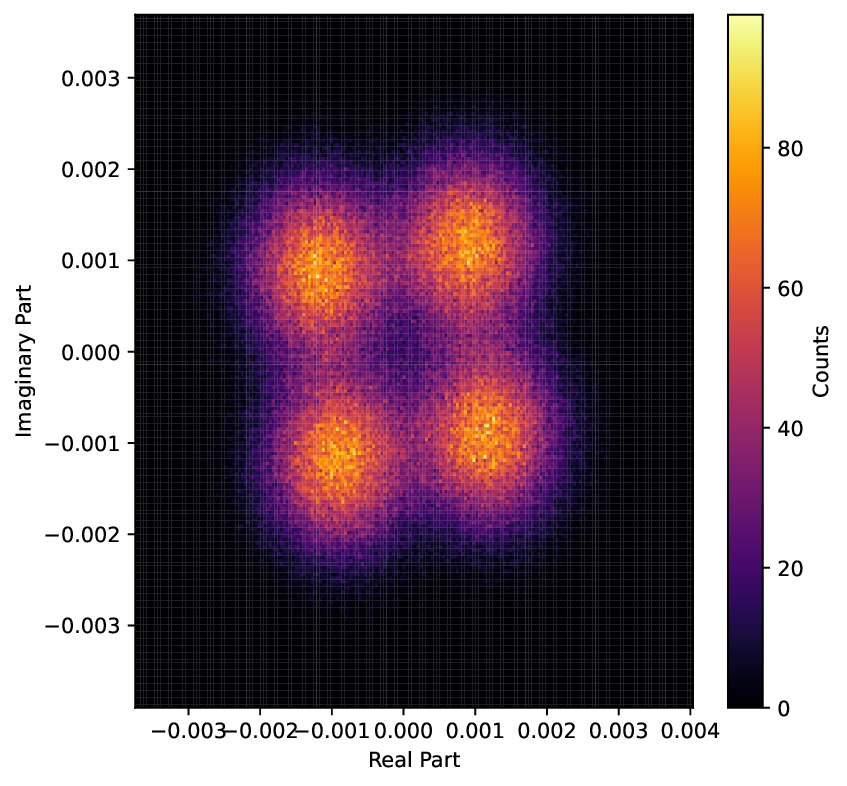}
        \caption{Inverse Phases,\\ SNR = 5.22 dB}
        \label{fig:sim_9q_inv}
    \end{subfigure}
    \begin{subfigure}[b]{0.24\linewidth}
        \centering
        \includegraphics[width=\linewidth]{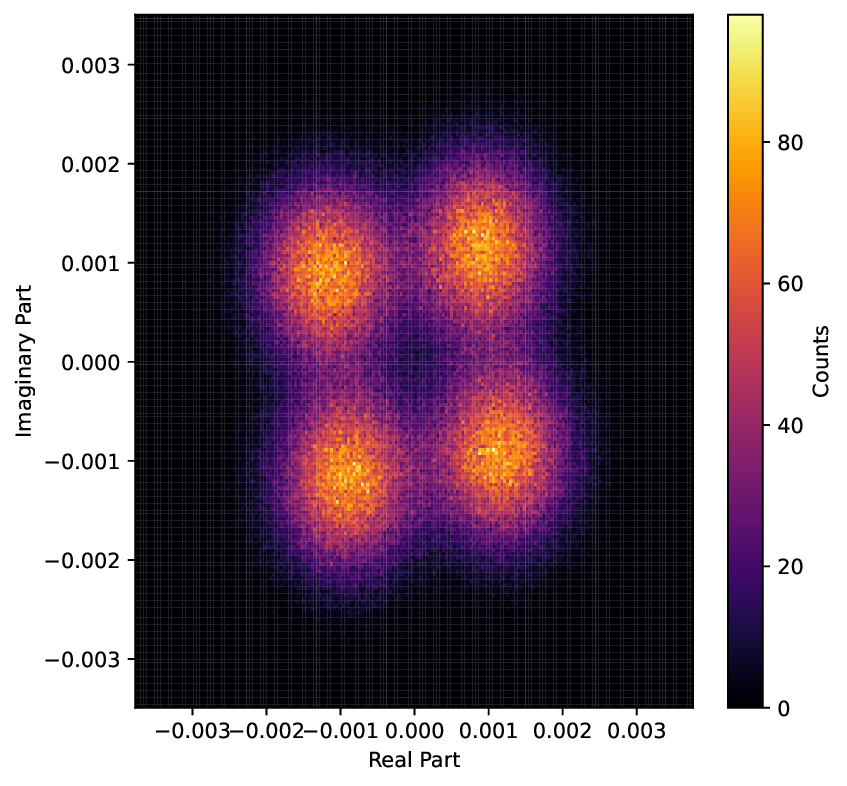}
        \caption{ARISE,\\ $\alpha_{\text{s}}=1.00$, SNR = 5.57 dB}
        \label{fig:sim_9q_SD1}
    \end{subfigure}
    \begin{subfigure}[b]{0.24\linewidth}
        \centering
        \includegraphics[width=\linewidth]{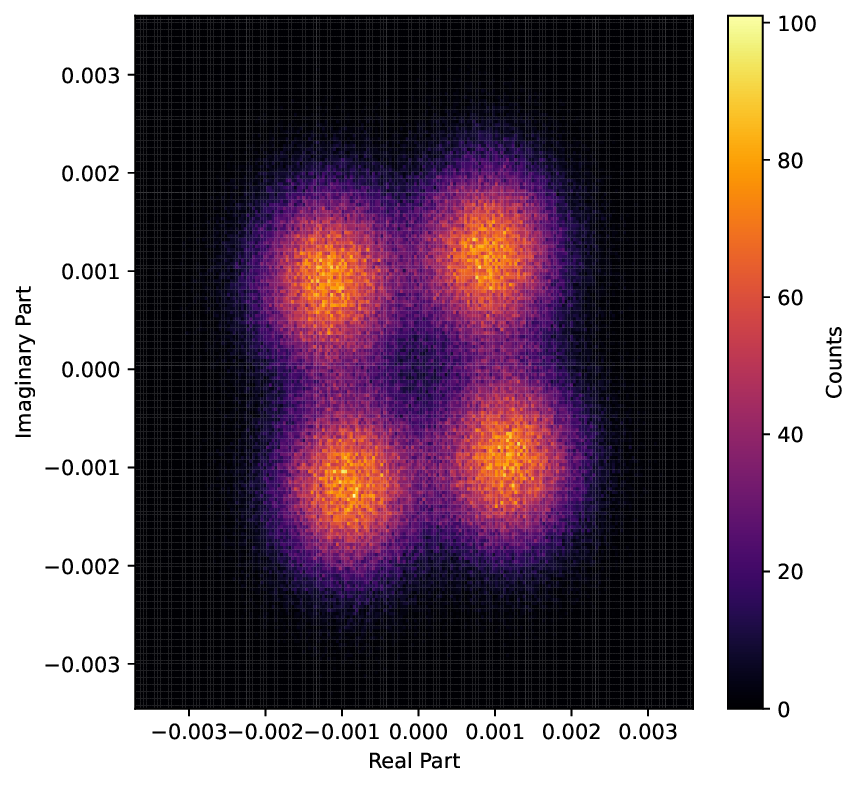}
        \caption{ARISE,\\ $\alpha_{\text{s}}=0.90$, SNR = 5.58 dB}
        \label{fig:sim_9q_SD2}
    \end{subfigure}
    \begin{subfigure}[b]{0.24\linewidth}
        \centering
        \includegraphics[width=\linewidth]{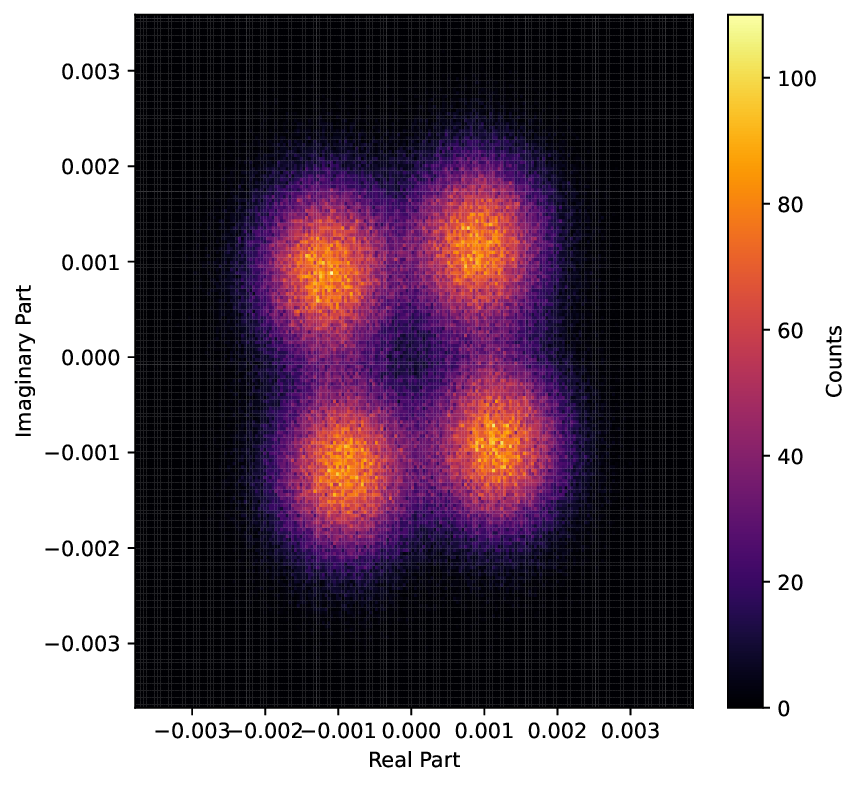}
        \caption{ARISE,\\ $\alpha_{\text{s}}=0.75$, SNR = 5.62 dB}
        \label{fig:sim_9q_SD3}
    \end{subfigure}
    \begin{subfigure}[b]{0.24\linewidth}
        \centering
        \includegraphics[width=\linewidth]{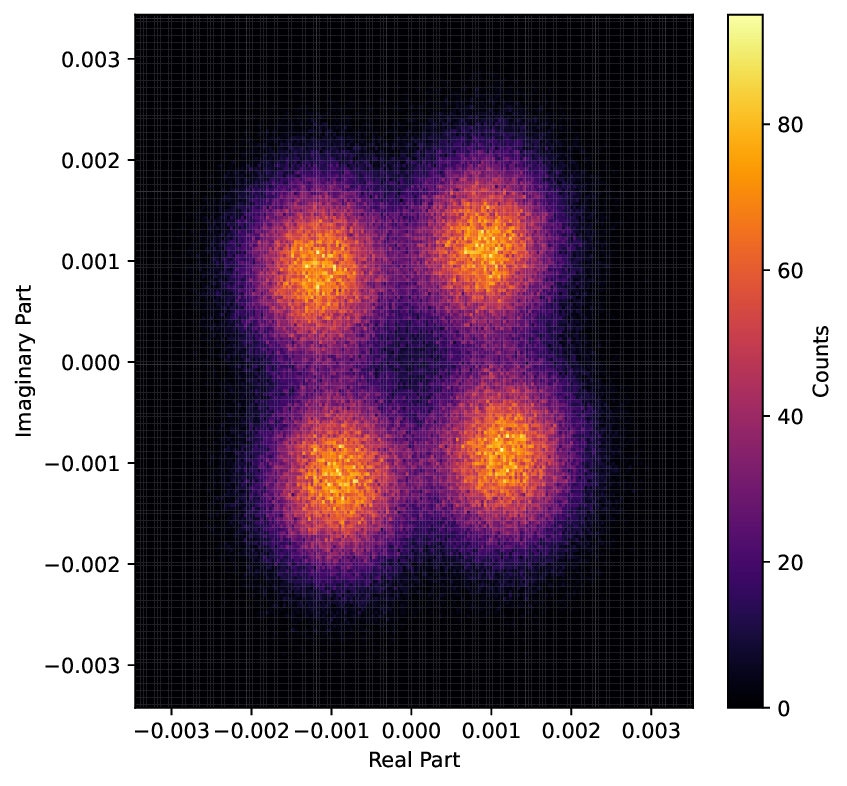}
        \caption{ARISE,\\ $\alpha_{\text{s}}=0.50$, SNR = 5.75 dB}
        \label{fig:sim_9q_SD4}
    \end{subfigure}
    \begin{subfigure}[b]{0.24\linewidth}
        \centering
        \includegraphics[width=\linewidth]{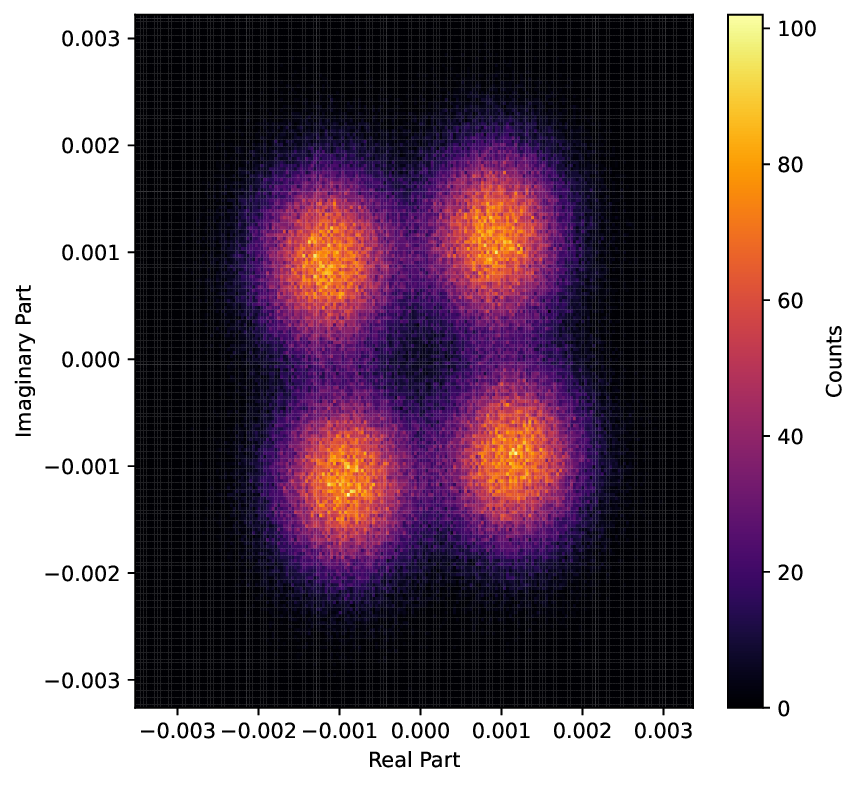}
        \caption{ARISE,\\ $\alpha_{\text{s}}=0.25$, SNR = 6.09 dB}
        \label{fig:sim_9q_SD5}
    \end{subfigure}
    \begin{subfigure}[b]{0.24\linewidth}
        \centering
        \includegraphics[width=\linewidth]{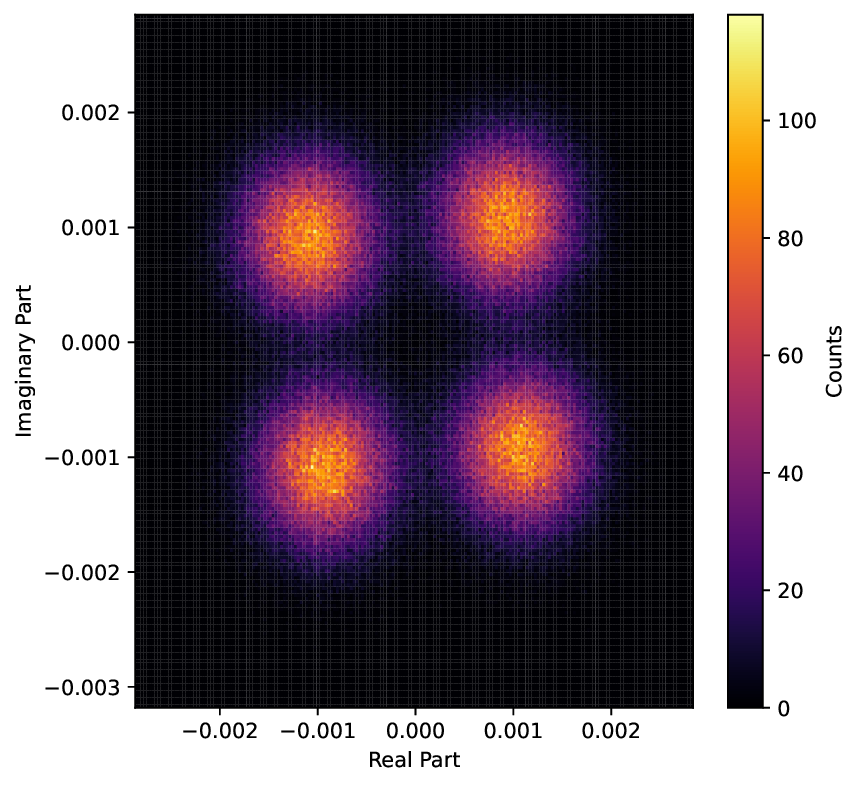}
        \caption{ARISE,\\ $\alpha_{\text{s}}=0.10$, SNR = 7.42 dB}
        \label{fig:sim_9q_SD6}
    \end{subfigure}
    \caption{Converged QPSK constellations at UE location (-20, -20), $M=100, \kappa=0,n_\text{r}=50$.}
    \label{fig:sim_9_qpsk}
\end{figure*}

\begin{figure*}[!htbp]
    \centering
    \begin{subfigure}[b]{0.49\linewidth}
        \centering
        \includegraphics[width=\linewidth]{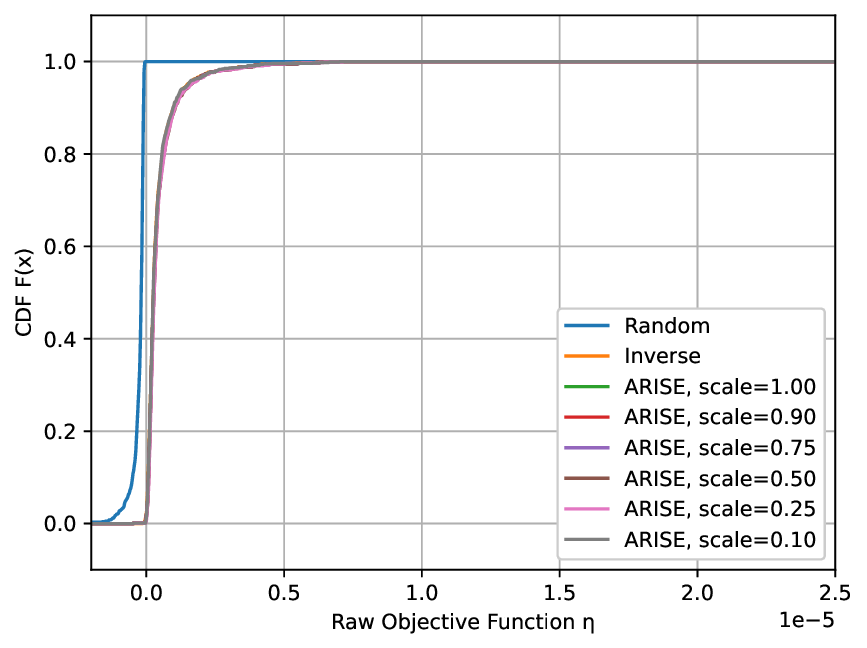}
        \caption{Converged $\eta$ values}
        \label{fig:sim_7_raw}
    \end{subfigure}
    \begin{subfigure}[b]{0.49\linewidth}
        \centering
        \includegraphics[width=\linewidth]{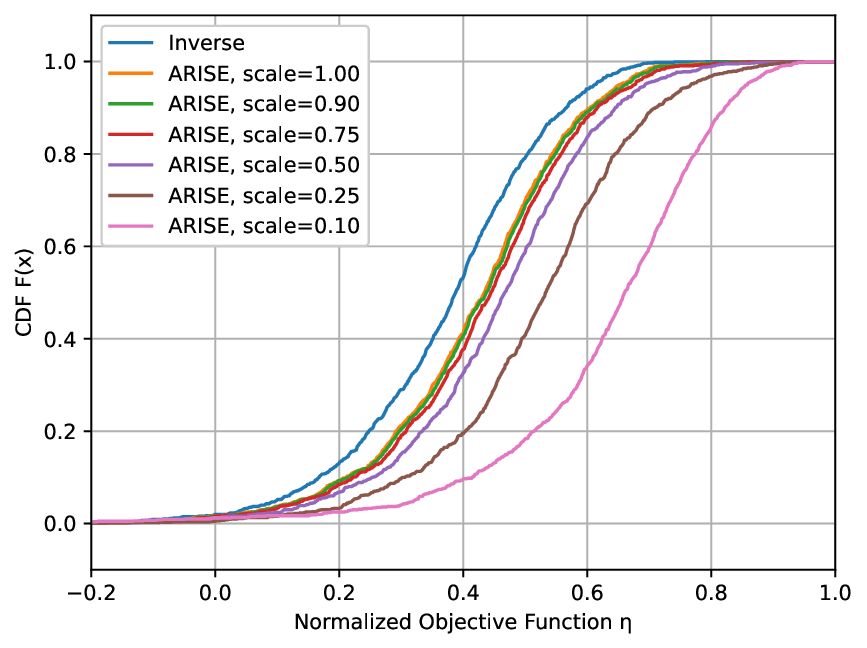}
        \caption{Converged normalized $\eta_{\text{n}}$ values}
        \label{fig:sim_7_norm}
    \end{subfigure}
    \caption{Converged CDFs over 1000 simulations, $M=100, \kappa=0,n_{\text{r}}=50$.}
    \label{fig:sim_7}
\end{figure*}

The first simulation compares the baseline schemes with the converged results from ARISE using various scaling factors $\alpha_{\text{s}}$. In this case, the Rician factor for the BS-RIS-UE link is $\kappa=10$ and the link between the BS and UE is characterized by Rayleigh fading due to obstacles in the path -- modeling pre-existing signal distortion and highlighting the dependency of the wireless environment on the RIS. The number of delayed paths is $n_{\text{r}}=10$, thus the number of ISI terms is $L=20$. The simulation was repeated using the random walk method over 1000 channel coherence blocks. The converged pulse responses are plotted in Fig.~\ref{fig:sim_8_pulse} and corresponding QPSK constellations for each algorithm, assuming the unit interval (UI) $T = 16.3$ ns, are plotted in Fig.~\ref{fig:sim_8_qpsk}, with CDF plots of the converged $\eta$ and normalized $\eta_{\text{n}}$ in Fig.~\ref{fig:sim_6}. The power of the ARISE algorithm is first shown in the pulse responses. It is observed that as the scaling factor $\alpha_{\text{s}}$ decreases and equalization is emphasized more than amplification, the ISI terms in the pulse responses are minimized until they are almost entirely eliminated. This is further demonstrated in the QPSK plots in Fig.~\ref{fig:sim_9_qpsk}, showing the SNR boost and variance reduction in the symbol constellations. The CDF plots in Fig.~\ref{fig:sim_6} further affirm this behavior, showing the ARISE and Inverse Phases schemes achieving near-equal $\eta$ values, with ARISE having strictly higher $\eta_{\text{n}}$ values, showcasing its superior equalization capability. These results are further reinforced in a more complicated scenario where all channels are entirely NLoS ($\kappa=0$) and there are many more ISI terms with $n_{\text{r}}=50$ in Figs.~\ref{fig:sim_9_pulse},~\ref{fig:sim_9_qpsk}, and~\ref{fig:sim_7}.

The feasibility of ARISE critically depends on obtaining sufficiently accurate channel state information (CSI), since its update rule in (\ref{eq:gamma_update_SD}) requires explicit knowledge of the cascaded channel matrix $\boldsymbol{h}_{\text{BRU}}$ to suppress the dominant ISI components. Estimating this high-dimensional channel matrix becomes increasingly burdensome as the RIS scales, and eliminating $L$ ISI taps in principle requires $M(L+1)$ coefficients associated with the $M$ RIS elements and the $(L+1)$ signal components (main tap and ISI). Although practical estimators may jointly recover the main and delayed channels using fewer than $M(L+1)$ pilot configurations, the design of such estimators for RIS-aided wideband channels with ISI lies outside the scope of this work \cite{9847080, 9765815, 9771077}. Importantly, ARISE cannot begin optimizing the RIS until channel estimation is completed, introducing additional delay and communication overhead due to the need for repeated pilot transmissions across multiple RIS configurations. Moreover, ARISE implicitly assumes a linear and frequency-flat RIS response, whereas realistic RIS hardware exhibits element coupling, amplitude losses, phase errors, and frequency-dependent reflection coefficients -- effects that complicate accurate modeling, especially in wideband scenarios \cite{11124274, 9774917, 10407046, 10949779}. We address the above problems by using the DRL-based methods discussed in the following sections, which avoid explicit channel estimation altogether and are inherently model-free. These algorithms only require a single pilot symbol transmission to obtain the pulse response at each time step, and the agent can begin optimizing the RIS as soon as a minimal number of samples are stored in its memory.

\section{DEEP REINFORCEMENT LEARNING}

In this section, we will review general concepts behind DRL, relate it to the RIS equalizer scenario, and introduce three different algorithms to solve the optimization problem (P1).

\subsection{BACKGROUND}

In the most general sense, reinforcement learning (RL) involves an agent that takes actions in a specific environment and receives different rewards for different actions. The goal is to learn a policy that guides the agent to choose actions that maximize long-term rewards through trial and error. In the case of DRL, we employ deep neural networks (DNNs) to model the environment and the agent. The key elements in RL, applied to our RIS case, are described as follows \cite{li2018deepreinforcementlearningoverview, 9110869}:

\begin{enumerate}
    \item \textbf{State:} A set of parameters characterizing the environment. We define the state at time $t$ as $s_t \in \mathcal{S}$ as a function of the received pulse response $\boldsymbol{y}$ at time $t$, denoted by $\boldsymbol{y}_t$, and $\mathcal{S}$ is the continuous state space. We define the notation $y_{t,k}$ to identify the sample at index $k$ from the received pulse response $\boldsymbol{y}_t$. Since the pulse response is complex due to reflection phase shifts in the channel, we expand the pulse of length $L+1$ to an array of length $2(L+1)$ which contains the real and imaginary parts of each sampled point in the pulse, to be used as inputs to the DNNs. The state is defined as the resulting array $s_t=\boldsymbol{y}_t'$ where $\boldsymbol{y}_t'=[\Re\{y_{t,0}\},\Im\{y_{t,0}\},\ldots,\Re\{y_{t,L}\},\Im\{y_{t,L}\}]$.
    \item \textbf{Action:} The choice $a_t \in \mathcal{A}$ of the agent at each time step to move from the current state of the environment $s_t$ to the next state $s_{t+1}$, sampled from the continuous action space $\mathcal{A}$. We define the action as the RIS reflection coefficients $\boldsymbol{\Gamma}$ at time $t$ that will aim to improve the quality of the pulse response, i.e., $a_t=\boldsymbol{\Gamma}_t'$. Here, we also split the real and imaginary parts of the $M$-dimensional reflection coefficient vector $\boldsymbol{\Gamma}_t$ into an array of length $2M$ as $\boldsymbol{\Gamma}_t'=[\Re\{\Gamma_{t,1}\},\Im\{\Gamma_{t,1}\},\ldots,\Re\{\Gamma_{t,M}\},\Im\{\Gamma_{t,M}\}]^{T}$.
    \item \textbf{Reward:} An instantaneous performance metric provided by the environment to inform the agent of the quality of its action $a_t$ at state $s_t$ as it moves to the next state $s_{t+1}$. Here, the reward $r_{t+1}$ is defined as the normalized version of $\eta$ according to~(\ref{eq:eta_norm}), $\eta_{\text{n}}$, expressed by
    \begin{equation}
        r_{t+1} = \frac{\eta}{\sum{_k}{|y_{t+1,k}|^2}}.
        \label{eq:reward_norm}
    \end{equation}
    This normalization ensures the rewards are always between $-1$ at the worst case and $+1$ at the best case, to provide stable gradients when the DNNs are trained. Note that per the definition of $\eta$ in (\ref{eq:eta}) and due to the normalization here, the reward function is penalized when the received signal $\boldsymbol{y}_t$ is dominated by ISI, when the real part of $y_{t,0}$ is small, and when the imaginary part of $y_{t,0}$ is significant, by accounting for the power at each sampled index of $\boldsymbol{y}_t$. Although this approach does not explicitly determine how much gain is achieved by beamforming using the RIS, generally a higher reward value would mean that the signal has either less ISI, a stronger real part of the main tap compared to the rest of the signal, or both. This also indicates that there will exist different local optima in RIS configurations, similar to how the ARISE algorithm performance depends on the scaling of the received signal. This problem invokes the tradeoffs between \textit{exploration} and $\textit{exploitation}$ of the DRL algorithm, which will be discussed later.
    \item \textbf{Experience:} Previous observations and interactions of the agent with the environment stored in memory as $(s_t, a_t, r_{t+1}, s_{t+1})$. 
    \item \textbf{Policy:} Determines the probability of choosing a specific action based on the current state, denoted by $\pi(a_t|s_t)$ and satisfying $\sum_{a_t \in \mathcal{A}}{\pi(a_t|s_t)}=1$. Our goal is to determine a policy that will quickly find a set of stable RIS reflection coefficients to reduce the ISI and improve the SNR of the received wireless signal.
    \item \textbf{State-action value function:} The quality of taking action $a_t$ in state $s_t$ based on the potential future rewards.
\end{enumerate}

We define $V_t$ as the future cumulative discounted reward
\begin{equation}
    V_t=\sum_{\tau=0}^{\infty}{\gamma^\tau r_{t+\tau+1}},
    \label{eq:future_discounted_rewards}
\end{equation}
where $\tau$ takes values from nonnegative integers and $0<\gamma<1$ is the discount rate. The state-action value function is given by the Q-function, which satisfies the Bellman equation
\begin{equation}
    \begin{split}
         Q_{\pi}(s_t,a_t) & =  E_{\pi}[V_t | s_t=s,a_t=a] \\ 
         & =  E_{\pi}[r_{t+1}|s_t=s,a_t=a] \\
         & +\gamma \sum_{s'\in \mathcal{S}}{P_{ss'}^{a} \left( \sum_{a'\in \mathcal{A}}{\pi(a'|s')Q_{\pi}(s',a')}\right)}
    \end{split}
    \label{eq:q_func}
\end{equation}
and can be learned recursively to find the optimal policy by utilizing the Q-learning algorithm \cite{10.5555/1893145}. In the above, $P_{ss'}^{a}$ is the transition probability from state $s$ to state $s'$ given action $a$ in the Markov decision process (MDP). A target Q-function is initialized and then learned to achieve the optimal Q-function, with the expectation taken over all possible next states $s_{t+1}$,
\begin{equation}
    Q^*(s_t,a_t) = E \left[ r_{t+1} +\gamma \max_{a_{t+1}} {Q^*(s_{t+1},a_{t+1})}\right],
    \label{eq:bellman}
\end{equation}
using the update rule
\begin{equation}
    \begin{split}
        Q^*(s_t,a_t) & \gets (1-\alpha)Q^*(s_t,a_t) \\ 
        & +\alpha \left(r_{t+1}+\gamma \max_{a_{t+1}} {Q_{\pi}(s_{t+1},a_{t+1})} \right),
    \end{split}
\label{eq:q_update}
\end{equation}
where $\alpha$ is the learning rate. By definition, the Q-function determines the future cumulative discounted rewards based on a given policy, the current state, and the choice of action. The optimal Q-function -- the one that would yield the highest future rewards for the current state and action -- would follow the desired optimal policy, $\pi^*$.

Considering the context of this work, the action space is high-dimensional and scales with the size of the RIS, prompting the use of machine learning to estimate the Q-function and optimal policy using two DNNs. The \textit{critic} DNN estimates the Q-function based on the states, actions, and rewards as the agent acts in the environment. Using the symbol $\theta$ to represent the critic DNN parameters (neuron weights, activation functions, number of layers, etc.), we have $Q(s_t,a_t)=Q_{\theta}(s_t,a_t)$. The \textit{actor} DNN seeks the optimal policy, taking the current state as the input and estimating an action at the output. Depending on the algorithm choice, each of the two DNNs may also split into training and target DNNs \cite{Silver2014DeterministicPG}:
\begin{itemize}
    \item \textbf{Training critic} $Q_{\theta}( s_t,a_t)$ learns the Q-function via gradient descent.
    \item \textbf{Target critic} $Q_{\theta_{\text{targ}}}(s_t,a_t)$ provides a slowly updated copy of the critic used to provide stable time-difference (TD) targets.
    \item \textbf{Training actor} $\mu_{\phi}(s_t)$ is the policy network and is optimized using the deterministic policy gradient.
    \item \textbf{Target actor} $\mu_{\phi_{\text{targ}}}(s_t)$ is a slowly updated copy of the actor used to generate stable next-state actions for loss function computations during training.
\end{itemize}
Note that the DNNs can now be trained by updating their weights using stochastic optimization algorithms and backpropagation as 
\begin{equation}
    \theta_{t+1}=\theta_{t}-\mu \nabla_{\theta} l(\theta),
\end{equation}
where $\mu$ is the learning rate and $\nabla_{\theta} l(\theta)$ is the gradient of the loss function $l(\theta)$ \cite{rumelhart1986learning}. Modern machine learning frameworks like TensorFlow \cite{tensorflow2015} automatically compute gradients through backpropagation, so we do not need to derive these expressions manually.

Another key component of DRL is the \textit{experience replay buffer}, which stores the interactions of the agent with the environment. Training the DNNs on randomly sampled batches from this buffer, rather than on consecutive states, reduces the strong temporal correlations present in the data. This increased stochasticity helps prevent the networks from overfitting to recent experiences and becoming trapped in local optima, thereby improving overall learning performance. This mechanism also interacts with the broader concept of \textit{exploration versus exploitation}: exploration seeks out new states, often through noisy or randomized actions, while exploitation leverages previously collected experiences, such as those stored in the replay buffer, to refine the policy.

The principles discussed in this subsection provide the foundation for the DRL algorithms presented next: DDPG, TD3, and SAC, where these ideas are applied in concrete algorithmic frameworks.

\begin{figure}[!htbp]
    \centering
    \includegraphics[width=1.0\linewidth]{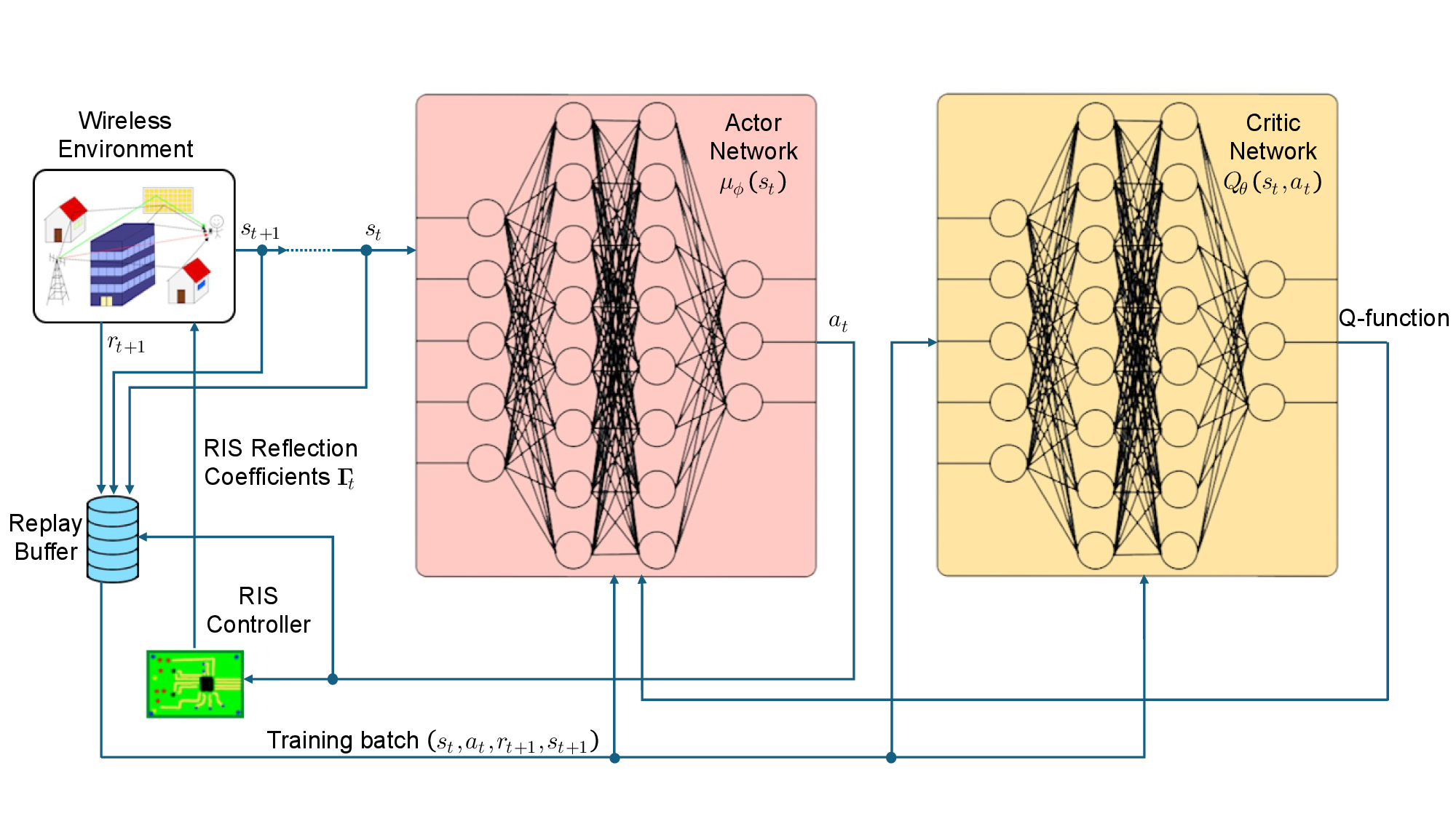}
    \caption{Illustration of the DRL-based RIS equalizer implementation. The actor (policy) DNN takes the state from the environment and outputs the action corresponding to RIS reflection coefficients that should improve the signal integrity. The RIS controller updates the RIS reflection coefficients and the environment changes accordingly. The environmental states (pulse responses), actions (RIS reflection coefficients), and rewards (normalized $\eta_{\text{n}}$ metrics) are stored in a replay buffer, from which training batches are extracted to optimize the DNN hyperparameters. The actor is also trained using the output of the critic (Q-network).}
    \label{fig:drl_model}
\end{figure}

\subsection{DDPG}

Deep Deterministic Policy Gradient (DDPG) is an off-policy (learns a target policy from data generated by a different behavior policy \cite{lillicrap2019continuouscontroldeepreinforcement}), model-free actor-critic algorithm designed for continuous control tasks. It combines the deterministic policy gradient framework with techniques from DRL, using separate actor and critic networks along with target networks and a replay buffer. The critic learns an action-value function, while the actor is updated by backpropagating the gradient of the critic with respect to the action, enabling the learning of a deterministic policy in high-dimensional continuous action spaces \cite{lillicrap2019continuouscontroldeepreinforcement, Silver2014DeterministicPG}. The actor DNN takes the state as its input and outputs the continuous action based on its learned policy, and the critic takes the state and the action, and outputs the estimated Q-function $Q(s_t,a_t)$ according to (\ref{eq:q_func}) \cite{9110869}. Referring to~(\ref{eq:q_update}), the role of the actor is to find the best action that will result in the $\max_{a_{t+1}}{Q_\pi (s_{t+1},a_{t+1})}$ term, thus eliminating the need for exhaustive search and non-convex optimization. DDPG uses a deterministic policy $\mu_{\phi}(s_t)$, which serves the same role as a stochastic policy $\pi(a_t|s_t)$ but outputs a single action rather than a distribution over actions.

The top-level algorithm is outlined as follows. We have an RIS-aided environment as described in the previous section, where the current state $s_t$ is defined as the received sampled pulse response at time $t$, $\boldsymbol{y}_t'$, and the current action $a_t=\mu_{\phi}(s_t)$ is the update to the reflection coefficients of the RIS by passing $\boldsymbol{y}_t'$ into the policy DNN, yielding $\boldsymbol{\Gamma}_t'$. We normalize the $\boldsymbol{y}_t'$ array to the limits $[-1,1]$ to produce stable gradients when training the DNNs using backpropagation \cite{rumelhart1986learning}. We define the current ``step'' as the current state of the environment with the current RIS configuration, and the current ``episode'' as the time frame during which the wireless channels remain coherent. At each state, we induce exploration by adding a random noise vector $\boldsymbol{\epsilon}$ to the RIS reflection coefficients, clip the reflection coefficients to $[-1,1]$, and assign unity magnitudes after combining the separated real and imaginary parts from $\boldsymbol{\Gamma}_t'$ into $\boldsymbol{\Gamma}_t$ and using~(\ref{eq:ris_mag_norm}). After updating the RIS reflection coefficients, we obtain the next (normalized) state $s_{t+1}=\boldsymbol{y}_{t+1}'$ and next reward $r_{t+1}$ from~(\ref{eq:reward_norm}), and store $(s_t,a_t,r_{t+1},s_{t+1})$ in the replay buffer $\mathcal{D}$.

To update the DNNs, we randomly sample a batch $\mathcal{B}$ of size $B$ from the replay buffer, compute and apply the gradients to all DNNs as shown in Algorithm~\ref{alg:ddpg}. The loss function used to train the critic with gradient descent is given by the mean-squared Bellman error (MSBE) function, used to determine how close $Q_{\theta}(s,a)$ comes to satisfying the Bellman equation from~(\ref{eq:bellman}) \cite{Mnih2015},
\begin{equation}
    l(\theta) = E \Big[ \big( Q_{\theta} (s,a)
    - (r' + \gamma Q_{\theta_{\text{targ}}}(s',\mu_{\phi_{\text{targ}}}(s'))) \big)^2 \Big],
    \label{eq:ddpg_critic_update}
\end{equation}
where the expectation is taken over a randomly selected batch of $(s,a,r',s')$ from the replay buffer $\mathcal{D}$.
The actor is trained using gradient ascent with the loss function
\begin{equation}
    l(\phi)=E \Big[ Q_{\theta}\big(s, \mu_{\phi}(s)\big) \Big],
    \label{eq:ddpg_actor_update}
\end{equation}
to learn a policy that maximizes the Q-function. The target DNNs are updated using Polyak averaging \cite{doi:10.1137/0330046} with a weighting factor $\tau$,
\begin{equation}
    \theta_{\text{targ}} \gets \tau \theta + (1-\tau)\theta_{\text{targ}}.
\end{equation}

Note that during each time step $t$, the actor and critic DNNs may be trained $N_{\text{train}}$ times by sampling random batches from the replay buffer. However, they will only start training once the replay buffer contains sufficiently many samples to form a batch.

\begin{algorithm}[!htbp]
    \caption{Deep Deterministic Policy Gradient for RIS Equalizer}
    \label{alg:ddpg}
    \begin{algorithmic}[1]
        \State {Initialize both training and target DNNs for both actor and critic, assign $\phi_{\text{targ}} \gets \phi, \quad \theta_{\text{targ}} \gets \theta$.}
        \For {$\text{episode}$ in $N_{\text{ep}}$}
            \For {$t$ in $N_{\text{steps}}$}
                \State {Observe state $s_t \gets \boldsymbol{y}_{t}'$ and set RIS coefficients \[\boldsymbol{\Gamma}_t'\gets\text{clip}(\mu_{\phi_{\text{targ}}}(\boldsymbol{y}_{t}')+\boldsymbol{\epsilon},-1,1), \quad \boldsymbol{\epsilon} \sim \mathcal{N}(0,\sigma_{\text{a}}^{2})^{2M \times 1}.\]}
                \State {Set $a_t \gets \boldsymbol{\Gamma}_t'$.}
                \State {Update RIS reflection coefficients as \[\boldsymbol{\Gamma}_t \gets [\Gamma_{t,1}'+j\Gamma_{t,2}',\ldots,\Gamma_{t,2M-1}'+j\Gamma_{t,2M}']^T\] and normalize using~(\ref{eq:ris_mag_norm}).}
                \State {Observe new state $s_{t+1}\gets\boldsymbol{y}_{t+1}'$.}
                \State {Calculate rewards $r_{t+1}$ using~(\ref{eq:reward_norm}).}
                \State {Store $(s_t,a_t,r_{t+1},s_{t+1})$ in the replay buffer $\mathcal{D}$.}
                \If {$|\mathcal{D}| \geq B$}
                    \For {$i$ in $N_{\text{train}}$}
                        \State {Sample batch $\mathcal{B}=\{(s,a,r',s')\}$ from $\mathcal{D}$.}
                        \State {Calculate $l(\theta)$ using~(\ref{eq:ddpg_critic_update}).}
                        \State {Update critic: \[ \theta \gets \theta - \mu_{\text{c}} \nabla_{\theta} l(\theta). \]}
                        \State {Calculate $l(\phi)$ using~(\ref{eq:ddpg_actor_update}).}
                        \State {Update actor: \[ \phi \gets \phi - \mu_{\text{a}} \nabla_{\phi} l(\phi). \]}
                        \State {Soft-update target critic: \[\theta_{\text{targ}} \gets \tau \theta + (1-\tau)\theta_{\text{targ}}.\]}
                        \State {Soft-update target actor: \[\phi_{\text{targ}} \gets \tau \phi + (1-\tau)\phi_{\text{targ}}.\]}
                    \EndFor
                \EndIf
            \EndFor
        \EndFor
    \end{algorithmic}
\end{algorithm}

\subsection{TD3}

Although DDPG can achieve strong performance, it often suffers from overestimation bias in the learned Q-function, which can destabilize training and increase sensitivity to hyperparameters \cite{fujimoto2018addressingfunctionapproximationerror}. Twin Delayed DDPG (TD3) is an algorithm that builds on the foundation of DDPG and seeks to address the above issues. It introduces the concept of \textit{clipped double Q-learning}, where two ``twin'' critic networks are learned (each with its own training and target networks), $Q_{\theta_1}$ and $Q_{\theta_2}$. During the gradient update, we modify the loss function in~(\ref{eq:ddpg_critic_update}) to take the minimum between the two target critic Q-function estimators
\begin{equation}
\begin{split}
    l(\theta_{i}) & = E \Bigg[ \Big( Q_{\theta_{i}} (s,a) \\ 
    &-(r' + \gamma \min_{i=1,2}{Q_{\theta_{\text{targ},i}}(s',\mu_{\phi_{\text{targ}}}(s')))} \Big)^2 \Bigg],
\end{split}
\label{eq:td3_critic_update}
\end{equation}
thus accounting for the overestimation bias.

Additionally, TD3 introduces \textit{delayed policy updates} to reduce the error in the value function estimates and decrease the risk of divergence, by updating the actor network and target networks every $d_{\text{TD3}}$ steps. This allows the critic to make more progress among policy updates, reducing the propagation of critic estimation errors into the policy.

Finally, TD3 uses \textit{target policy smoothing regularization} during training. When computing the target value to train the critic, TD3 adds clipped noise to the output of the target actor $\mu_{\phi_{\text{targ}}}(s)$, encouraging the critic to learn a smoother Q-function and preventing exploitation of sharp value peaks.

The pseudocode for the TD3 algorithm is given in Appendix B as Algorithm~\ref{alg:td3}.

\subsection{SAC}

Soft Actor-Critic (SAC) extends the actor-critic framework by introducing \textit{maximum entropy reinforcement learning}, which augments the standard objective with an entropy term that encourages exploration and robustness \cite{sac1}. Unlike DDPG and TD3, which rely on deterministic policies, SAC learns a stochastic policy $\pi_{\phi}(a|s)$ that maximizes both expected return and policy entropy. The resulting objective is
\begin{equation}
    J(\pi) = E_{\pi} \left[ \sum_{t=0}^{\infty}{\gamma^t \left( r(s_t,a_t) + \alpha H(\pi(\cdot | s_t)) \right)} \right],
\end{equation}
where $\alpha$ is the temperature parameter that controls the tradeoff between reward maximization and the entropy $H(\cdot)$, and $\gamma$ is the discount rate from (\ref{eq:future_discounted_rewards}). Note that the exploration in DDPG and TD3 solely depends on adding noise to the RIS reflection coefficients that are selected from the policy DNN. In the case of SAC, the stochastic policy yields different actions by producing Gaussian distributions for each RIS reflection coefficient and randomly sampling from them.

SAC employs three main DNNs: a stochastic policy network $\pi_{\phi}$ as the actor, and two Q-function approximators $Q_{\theta_1}$ and $Q_{\theta_2}$ as the critics (with their training and target networks). The two critics resemble the double Q-learning introduced in TD3 to mitigate overestimation bias by taking the minimum of their outputs when forming the target for the loss function. However, SAC differs in that the loss function incorporates both the next-state value and the entropy of the policy. The new MSBE loss function to train the critics thus becomes
\begin{equation}
\begin{split}
    & l(\theta_{i}) = E \Bigg[ \bigg( Q_{\theta_{i}} (s,a)- \\ 
    & \Big(r' + \gamma\big( \min_{i=1,2}{Q_{\theta_{\text{targ},i}}(s',\mu_{\phi_{\text{targ}}}(s')) - \alpha \log {\pi_{\phi}(a'|s')}\big)\Big)} \bigg)^2 \Bigg].
\end{split}
\label{eq:sac_critic_update}
\end{equation}

The actor aims to maximize both the expected future rewards and the expected future entropy. The actor loss function is
\begin{equation}
    l(\phi)=E_{a \sim \pi}[\min_{i=1,2} {Q_{\theta_{i}}(s,a)}-\alpha \log {\pi_{\phi}(a|s)}].
    \label{eq:sac_actor_update}
\end{equation}

A key feature of SAC is the \textit{automatic temperature tuning mechanism}, which adjusts $\alpha$ during training to maintain a target entropy level \cite{sac2}. This allows the algorithm to adaptively balance exploration and exploitation without manual tuning of the entropy weight, using the loss function for gradient descent
\begin{equation}
    l(\log{\alpha})=E_{a \sim \pi}[-\log{\alpha}( \log {\pi_{\phi}(a|s)} - \mathcal{H})],
    \label{eq:sac_alpha_update}
\end{equation}
where $\mathcal{H}=-\dim{\mathcal{A}}=-2M$ is the target entropy and $\mathcal{A}$ is the action space corresponding to the $2M$ real and imaginary parts of RIS reflection coefficients. We update the log value of $\alpha$ to avoid accidental negative $\alpha$ values and ensure smooth and stable gradients. The pseudocode for SAC is given in Appendix B as Algorithm~\ref{alg:sac}.

Overall, SAC provides several advantages over deterministic methods such as DDPG and TD3, mainly due to its improved exploration through stochasticity, reduced sensitivity to hyperparameters, and enhanced stability due to entropy regularization and double Q-learning. All the above algorithms are effective contestants for high-dimensional continuous control tasks such as the RIS equalization problem discussed in this work. In the following subsection, we discuss the computational and communication costs of each of these algorithms to provide a fair comparison between them and the ARISE algorithm.

\subsection{COMPLEXITY ANALYSIS}

Following the structure of the update rules in DDPG, TD3, and SAC, we derive the per-update computational cost in terms of the number of forward/backward passes through the actor and critic networks. To our knowledge, no prior work provides a formal asymptotic complexity analysis of these algorithms. We define the cost of a single forward/backward pass per batch through a single DNN as $C$, which depends on the architecture and hyperparameters of each DNN and the batch size. We assume all batch sizes are the same and the DNN architectures across the three different algorithms are similar such that $C$ does not vary significantly among the algorithms. Note that the overall computational complexity until convergence of each algorithm would vary depending on its performance for the RIS equalization problem -- whether training is required over a long period of time, or if the reflection coefficients can be optimized quickly. The per-step complexities of all the algorithms are summarized in Table~\ref{tab:complexity}.

For DDPG, we update the actor and critic DNNs $N_{\text{train}}$ times per step, yielding $T_{\text{DDPG}}=\mathcal{O}(N_{\text{train}}(C_{\text{critic}}+C_{\text{actor}}))$, where $C_{\text{critic}}$ is the cost per critic update, and $C_{\text{actor}}$ is the cost per actor update. We leave the target DNN updates out of this expression, since the complexity of Polyak averaging is insignificant compared to that of a forward/backward pass through a DNN.

In the case of TD3, we have two critic DNNs and a single actor. The actor updates every $d_{\text{TD3}}$ iterations, yielding $T_{\text{TD3}}=\mathcal{O}(N_{\text{train}}(2C_{\text{critic}}+\frac{1}{d_{\text{TD3}}}C_{\text{actor}}))$. SAC also has two critic DNNs and a single actor, with an actor update every iteration. Assuming the temperature and target updates are negligible compared to the backpropagation required for DNN training, we obtain $T_{\text{SAC}} = \mathcal{O}(N_{\text{train}}(2C_{\text{critic}}+C_{\text{actor}}))$.

To derive the per-update complexity of ARISE, we refer to the update rule in~(\ref{eq:gamma_update_SD}). We perform vector multiplications on $M$-dimensional vectors $L+1$ times due to the length of the captured pulse response. Under the assumption that each time step introduces a new pulse response with $L+1$ taps to equalize, we take the expectation over $L+1$ signal taps to create the ensemble average, yielding a final per-step cost of $T_{\text{ARISE}}=\mathcal{O}(M(L+1)^2)$ \cite{sayed2011adaptive}. Note that for the stochastic gradient descent (SGD) version of ARISE in~(\ref{eq:gamma_update_SGD}), the cost becomes $T_{\text{ARISE-SGD}}=\mathcal{O}(M(L+1))$ as the pulse is processed only a single time per RIS update, without averaging an ensemble.

\begin{table}[!htbp]
    \centering
    \begin{tabular}{|p{0.3\linewidth}|p{0.6\linewidth}|}
        \hline
        \textbf{Algorithm} & \textbf{Per-Step Complexity} \\
        \hhline{|=|=|}
        ARISE & $\mathcal{O}(M(L+1)^2)$ \\
        \hline
        ARISE-SGD & $\mathcal{O}(M(L+1))$ \\
        \hline
        DDPG & $\mathcal{O}(N_{\text{train}}(C_{\text{critic}}+C_{\text{actor}}))$ \\
        \hline
        TD3 & $\mathcal{O}(N_{\text{train}}(2C_{\text{critic}}+\frac{1}{d}C_{\text{actor}}))$ \\
        \hline
        SAC & $\mathcal{O}(N_{\text{train}}(2C_{\text{critic}}+C_{\text{actor}}))$\\
        \hline
    \end{tabular}
    \caption{Per-step computational complexities of the proposed algorithms.}
    \label{tab:complexity}
\end{table}

The per-step complexity ultimately depends on the size of the DNNs and their activation functions that yield $C_{\text{critic}}$ and $C_{\text{actor}}$. Depending on the available hardware at the RIS controller, the training of the DNNs may be accelerated to reduce the amount of time required per step.

Furthermore, the communication complexities associated with ARISE and DRL are significantly different from each other. As previously discussed, the success of our ARISE algorithm primarily depends on available CSI of the $\boldsymbol{h}_{\text{BRU}}$ cascaded channel matrix for the gradient descent rule to work correctly. If we have $M$ channels and $L+1$ samples in the pulse response, we would need to estimate $M(L+1)$ channel coefficients, which would require highly complex channel estimation procedures that scale with the size of the RIS and amount of equalization required, as documented in recent works such as \cite{9328485,9133156,9366894,9614196,9839429}. This creates a bottleneck in RIS optimization because the channels must be estimated before ARISE can begin -- a process which usually requires configuring the RIS in enough states such that the channel matrix can be determined accurately. This estimation process would take time and may degrade the signal quality during its runtime. Furthermore, to capture the pulse responses after each RIS update in practice, we either need to calculate the expected pulse offline (which requires channel estimation for the BS-UE link $\boldsymbol{h}_{\text{BU}}$), or reconfigure the RIS and obtain the pulse response from the receiver. For ARISE-SGD, more time steps would be required until convergence due to noisy gradients. In contrast, the DRL algorithms analyzed above eliminate the need for explicit channel estimation and can operate directly on the received pulse responses. DRL can start the optimization of the RIS to improve the signal quality at any given point in time, as long as it is provided with the pulse response after each RIS update. This pulse response information transfer may be achieved reliably via out-of-band (OOB) communication without disrupting live signal traffic, with low latency, and independently of the RIS configuration \cite{8114345}. Furthermore, the model-free operation enabled by DRL allows it to naturally accommodate RIS nonlinearities, coupling effects, and frequency-dependent behavior, making it a more scalable and robust approach for wideband RIS-aided equalization.

Overall, although ARISE may be computationally lightweight, its practicality is dominated by the significantly higher complexity of the channel estimation procedures it relies on, as well as scaling of the RIS and amount of equalization required. In contrast, the per-update computational costs of the DRL algorithms analyzed above are dominated by the costs associated with forward/backward passes through their actor and critic networks, but they eliminate the need for explicit channel estimation and can operate directly on received pulse responses to optimize the RIS right away.
\section{COMPARATIVE SIMULATION RESULTS}

\begin{table}[!htbp]
    \centering
    \begin{tabular}{|l|p{3cm}|p{1.8cm}|}
        \hline
         \textbf{Variable Name} & \textbf{Description} & \textbf{Value} \\
         \hhline{|=|=|=|}
         $N_{\text{train}}$ & Initial actor and critic updates per step & 4 \\ \hline
         $B$ & Batch size & 4 \\ \hline
         $D_{\text{max}}$ & Max replay buffer size & 400 \\ \hline
         $d_{\text{TD3}}$ & SAC actor update period & 2 \\ \hline
         $\sigma_{\text{a}}$ & DDPG and TD3 initial action noise standard deviation & 0.3 \\ \hline
         $\tau_{\text{d}}$ & Noise decay rate & 0.999 \\ \hline
         $\sigma^2_{\text{t}}$ & TD3 training action standard deviation & 0.3 \\ \hline
         $c_{\text{e}}$ & TD3 training action noise bound & 0.5 \\ \hline
         $\mu_{\text{c}}$ & Critic learning rate & $10^{-3}$ \\ \hline
         $\mu_{\text{a}}$ & Actor learning rate & DDPG: $10^{-3}$; TD3 and SAC: $2\times10^{-3}$ \\ \hline
         $\mu_{\alpha}$ & SAC $\alpha$ learning rate & $10^{-3}$ \\ \hline
         $\tau$ & Target network Polyak averaging factor & 0.001 \\ \hline
    \end{tabular}
    \caption{DRL implementation parameters.}
    \label{tab:drl_params}
\end{table}

\begin{figure*}[!htbp]
    \centering
    \begin{subfigure}[b]{0.24\linewidth}
        \centering
        \includegraphics[width=\linewidth]{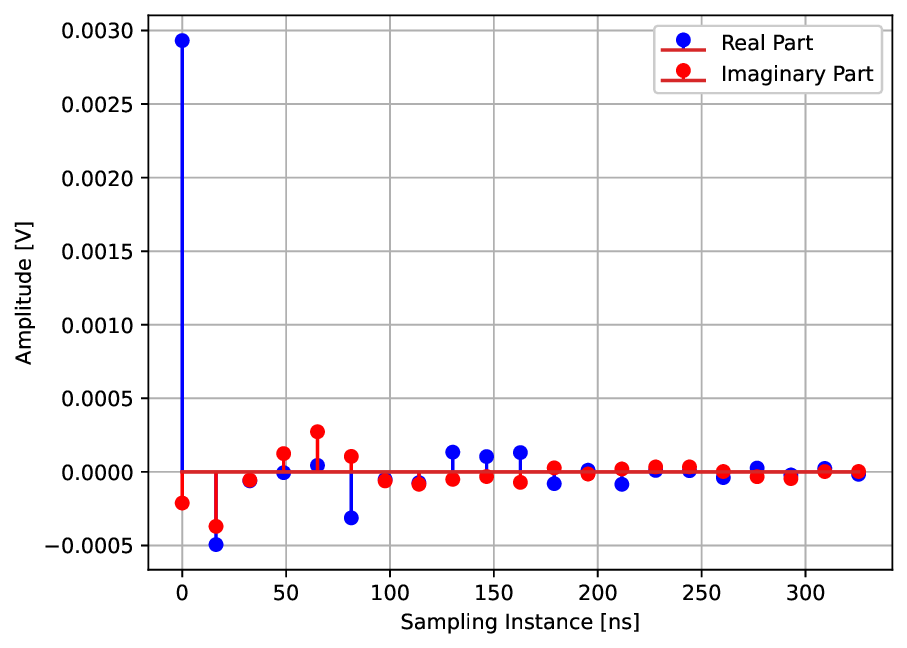}
        \caption{ARISE, $\eta_{\text{n}}$ = 0.848}
        \label{fig:sim_1p_SD}
    \end{subfigure}
    \begin{subfigure}[b]{0.24\linewidth}
        \centering
        \includegraphics[width=\linewidth]{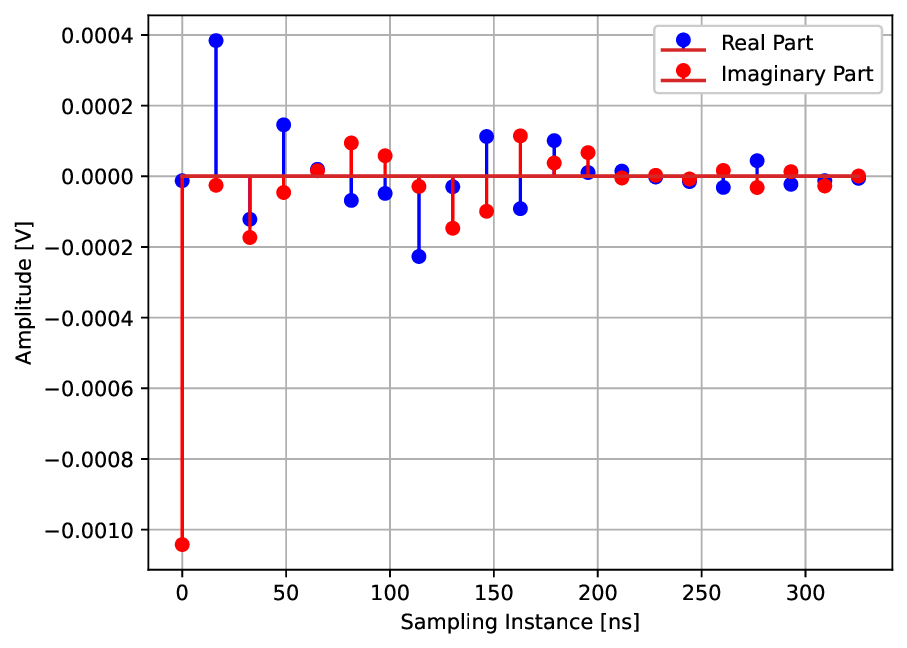}
        \caption{DDPG, $\eta_{\text{n}}$ = -0.258}
        \label{fig:sim_1p_ddpg}
    \end{subfigure}
    \begin{subfigure}[b]{0.24\linewidth}
        \centering
        \includegraphics[width=\linewidth]{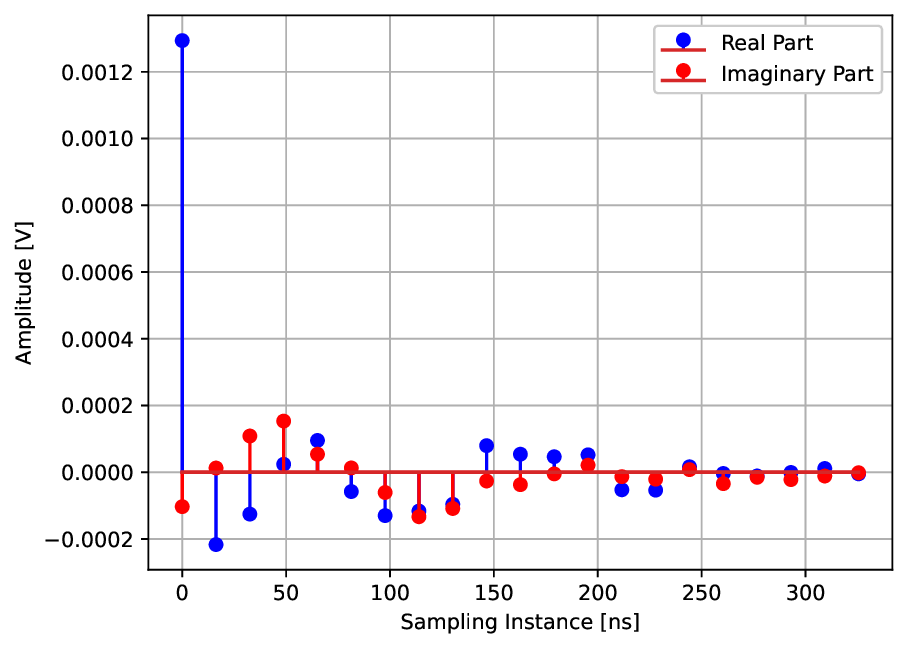}
        \caption{TD3, $\eta_{\text{n}}$ = 0.770}
        \label{fig:sim_1p_td3}
    \end{subfigure}
    \begin{subfigure}[b]{0.24\linewidth}
        \centering
        \includegraphics[width=\linewidth]{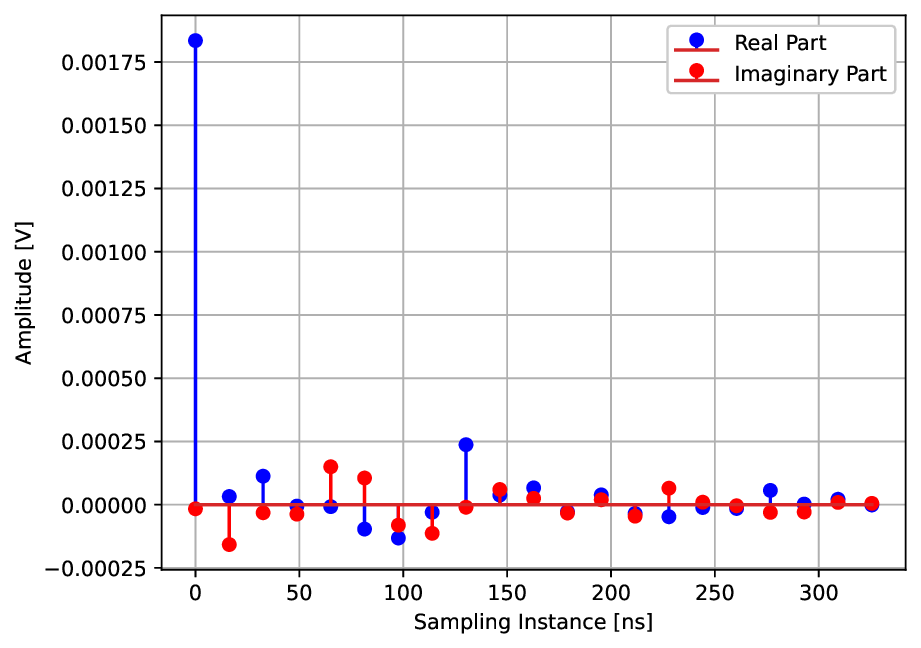}
        \caption{SAC, $\eta_{\text{n}}$ = 0.883}
        \label{fig:sim_1p_sac}
    \end{subfigure}
    \caption{Converged pulse responses real and imaginary parts at UE location (-20, -20), $M=100, \kappa=10,n_{\text{r}}=10$.}
    \label{fig:sim_1_pulse}
\end{figure*}

\begin{figure*}[!htbp]
    \centering
    \begin{subfigure}[b]{0.24\linewidth}
        \centering
        \includegraphics[width=\linewidth]{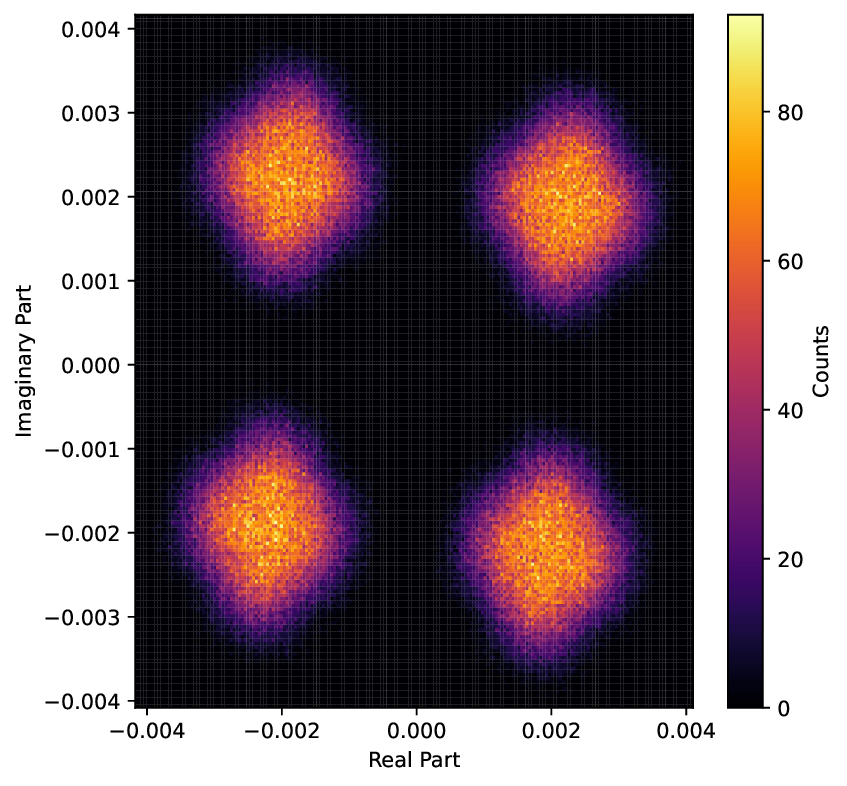}
        \caption{ARISE, SNR = 11.01 dB}
        \label{fig:sim_1q_SD}
    \end{subfigure}
    \begin{subfigure}[b]{0.24\linewidth}
        \centering
        \includegraphics[width=\linewidth]{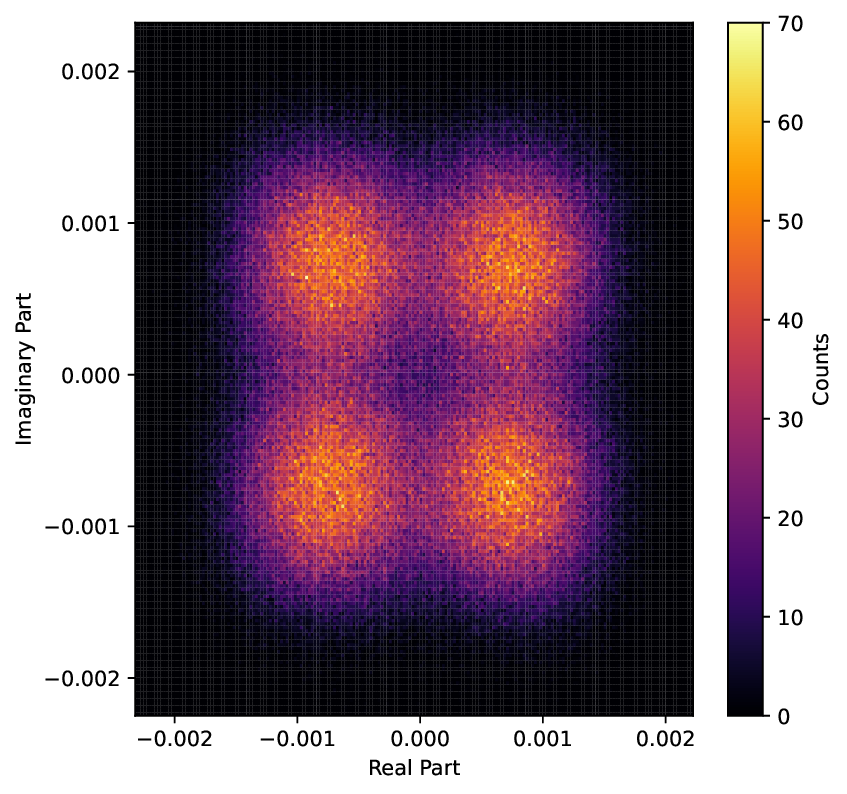}
        \caption{DDPG, SNR = 5.20 dB}
        \label{fig:sim_1q_ddpg}
    \end{subfigure}
    \begin{subfigure}[b]{0.24\linewidth}
        \centering
        \includegraphics[width=\linewidth]{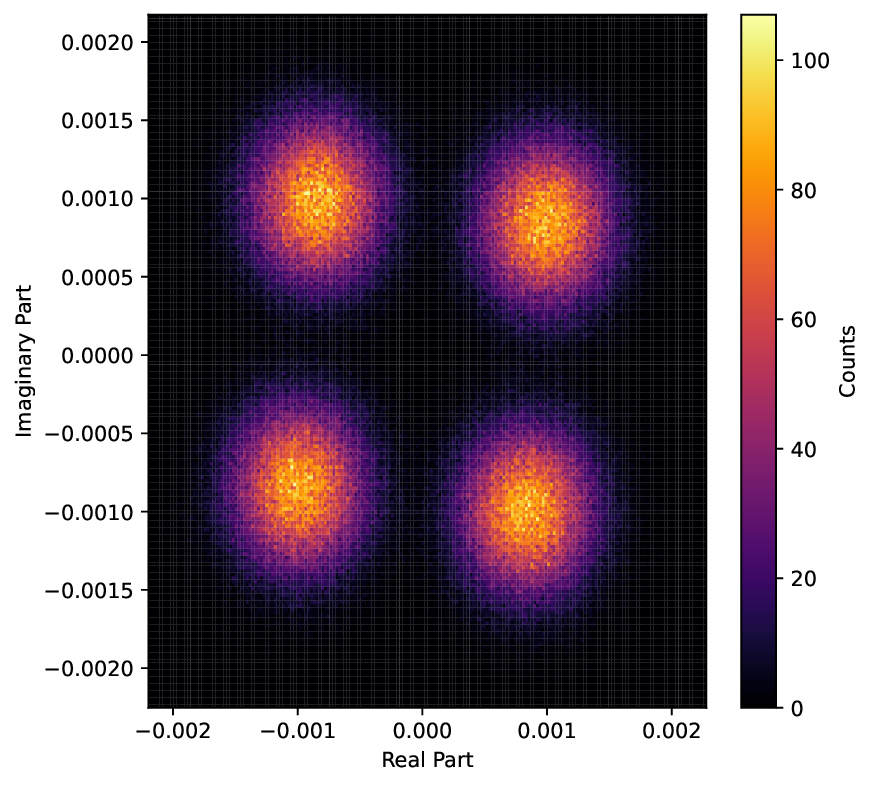}
        \caption{TD3, SNR = 9.00 dB}
        \label{fig:sim_1q_td3}
    \end{subfigure}
    \begin{subfigure}[b]{0.24\linewidth}
        \centering
        \includegraphics[width=\linewidth]{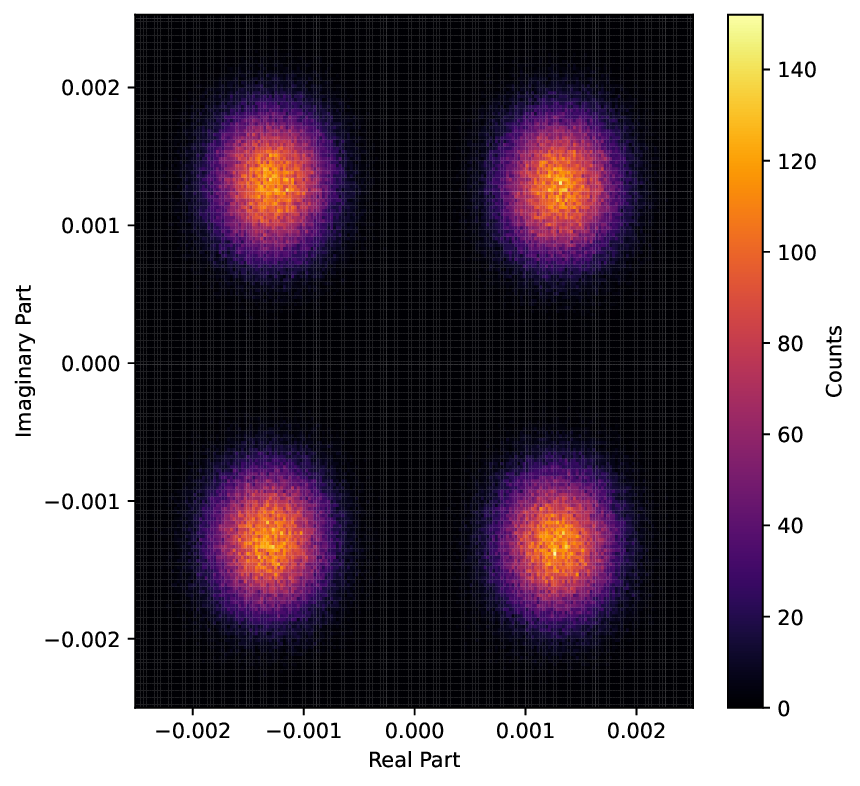}
        \caption{SAC, SNR = 12.11 dB}
        \label{fig:sim_1q_sac}
    \end{subfigure}
    \caption{Converged QPSK constellations at UE location (-20, -20), $M=100, \kappa=10,n_\text{r}=10$.}
    \label{fig:sim_1_qpsk}
\end{figure*}

\begin{figure*}[!htbp]
    \centering
    \begin{subfigure}[b]{0.49\linewidth}
        \centering
        \includegraphics[width=\linewidth]{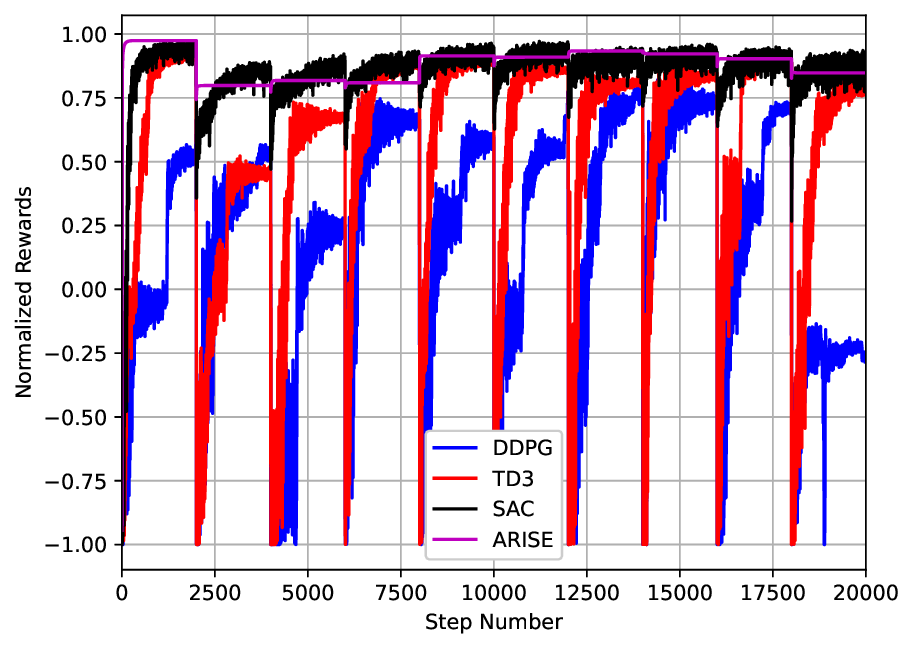}
        \caption{Normalized rewards $\eta_{\text{n}}$}
        \label{fig:sim_1_rewards_a}
    \end{subfigure}
    \begin{subfigure}[b]{0.49\linewidth}
        \centering
        \includegraphics[width=\linewidth]{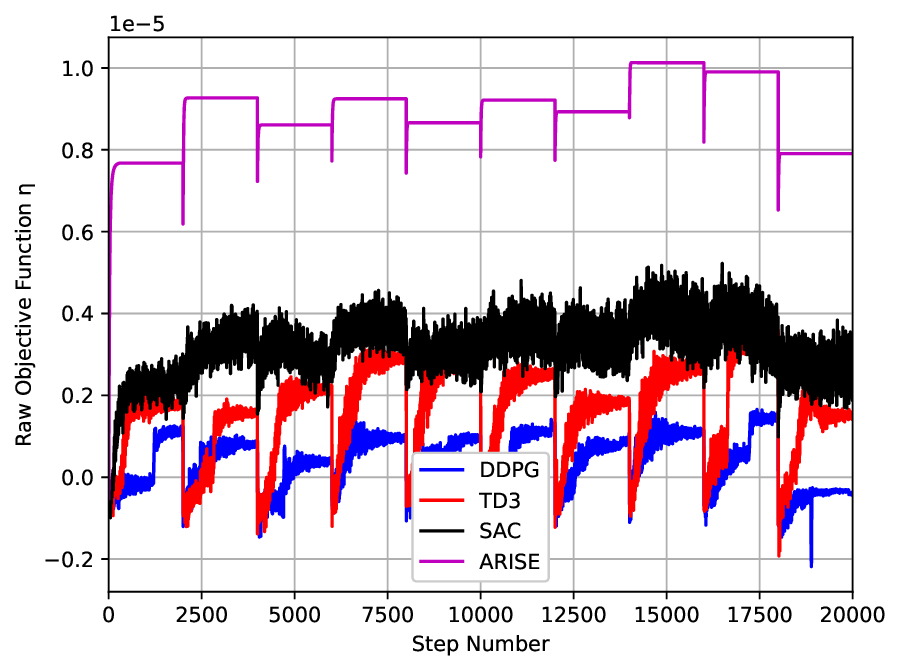}
        \caption{Objective function $\eta$}
        \label{fig:sim_1_rewards_b}
    \end{subfigure}
    \begin{subfigure}[b]{0.49\linewidth}
        \centering
        \includegraphics[width=\linewidth]{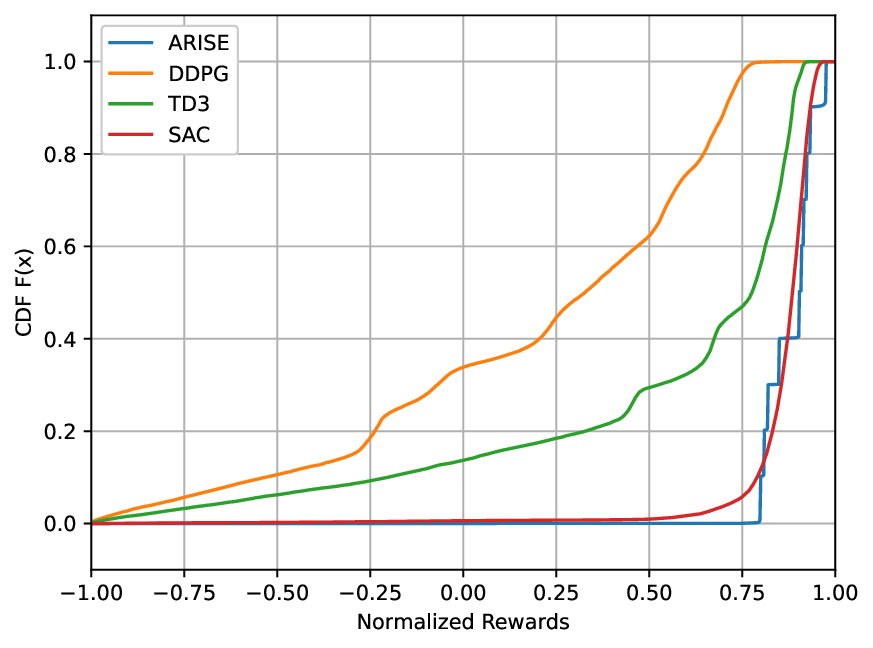}
        \caption{Normalized rewards $\eta_{\text{n}}$ CDF over time}
        \label{fig:sim_1_rewards_cdf_a}
    \end{subfigure}
    \begin{subfigure}[b]{0.49\linewidth}
        \centering
        \includegraphics[width=\linewidth]{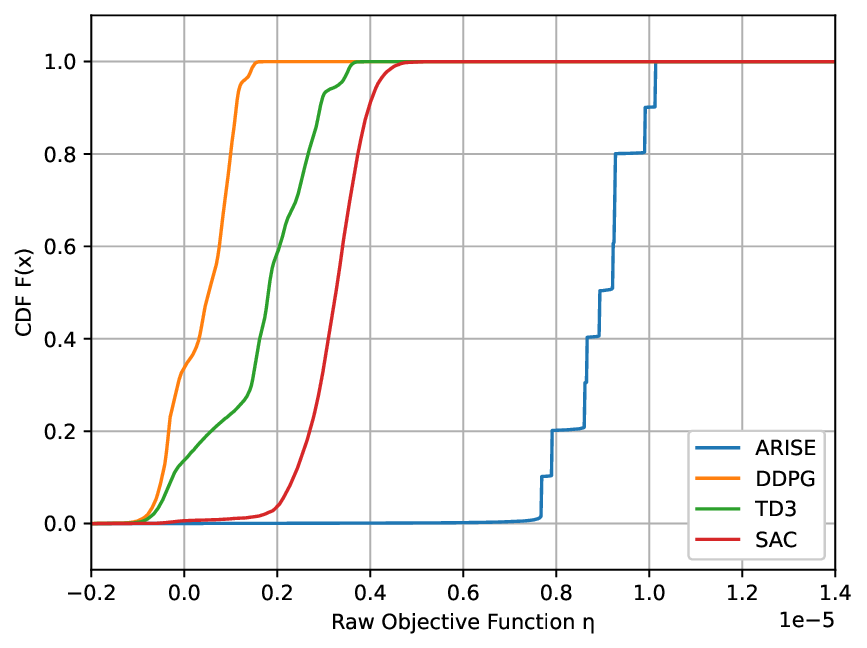}
        \caption{Objective function $\eta$ CDF over time}
        \label{fig:sim_1_rewards_cdf_b}
    \end{subfigure}
    \caption{Objective functions over time, $M=100, \kappa=10,n_{\text{r}}=10$.}
    \label{fig:sim_1_rewards}
\end{figure*}

We use the same simulation setup as in Section III-B to compare the performance of the DRL algorithms with that of ARISE. Based on the results in Section III-B, we use a scaling factor $\alpha_{\text{s}}=0.10$ for the ARISE algorithm to enhance its equalization capability and provide a strong baseline to compare against the DRL algorithms. Here, we define an ``episode'' as a channel coherence block. At the beginning of each episode, the UE moves according to a random walk. A ``time step'' corresponds to a change in the RIS reflection coefficients, after which the resulting pulse response and the corresponding reward are obtained. Each episode consists of 2000 time steps. We use reward scaling by multiplying each $r_{t+1}$ by 100 before saving it in the replay buffer to boost the learning rates of the networks. The replay buffers are updated in a first-in-first-out (FIFO) policy with a maximum buffer size $D_{\text{max}}$. At the beginning of each episode, we reset the replay buffers and learning rates of all DNNs, as well as the SAC temperature coefficient $\alpha$. Exclusively for DDPG, we also randomize the weights of the actor and critic networks to avoid remaining stuck at local optima. 

\begin{figure}[!t]
    \centering
    \includegraphics[width=0.7\linewidth]{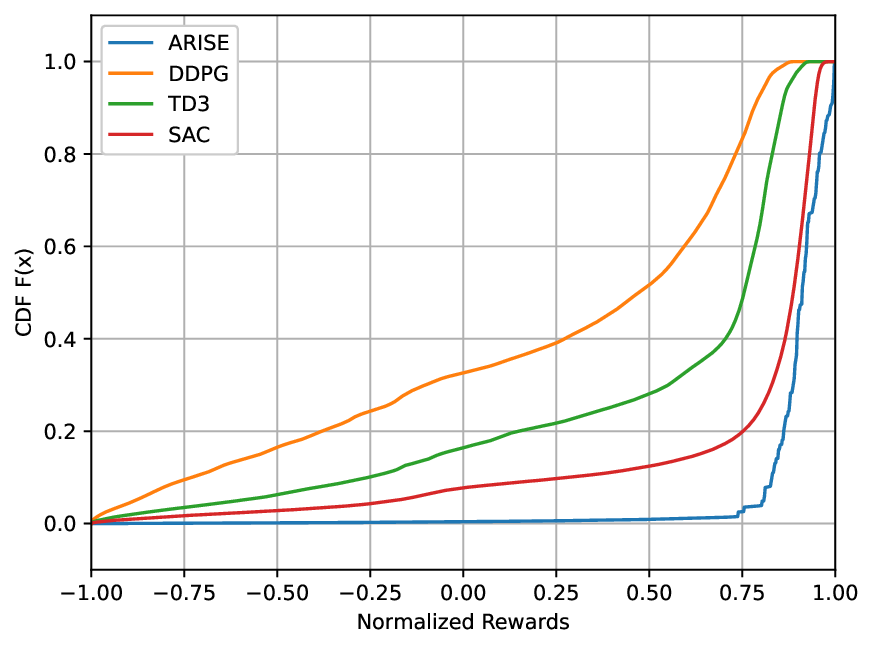}
    \caption{Normalized rewards $\eta_{\text{n}}$ for each step over the duration of 100 episodes, $M=100$, $\kappa=10$, $n_{\text{r}}=10$.}
    \label{fig:sim_2}
\end{figure}

The noise added to the actions for DDPG and TD3 is scaled each time step by $10^{-\tau_{\text{d}}t}$ to provide a decaying noise schedule which promotes exploration at the beginning of each episode, and more exploitation at the end of each episode to promote convergence stability, where $t$ is the current time step index, $t=0,1,\ldots,1999$, and resets at the beginning of each episode. Furthermore, we reduce the number of trainings per time step and actor learning rates depending on the average of the latest five consecutive rewards achieved by each algorithm. For instance, we start with $\mu_{\text{a}}=10^{-3}$ and $N_{\text{train}}=4$. When the average of 5 consecutive rewards are over 0.75, we set $\mu_{\text{a}} \gets 0.8\mu_{\text{a}}$, $N_{\text{train}} \gets N_{\text{train}}-1$, and increase the reward target by 0.05. We repeat the same for rewards being over 0.8, then 0.85, etc., while setting the minimum bound for $N_{\text{train}}$ to 1, to ensure the networks continue learning while promoting stability and reducing training iterations at higher reward values. All the other relevant implementation parameters are shown in Table~\ref{tab:drl_params}.

All actor and critic DNNs are implemented as shown in Fig.~\ref{fig:drl_model} with two hidden layers, with 512 nodes per hidden layer and ReLu activation functions \cite{8407425}. Specifically for DDPG, we employ layer normalization after each hidden layer of the actor and critic networks to improve the training stability of the algorithm. All layers are initialized using the Random Normal initializer with zero mean and 0.1 standard deviation \cite{Narkhede2022}. The networks are trained using TensorFlow Keras \cite{tensorflow2015} in Python, using the Adam optimizer \cite{kingma2017adammethodstochasticoptimization}.

We first analyze the performance of the DRL algorithms compared with the ARISE algorithm for a stationary UE located at $(-20,-20)$, with the assumption of Rayleigh fading for the BS-UE link, Rician factor $\kappa=10$ for the BS-RIS-UE link, and $n_{\text{r}}=10$ delayed paths (such that $L=20$ ISI components). The simulated scenario has 10 episodes where at the beginning of each episode, the NLoS channels change randomly. For fairness, we assume ARISE has access to the full cascaded CSI at the start of each episode, even though in practice this would require a dedicated channel estimation phase involving multiple RIS configurations before optimization could begin. The converged pulse responses after 10 episodes are plotted in Fig.~\ref{fig:sim_1_pulse} and their corresponding QPSK constellations in Fig.~\ref{fig:sim_1_qpsk}. Additionally, we show the normalized rewards $r_{t+1}$ and the objective function $\eta$ in Fig.~\ref{fig:sim_1_rewards}. We observe that the DRL algorithms all perform very differently from each other. SAC achieves equalization performance comparable and at times exceeding that of ARISE for the same channel coefficients, with very rapid convergence, taking less than 1000 steps at the beginning to converge to high reward values and staying at consistently high rewards over time afterwards, even when the channels are changing. Next, TD3 takes longer to converge than SAC, but seems to converge to high reward values after a few episodes and remain stable. DDPG is trailing behind and gets stuck at local optima, often unable to achieve the performance of TD3. Comparing the objective function $\eta$ realized by each algorithm, ARISE consistently attains the highest objective mainly due to its amplification capabilities when attempting to match an attenuated signal with an amplified and equalized version of itself. The rewards defined for the DRL algorithms did not explicitly require signal amplification because they are normalized with respect to the sum of the absolute pulse response samples. On average, the RIS optimized using ARISE has around 3 dB more power gain than the RIS optimized using SAC, even though both achieve near-identical equalization capabilities for the given scenario. This is particularly notable because SAC attains ARISE-level equalization without requiring the extensive pilot transmissions and full cascaded channel estimation that ARISE depends on, thereby reducing the communication overhead associated with RIS optimization.

\begin{figure}[!t]
    \centering
    \includegraphics[width=0.7\linewidth]{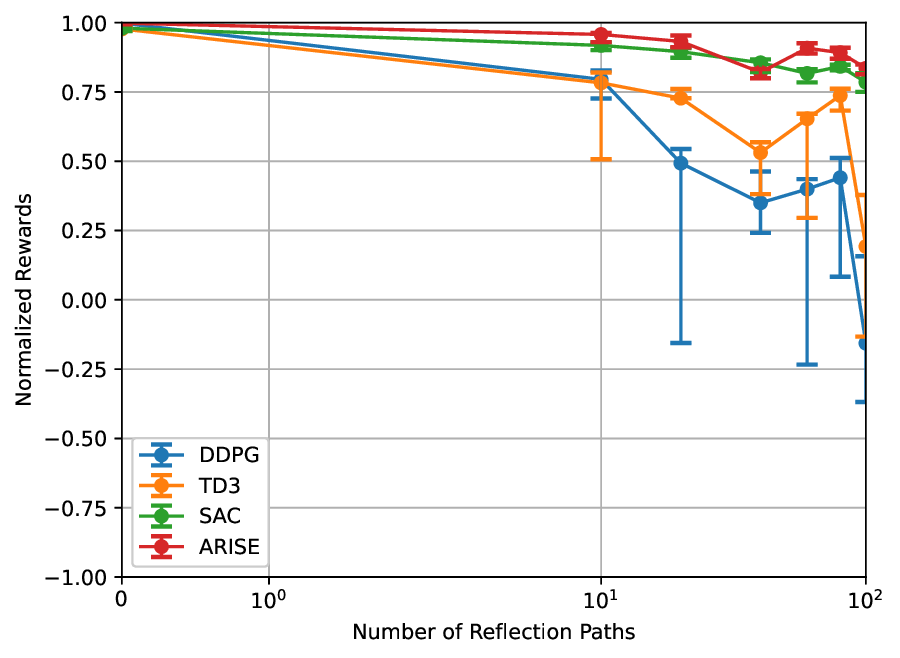}
    \caption{Converged normalized rewards $\eta_{\text{n}}$ versus number of delayed paths $n_{\text{r}}$, $M=100$, $\kappa=10$.}
    \label{fig:sim_3}
\end{figure}

Next, we employ the random walk model and compare the performance of each algorithm over more rigorous cases to evaluate their applicability to different RIS-based scenarios. In Fig.~\ref{fig:sim_2}, we use the random walk model of the UE for the duration of 100 episodes and compare the normalized rewards $\eta_{\text{n}}$ over time, still using the same RIS size and environment as in the previous case. Even though both the LoS and NLoS components change entirely at the beginning of each episode, SAC consistently maintains high and rapid equalization capabilities with normalized rewards higher than 0.75 for more than 80\% of the time, approaching the performance of ARISE. Unlike ARISE, which would require repeated channel re-estimation as the UE moves, SAC adapts directly from received pulse responses, avoiding this additional communication burden. TD3 converges slower than SAC but maintains high rewards most of the time, while DDPG trails behind.

\begin{figure}[!t]
    \centering
    \includegraphics[width=0.7\linewidth]{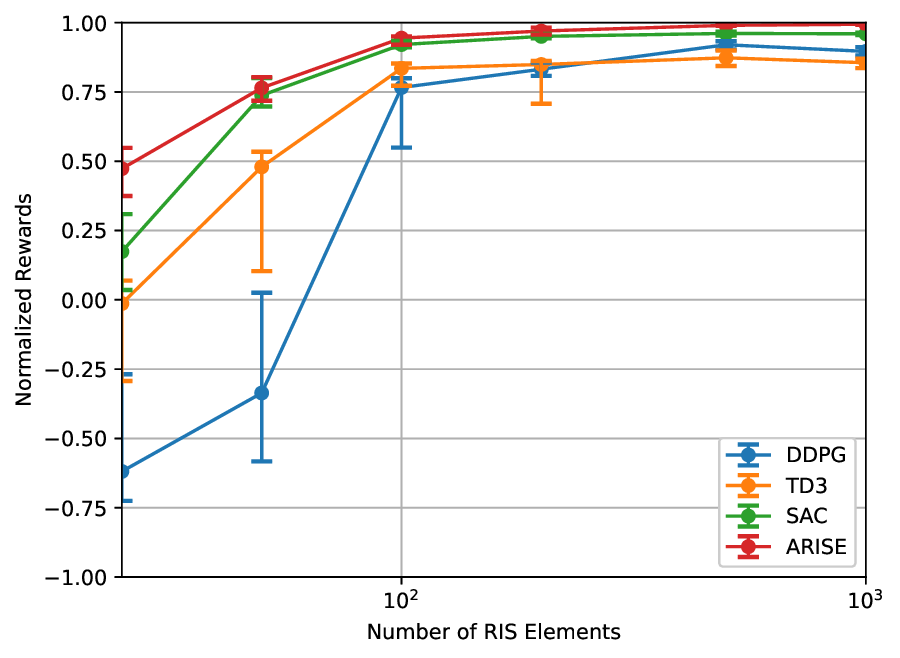}
    \caption{Converged normalized rewards $\eta_{\text{n}}$ versus number of RIS elements $M$, $n_{\text{r}}=10$, $\kappa=10$.}
    \label{fig:sim_4}
\end{figure}

Next, we compare the converged normalized rewards versus the number of delayed paths $n_{\text{r}}$ that the signals have, with a total number of ISI components given by $L=2n_{\text{r}}$ in Fig.~\ref{fig:sim_3}. Each data point represents the average normalized rewards from the last 10 steps of 10 consecutive episodes, with the error bars showing the standard deviations above and below the mean value. Here, SAC and ARISE still achieve similar performance with high stability, shown by the small standard deviations and high average normalized rewards, while TD3 and SAC have smaller rewards with SAC having more variance among episodes. Importantly, SAC maintains this performance without the need to estimate the increasing number of cascaded channel coefficients required by ARISE as the number of ISI components grows.

\begin{figure}[!t]
    \centering
    \includegraphics[width=0.7\linewidth]{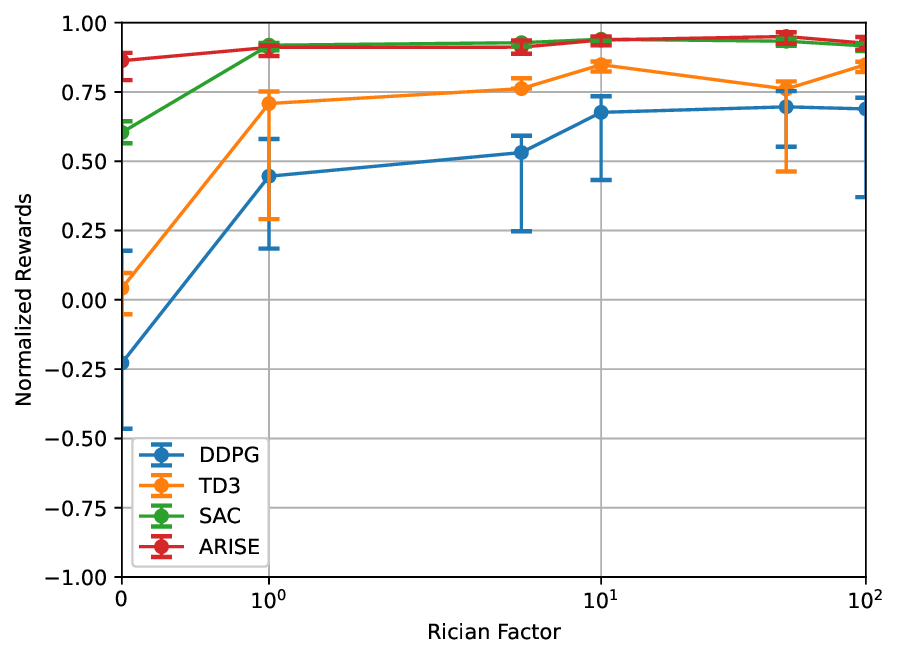}
    \caption{Converged normalized rewards $\eta_{\text{n}}$ versus Rician factor $\kappa$, $M=100$, $n_{\text{r}}=10$.}
    \label{fig:sim_5}
\end{figure}

The overall performance of the 100-element RIS as an equalizer remains consistent as more ISI terms are introduced to the signal. We evaluate the equalization capability of the DRL algorithms as the RIS scales in Fig.~\ref{fig:sim_4}. We note that modifying the number of RIS elements $M$ would alter the output dimension of the actor DNN, while changing the number of delayed paths $n_{\text{r}}$ may alter the input dimensions of the actor and critic DNNs (depending on the choice of the RIS operator to equalize a specific number of ISI terms). In our experiments, we modify the input and output dimensions of the actor and critic DNNs to correspond with each variable change, while keeping their hidden layers fixed. In the scenario in Fig.~\ref{fig:sim_4}, we observe that when the RIS size is small, its equalization capability is considerably lower than the previous simulated cases, across all algorithms, with ARISE leading. As the RIS size grows, the algorithms systematically achieve greater performance and increasing stability. Additionally, ARISE continues to improve marginally as the number of RIS elements increases above 200, whereas the DRL algorithms seem to plateau. This is most likely a consequence of the hidden layers of the DNNs maintaining the same sizes, struggling to adhere to the increasing number of random channels while being able to generalize enough to eliminate most of the ISI.

Finally, we assess the performance of each algorithm with respect to the Rician factor of the BS-RIS-UE link $\kappa$, in Fig.~\ref{fig:sim_5}. For the Rayleigh case, ARISE attains a considerably better performance than the DRL algorithms with $\eta_{\text{n}}\approx 0.85$, with DDPG being completely unable to equalize the signal, TD3 slightly better, and SAC consistently achieving an average value of $\eta_{\text{n}} \approx 0.6$. As the Rician factor increases and the DNNs are able to learn the geometry of the LoS components of the wireless channels, their performance improves significantly, with SAC achieving the same equalization performance as ARISE. The performance parity is achieved despite SAC operating without explicit CSI acquisition and RIS model, further underscoring its advantage in communication efficiency.

We overall demonstrate the ability of the DRL algorithms to successfully configure the RIS as an equalizer, with SAC achieving the same performance as ARISE in most cases and a comparable performance in others. Though requiring a higher per-step computational cost than TD3 and DDPG, it provides a means to quickly and efficiently boost the signal integrity of the link without requiring excessive channel estimation. Taken together, these results show that SAC achieves ARISE-level equalization accuracy while incurring substantially lower communication complexity and only moderate computational overhead, making it a more scalable solution for practical RIS deployments. 
\section{CONCLUSION}

In this work, we demonstrated the utility of RIS technology as both a signal booster and an equalizer across diverse channel conditions and wireless environments. We developed a steepest descent-based approach -- ARISE -- that leverages the estimated cascaded BS-RIS-UE channels and showed its ability to configure the RIS coefficients to equalize the received pulse response and enhance the SNR. We also examined the challenges of applying ARISE to passive RISs, particularly the extensive channel estimation required for accurate equalization and the resulting tradeoff between signal gain and equalization performance -- an effect that may necessitate iterative optimization or deeper SNR analysis.

As an alternative, we explored several DRL methods -- DDPG, TD3, and SAC -- that operate without channel estimation and rely solely on the received pulse response to optimize the RIS coefficients. Through comparative analysis of computational complexity and achievable performance, we found SAC to be the most effective, converging rapidly to stable solutions across varying channels and RIS sizes, and achieving performance comparable to ARISE. Crucially, SAC attains this performance without the heavy channel-estimation requirements of ARISE, resulting in significantly reduced communication complexity and a more practical overall optimization pipeline. Its model-free nature makes SAC inherently robust to RIS nonlinearities, element coupling, and frequency-dependent reflection behavior, which are difficult to capture in analytical channel models. Moreover, unlike ARISE -- which must wait for full CSI acquisition before optimization can begin -- SAC can refine the RIS configuration immediately from raw pulse response observations, enabling faster adaptation and reduced latency in dynamic environments. Overall, DRL emerges as a promising and potentially integral component of future RIS-assisted wireless systems due to its implementation simplicity, robustness, and strong performance.

Following the contributions of this study, future work should analyze the performance of ARISE and the DRL algorithms in the presence of multiple users, multiple-antenna base stations, and extend the RIS from a uniform linear array (ULA) to a uniform planar array (UPA), while including physical RIS models with hardware limitations and nonlinearities. These will generalize the concept of using RIS as an equalizer for a broader range of scenarios and advance its practicality in future wireless multiple-input multiple-output (MIMO) systems.
\appendices
\section{ARISE ALGORITHM DERIVATION}
We wish to minimize the mean squared error (MSE) between a desired signal $\boldsymbol{s}$ and the received signal $\boldsymbol{y}$
\begin{equation}
    \boldsymbol{y}=\left(\boldsymbol{h}_{\text{BU}} + \sum_{m=1}^{M}{\Gamma_m \boldsymbol{h}_{\text{BRU},m}}\right)*\boldsymbol{s}
\end{equation}
via gradient descent on the RIS reflection coefficients $\Gamma_m$ for $m=1,2,\ldots,M$. The cost function is defined by $J=E[\epsilon_k\epsilon_k^*]$ with error signal $\boldsymbol{\epsilon}=\boldsymbol{s}-\boldsymbol{y}$. Here, $\boldsymbol{s}=[1,0,\ldots,0]$ is the transmitted discrete-time pulse. We can express $\Gamma_m$ as the composition of its real and imaginary parts
\begin{equation}
    \Gamma_m=a_m+jb_m,
\end{equation}
thus expressing $\boldsymbol{y}$ as
\begin{equation}
    \boldsymbol{y}=\left( \boldsymbol{h}_{\text{BU}} + \sum_{m=1}^{M}{(a_m+jb_m) \boldsymbol{h}_{\text{BRU},m}}\right)*\boldsymbol{s}.
\end{equation}
The error gradient with respect to $\Gamma_m$ is defined as
\begin{equation}
    \frac{\partial\boldsymbol{\epsilon}}{\partial \Gamma_m}=\frac{\partial\boldsymbol{\epsilon}}{\partial a_m}+j\frac{\partial\boldsymbol{\epsilon}}{\partial b_m}
\end{equation}
and is related to $\boldsymbol{y}$ by
\begin{equation}
    \frac{\partial\boldsymbol{\epsilon}}{\partial a_m} = -\frac{\partial\boldsymbol{y}}{\partial a_m}; \quad \frac{\partial\boldsymbol{\epsilon}}{\partial b_m}=-\frac{\partial\boldsymbol{y}}{\partial b_m}.
\end{equation}
Expanding the convolution operator and considering each term in $\boldsymbol{y}$, $y_k$, we obtain
\begin{equation}
    \frac{\partial y_k}{\partial a_m}=\sum_{l=0}^{L}{h_{\text{BRU},m,l} s_{k-l}}
\end{equation}
and
\begin{equation}
    \frac{\partial y_k}{\partial b_m}=j\sum_{l=0}^{L}{h_{\text{BRU},m,l} s_{k-l}}.
\end{equation}
Differentiating the cost function yields
\begin{equation}
\begin{split}
    \frac{\partial J}{\partial a_m} & = E \left[ \frac{\partial \epsilon_k}{\partial a_m}\epsilon_k^* + \frac{\partial \epsilon_k^*}{\partial a_m}\epsilon_k \right] \\
    & = E \left[ -\sum_{l=0}^{L}{h_{\text{BRU},m,l} s_{k-l}}\epsilon_k^* - \sum_{l=0}^{L}{h_{\text{BRU},m,l}^* s_{k-l}^*}\epsilon_k \right]
\end{split}
\end{equation}
and
\begin{equation}
\begin{split}
    \frac{\partial J}{\partial b_m} & = E \left[ \frac{\partial \epsilon_k}{\partial b_m}\epsilon_k^* + \frac{\partial \epsilon_k^*}{\partial b_m}\epsilon_k \right] \\
    & = E \left[ -j\sum_{l=0}^{L}{h_{\text{BRU},m,l} s_{k-l}}\epsilon_k^* + j\sum_{l=0}^{L}{h_{\text{BRU},m,l}^* s_{k-l}^*}\epsilon_k \right].
\end{split}
\end{equation}
Combining, we get
\begin{equation}
    \frac{\partial J}{\partial \Gamma_m} = E \left[ -2 \sum_{l=0}^{L}{h_{\text{BRU},m,l}^* s_{k-l}^*}\epsilon_k \right]
\end{equation}
and the final gradient is
\begin{equation}
    \nabla_{\boldsymbol{\Gamma}}=-2 E\left[ \begin{array}{c}
         \sum_{l=0}^{L}{h_{\text{BRU},1,l}^* s_{k-l}^*} \\
         \sum_{l=0}^{L}{h_{\text{BRU},2,l}^* s_{k-l}^*} \\
         \vdots \\
         \sum_{l=0}^{L}{h_{\text{BRU},M,l}^* s_{k-l}^*}
    \end{array}\right]\epsilon_k.
\end{equation}
Finally, we have the update rule for $\boldsymbol{\Gamma}$ given by gradient descent
\begin{equation}
\begin{split}
    \boldsymbol{\Gamma}_{t+1} &=\boldsymbol{\Gamma}_t-\frac{1}{2}\mu \nabla_{\boldsymbol{\Gamma}_t} \\
    &= \boldsymbol{\Gamma}_t+\mu E \left[\sum_{l=0}^{L}{\boldsymbol{h}_{\text{BRU},l}^* s_{k-l}^*}\epsilon_k \right],
\end{split}
\end{equation}
where $\mu$ is the step size. To accommodate the passivity requirement of the RIS, we normalize the magnitudes of the reflection coefficients after every update by
\begin{equation}
    \Gamma_{t+1,m} \gets \frac{\Gamma_{t+1,m}}{|\Gamma_{t+1,m}|}
\end{equation}
for all $m=1,2,\ldots,M$.

\section{TD3 AND SAC ALGORITHMS PSEUDOCODES}

The pseudocode for the TD3 algorithm is given as Algorithm~\ref{alg:td3} and the pseudocode for the SAC algorithm is given as Algorithm~\ref{alg:sac} in this appendix.

\begin{algorithm}
    \caption{Twin Delayed DDPG for RIS Equalizer}
    \label{alg:td3}
    \begin{algorithmic}[1]
        \State {Initialize both training and target DNNs for both actor and critic, assign $\phi_{\text{targ}} \gets \phi, \quad \theta_{\text{targ}} \gets \theta$.}
        \For {$\text{episode}$ in $N_{\text{ep}}$}
            \For {$t$ in $N_{\text{steps}}$}
                \State {Observe state $s_t \gets \boldsymbol{y}_{t}'$ and set RIS coefficients \[\boldsymbol{\Gamma}_t'\gets\text{clip}(\mu_{\phi_{\text{targ}}}(\boldsymbol{y}_{t}')+\boldsymbol{\epsilon},-1,1), \quad \boldsymbol{\epsilon} \sim \mathcal{N}(0,\sigma_{\text{a}}^{2})^{2M \times 1}.\]}
                \State {Set $a_t \gets \boldsymbol{\Gamma}_t'$.}
                \State {Update RIS reflection coefficients as \[\boldsymbol{\Gamma}_t \gets [\Gamma_{t,1}'+j\Gamma_{t,2}',\ldots,\Gamma_{t,2M-1}'+j\Gamma_{t,2M}']^T\] and normalize using~(\ref{eq:ris_mag_norm}).}
                \State {Observe new state $s_{t+1}\gets\boldsymbol{y}_{t+1}'$.}
                \State {Calculate rewards $r_{t+1}$ using~(\ref{eq:reward_norm}).}
                \State {Store $(s_t,a_t,r_{t+1},s_{t+1})$ in the replay buffer $\mathcal{D}$.}
                \If {$|\mathcal{D}| \geq B$}
                    \For {$i$ in $N_{\text{train}}$}
                        \State {Sample batch $\mathcal{B}=\{(s,a,r',s')\}$ from $\mathcal{D}$.}
                        \State{Add noise to $a$: \[a \gets \text{clip}(a+\text{clip}(\epsilon,-c_e,c_e),-1,1)\quad \epsilon \sim \mathcal{N}(0,\sigma_{\text{t}}^2).\]}
                        \State {Calculate $l(\theta_{i})$ using~(\ref{eq:td3_critic_update}) for $i=1,2$.}
                        \State {Update critics: \[ \theta_i \gets \theta_i - \mu_{\text{c}} \nabla_{\theta_i} l(\theta_i). \]}
                        \State {Calculate $l(\phi)$ using~(\ref{eq:ddpg_actor_update}).}
                        \If {$i \mod{d_{\text{p}}} == 0$}
                            \State {Update actor: \[ \phi \gets \phi - \mu_{\text{a}} \nabla_{\phi} l(\phi). \]}
                            \State {Soft-update target critics: \[\theta_{\text{targ},i} \gets \tau \theta_{i} + (1-\tau)\theta_{\text{targ},i},\quad i=1,2.\]}
                            \State {Soft-update target actor: \[\phi_{\text{targ}} \gets \tau \phi + (1-\tau)\phi_{\text{targ}}.\]}
                        \EndIf
                    \EndFor
                \EndIf
            \EndFor
        \EndFor
    \end{algorithmic}
\end{algorithm}

\begin{algorithm}
    \caption{Soft Actor-Critic (SAC) for RIS Equalizer} \label{alg:sac}
    \begin{algorithmic}[1]
        \State {Initialize actor DNN hyperparameters $\phi$, training and target critic DNNs, assign $\theta_{\text{targ}} \gets \theta$.}
        \State {Initialize temperature parameter $\alpha$.}
        \For {$\text{episode}$ in $N_{\text{ep}}$}
            \For {$t$ in $N_{\text{steps}}$}
                \State {Observe state $s_t \gets \boldsymbol{y}_t'$.}
                \State {Sample action from stochastic policy: \[ a_t \sim \pi_{\phi}(a_t \mid s_t). \]}
                \State {Clip RIS coefficients: $\boldsymbol{\Gamma}_t' \gets \text{clip}(a_t,-1,1)$.}
                \State {Update RIS reflection coefficients as \[\boldsymbol{\Gamma}_t \gets [\Gamma_{t,1}'+j\Gamma_{t,2}',\ldots,\Gamma_{t,2M-1}'+j\Gamma_{t,2M}']^T\] and normalize using~(\ref{eq:ris_mag_norm}).}
                \State {Observe new state $s_{t+1}\gets\boldsymbol{y}_{t+1}'$.}
                \State {Calculate rewards $r_{t+1}$ using~(\ref{eq:reward_norm}).}
                \State Store $(s_t,a_t,r_{t+1},s_{t+1})$ in replay buffer $\mathcal{D}$.
                \If {$|\mathcal{D}| \geq B$}
                    \For {$i$ in $N_{\text{train}}$}
                        \State {Sample batch $\mathcal{B}=\{(s,a,r',s')\}$ from $\mathcal{D}.$}
                        \State {Sample next actions and log-probabilities: \[ a' \sim \pi_{\phi}(a' \mid s'), \quad \log\pi_{\phi}(a' \mid s'). \]}
                        \State {Calculate $l(\theta_{i})$ using~(\ref{eq:sac_critic_update}) for $i=1,2$.}
                        \State {Update critics: \[ \theta_i \gets \theta_i - \mu_{\text{c}} \nabla_{\theta_i} l(\theta_i) .\]}
                        \State {Sample actions for actor update: \[ a \sim \pi_{\phi}(a \mid s), \quad \log\pi_{\phi}(a \mid s). \]}
                        \State {Calculate $l(\phi)$ using~(\ref{eq:sac_actor_update}).}
                        \State {Update actor: \[ \phi \gets \phi - \mu_{\text{a}} \nabla_{\phi} l(\phi). \]}
                        \State {Calculate $l(\log{\alpha})$ using~(\ref{eq:sac_alpha_update}).}
                        \State {Update temperature: \[ \log\alpha \gets \log\alpha - \mu_{\alpha} \nabla_{\log\alpha} l(\log\alpha). \]}
                        \State {$\alpha \gets \exp{(\log{\alpha})}$.}
                        \State {Soft-update target critics: \[\theta_{\text{targ},i} \gets \tau \theta_{i} + (1-\tau)\theta_{\text{targ},i},\quad i=1,2.\]}
                    \EndFor
                \EndIf
            \EndFor
        \EndFor 
    \end{algorithmic} 
\end{algorithm}
\bibliographystyle{IEEEtran}
\bibliography{IEEEabrv,bibJournalList,bibfile}
\begin{IEEEbiography}[{
\vspace{-5mm}\includegraphics[width=1.0\textwidth]{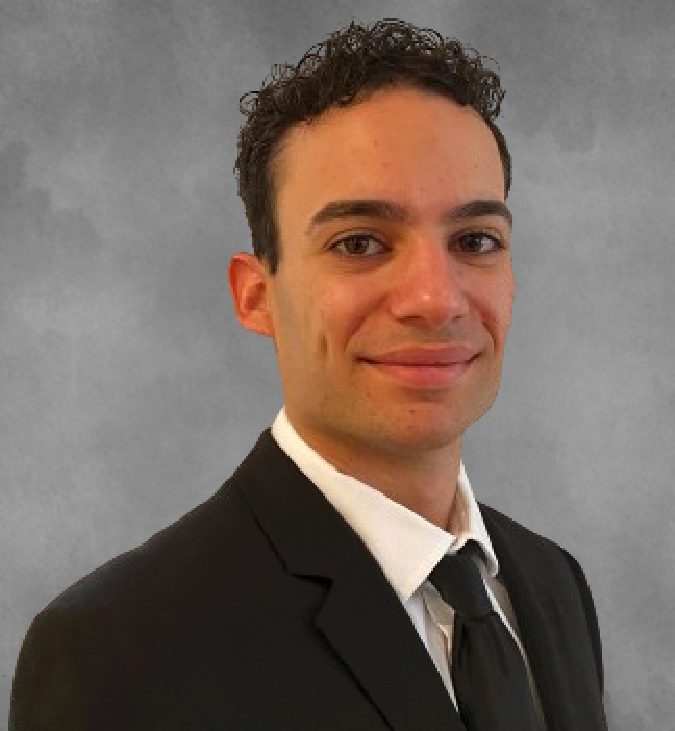}}]
{GAL BEN-ITZHAK}
\hspace{1em}(Graduate Student Member, IEEE) received the B.S. and M.S.
degrees in electrical engineering from the University of California, Irvine
(UCI), Irvine, CA, USA, in 2023 and 2024, respectively. He is currently
pursuing his Ph.D. degree in electrical engineering at UCI. His current
research and professional interests include high-speed communications, optimization, digital signal processing, high-speed circuit design, and signal integrity.
\end{IEEEbiography}

\begin{IEEEbiography}[{
\includegraphics[width=1.0\textwidth]{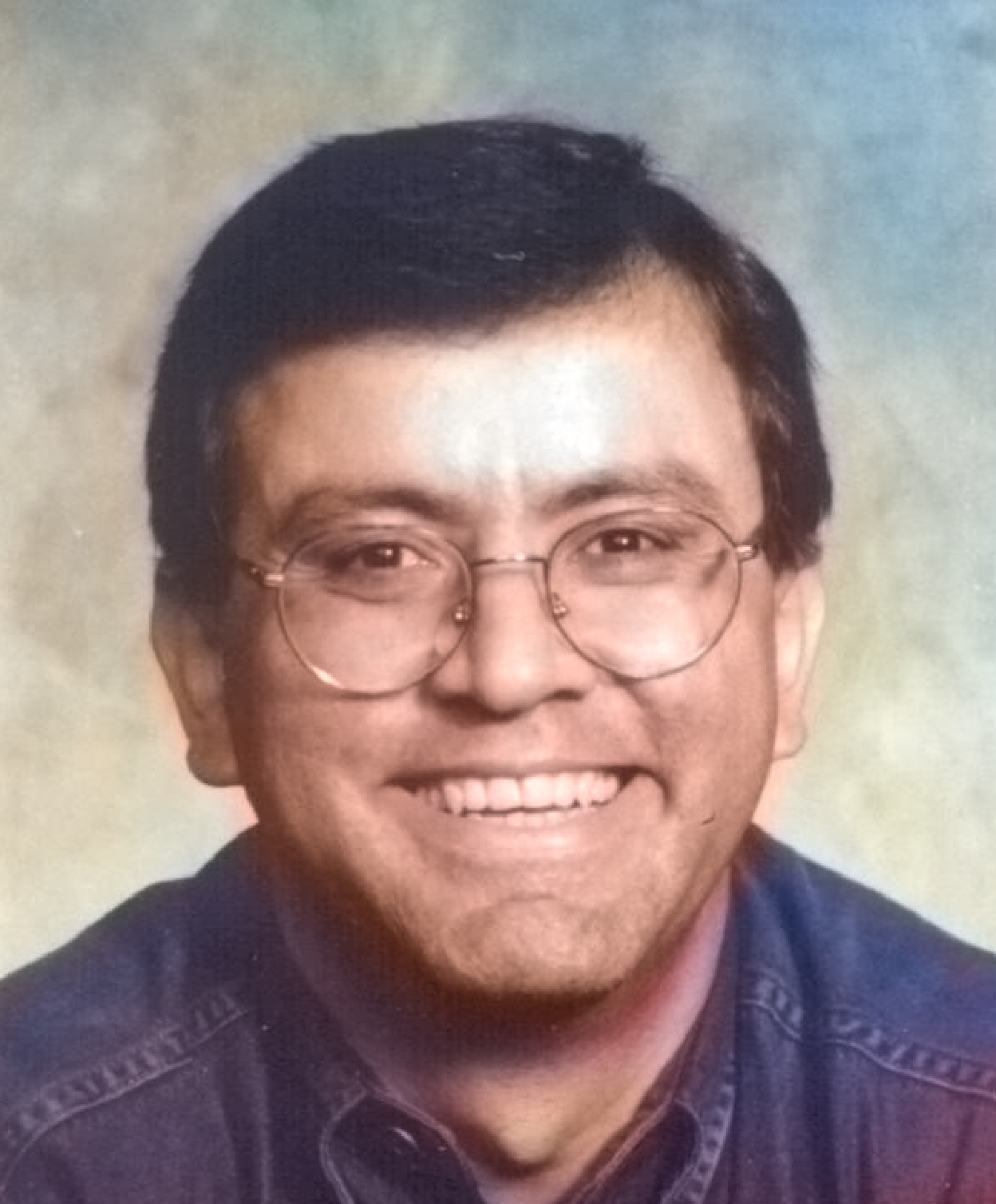}}]
{ENDER AYANOGLU}\hspace{1em}(Fellow, IEEE) received
the Ph.D. degree in electrical engineering from
Stanford University, Stanford, CA, USA, in 1986.
He was with the Bell Laboratories Communications
Systems Research Laboratory, Holmdel, NJ, USA.
From 1999 to 2002, he was a Systems Architect
with Cisco Systems Inc., San Jose, CA. Since 2002,
he has been a Professor with the Department of
Electrical Engineering and Computer Science,
University of California, Irvine, CA, where he was
the Director of the Center for Pervasive Communications
and Computing and the Conexant-Broadcom Endowed Chair,
from 2002 to 2010. He was a recipient of the IEEE Communications Society
Stephen O. Rice Prize Paper Award in 1995, the IEEE Communications
Society Best Tutorial Paper Award in 1997, and the IEEE Communications
Society Communication Theory Technical Committee Outstanding Service
Award in 2014. From 2000 to 2001, he was the Founding Chair of the IEEE-ISTO
Broadband Wireless Internet Forum, an industry standards organization.
He served on the Executive Committee for the IEEE Communications
Society Communication Theory Committee, from 1990 to 2002, and its
Chair, from 1999 to 2002. From 1993 to 2014, he was an Editor of IEEE
TRANSACTIONS ON COMMUNICATIONS. He was the Editor-in-Chief of IEEE
TRANSACTIONS ON COMMUNICATIONS, from 2004 to 2008, and the IEEE
JOURNAL ON SELECTED AREAS IN COMMUNICATIONS-Series on Green
Communications and Networking, from 2014 to 2016. He was the Founding
Editor-in-Chief of IEEE TRANSACTIONS ON GREEN COMMUNICATIONS
AND NETWORKING, from 2016 to 2020. He served as an IEEE Communications Society Distinguished
Lecturer in 2022-2023 and 2024-2025. In 2023, he received the IEEE Communications Society Joseph L. LoCicero
Award for outstanding contributions to IEEE Communications Society journals as Editor, Editor-in-Chief (EiC), and Founding EiC.
\end{IEEEbiography}
\vfill
\end{document}